\documentclass[a4paper,11pt]{article}
\pdfoutput=1 

\usepackage{jcappub} 

\usepackage[T1]{fontenc} 
\usepackage{subcaption} 
\usepackage{graphicx}
\usepackage{multirow}
\usepackage{epstopdf}
\usepackage[normalem]{ulem}
\usepackage{siunitx}
\DeclareSIUnit \parsec {pc}
\DeclareSIUnit \year {yr}
\usepackage{amsmath}
\DeclareMathOperator{\sinc}{sinc}

\title{\boldmath Measuring the spectrum of primordial gravitational waves with CMB, PTA and Laser Interferometers}

\author[a,b,c]{Paolo Campeti,\note{Corresponding author.}}
\author[d,e]{Eiichiro Komatsu,}
\author[a,b,c]{Davide Poletti}
\author[a,b,c]{and Carlo Baccigalupi}

\affiliation[a]{SISSA - Scuola Internazionale Superiore di Studi Avanzati,\\Via Bonomea 265, 34136, Trieste, Italy}
\affiliation[b]{IFPU - Institute for Fundamental Physics of the Universe, \\Via Beirut 2, 34014, Trieste, Italy}
\affiliation[c]{INFN - National Institute for Nuclear Physics, Sezione di Trieste,\\ Via Valerio 2, 34127, Trieste, Italy}
\affiliation[d]{Max Planck Institute for Astrophysics, Karl-Schwarzschild-Str.1, 85741 Garching, Germany}
\affiliation[e]{Kavli Institute for the Physics and Mathematics of the Universe (Kavli IPMU, WPI),
UTIAS, The University of Tokyo, Chiba, 277-8583, Japan
}

\emailAdd{paolo.campeti@sissa.it}
\emailAdd{komatsu@mpa-garching.mpg.de}
\emailAdd{davide.poletti@sissa.it}
\emailAdd{carlo.baccigalupi@sissa.it}

\abstract{We investigate the possibility of measuring the primordial gravitational wave (GW) signal across 21 decades in frequencies, using the cosmic microwave background (CMB), pulsar timing arrays (PTA), and laser and atomic interferometers. For the CMB and PTA experiments we consider the LiteBIRD mission and the Square Kilometer Array (SKA), respectively. For the interferometers we consider space mission proposals including the Laser Interferometer Space Antenna (LISA), the Big Bang Observer (BBO), the Deci-hertz Interferometer Gravitational wave Observatory (DECIGO), the $\mu$Ares experiment, the Decihertz Observatory (DO), and the Atomic Experiment for Dark Matter and Gravity Exploration in Space (AEDGE), as well as the ground-based Einstein Telescope (ET) proposal. We implement the mathematics needed to compute sensitivities for both CMB and interferometers, and derive the response functions for the latter from the first principles. We also evaluate the effect of the astrophysical foreground contamination in each experiment. We present binned sensitivity curves and error bars on the energy density parameter, $\Omega_{GW}h^2$, as a function of frequency for two representative classes of models for the stochastic background of primordial GW: the quantum vacuum fluctuation in the metric from single-field slow-roll inflation, and the source-induced tensor perturbation from the spectator axion-SU(2) inflation models. We find excellent prospects for joint measurements of the GW spectrum by CMB and space-borne interferometers mission proposals.
}

\begin{document}
\maketitle
\flushbottom

\section{Introduction}\label{sec:intro}
The cosmic inflation paradigm \cite{guth:1981,sato:1981,linde:1982,albrecht/steinhardt:1982,starobinsky:1980} predicts the primordial Stochastic Background of Gravitational Waves (hereafter SGWB) \cite{grishchuk:1975,starobinsky:1979}. 
In the standard picture the scalar and tensor perturbations are generated by the  quantum vacuum fluctuations during inflation \cite{starobinsky:1979,mukhanov/chibisov:1981,hawking:1982,guth/pi:1982,starobinsky:1982,abbott/wise:1984}.
The scalar modes are the seeds for the large-scale structure of the Universe and
have been subject to meticulous measurements (see e.g. \cite{planck_2018}), while the primordial tensor modes still remain undetected. The importance of their detection cannot be overstated, since the primordial SGWB would contain an unparalleled information on the very early Universe physics. If the single-field slow-roll inflationary scenario is confirmed, a detection of the tensor-to-scalar ratio $r$, i.e., the ratio of the tensor and scalar power spectra, can be used to directly infer the energy scale of inflation, allowing us to probe the ultra-high energy scales not accessible by terrestrial particle colliders \cite{Lyth:1998xn}.

There are (at least) three ways to search for the SGWB at widely separated frequencies: the cosmic microwave background (CMB) at $f\approx 10^{-18}-10^{-16}$~Hz, pulsar timing arrays at $f\approx 10^{-9}-10^{-7}$~Hz, and laser and atomic interferometers at $f\gtrsim 10^{-7}$~Hz (see \cite{Kamionkowski_Kovetz,Kramer:2013kea,Bartolo:2016ami} for reviews). 

For CMB, the primordial SGWB would imprint its signature in the B-mode polarization \cite{kamionkowski_kosowski_stebbins_1997,seljak_zaldarriaga_1997}, which is currently the most promising channel for a near-future detection. Numerous ground-based experiments are currently scanning the microwave sky in search of the primordial B-mode, among them the Background Imaging of Cosmic Extragalactic Polarization 2 (BICEP2)/Keck Array \citep{bicep_b_modes}, POLARBEAR/Simons Array \citep{polarbear_b_modes, sa_ref}, the Atacama Cosmology Telescope (ACT) \citep{act_b_modes}, the South Pole Telescope (SPT) \citep{spt_b_modes}, and the Cosmology Large Angular Scale Surveyor (CLASS) \citep{CLASS_paper}. Furthermore, the next decade will see a great increase in the efforts for detection with a new generation of experiments including the Simons Observatory (SO) \citep{so_ref}, the South Pole Observatory (SPO) and the Stage-IV network of ground-based observatories (CMB-S4) \citep{cmbs4_book, CMB-S4_project, CMB-S4_decadal}. As for space-borne experiments, the Japan Aerospace Exploration Agency has selected the LiteBIRD \citep{litebird_ref} as the second Strategic Large-class mission.

For pulsar timing arrays (hereafter PTA), the current generation experiments such as the Nanohertz Observatory for Gravitational Waves (NANOGrav) \cite{Arzoumanian:2015liz,NANOGrav_2018}, the European PTA \cite{Lentati:2015qwp} and the Parkes PTA \cite{Lasky:2015lej} are placing limits on the SGWB. In future the Square Kilometre Array (SKA) \citep{SKA_paper} will add to this international network of PTA.

For interferometers, the current generation of ground-based laser interferometers (LIGO \cite{Harry:2010zz}, VIRGO \cite{TheVirgo:2014hva}, KAGRA \cite{Somiya:2011np}) will be succeeded by the Cosmic Explorer (CE) \citep{CE_paper} and Einstein Telescope (ET) \citep{ET_paper}, operating between a few Hertz and a few kilo-Hertz. The space-borne
Laser Interferometer Space Antenna (LISA) \citep{LISA_astro2020_whitepaper, LISAxcosmo} will be probing in the milli-Hertz band. In addition there are a host of proposals for future space missions including the $\mu$Ares \citep{Ares_paper} in the micro-Hertz band; the Advanced Millihertz Gravitational-wave Observatory (AMIGO) \cite{Baibhav:2019rsa} in the milli-Hertz band; the Big Bang Observer (BBO) \citep{Crowder_Cornish,Smith_and_Caldwell_2016}, the Deci-hertz Interferometer Gravitational wave Observatory (DECIGO) \citep{Seto:2001qf,current_DECIGO_2020}, the Decihertz Observatory (DO) \citep{DO_paper}, and the Atomic Experiment for Dark Matter and Gravity Exploration in Space (AEDGE) \citep{AEDGE_paper}  in the deci-Hertz bands.

Combining these experiments, we can measure the SGWB spectrum across  $21$ decades in frequency. If we include indirect probes using the Big Bang Nucleosynthesis (BBN) and the number of relativistic degrees of freedom, the range extends to 29 decades \cite{Lasky:2015lej,adshead/etal:2020a}. This combination enables a detailed characterization of the SGWB that goes beyond the simple detection of $r$, which will be of utmost importance to determine if the detected primordial SGWB was sourced by the quantum vacuum fluctuations in the metric tensor, as in the single-field slow-roll scenario, or from alternative scenarios that can also produce the SGWB. In this context, the possibility of SGWB production from gauge fields, both Abelian \cite{sorbo:2011,barnaby/peloso:2011,barnaby/etal:2012,cook/sorbo:2012,cook/sorbo:2013,namba/etal:2015,shiraishi/etal:2016,Domcke:2016bkh, Ozsoy_2020} and non-Abelian \cite{Maleknejad_etal,Adshead_etal,adshead/etal:2013,Dimastrogiovanni_etal,maleknejad/sheikh-jabbari:2011,maleknejad/sheikh-jabbari:2013,maleknejad:2016,Dimastrogiovanni_fasiello_fujita_2016,obata/soda:2016,adshead/martinec/sfakianakis:2016,adshead/sfakianakis:2017,Agrawal_etal,agrawal/fujita/komatsu:2018b}, has been investigated in the literature. 

These sourced gravitational waves come with distinct observational signatures: they can be non-scale-invariant, partially chiral (circularly polarized), and strongly non-Gaussian. In this paper, we focus on the first signature, i.e., the spectrum of the SGWB, which can be blue, red, or with a bump. See the above list of references for the other two signatures.\footnote{The high-frequency GW produced by the gauge field during (p)reheating after inflation contributes to the effective number of relativistic degrees of freedom, which provides further constraints on the axion-U(1) models \cite{adshead/giblin/weiner:2018,adshead/etal:2020a,adshead/etal:2020b}.} 
Specifically, we seek to gather in one resource the expectations on the SGWB from the most promising future experiments, covering the whole frequency range of the GW spectrum, and study how they distinguish between the single-field slow-roll prediction and the SU(2) gauge field predictions. We build on the work of Ref.~\cite{Thorne} whose focus was on detection of chirality of the SGWB from the SU(2) gauge field (also see Ref.~\cite{Domcke_2020} for prospects to detect chirality of the SGWB by LISA and ET).

To this end, we try to use coherent assumptions for each experiment and, whenever possible, to derive the relevant quantities from the first principles using the latest available information in the literature.
We provide therefore a quick reference for both communities of cosmologists and GW astronomers for the sensitivities of future experiments capable of detecting a SGWB, summarizing the mathematical tools needed to compute such sensitivities for both the CMB and the interferometers. Finally, we show our results in a coherent manner by plotting error bars representing the uncertainty on the binned tensor power spectrum for each experiment\footnote{Python code is available at the link \url{https://github.com/pcampeti/SGWBProbe}}. For example, we derive forecasts for the precision on the tensor-to-scalar ratio $r$ and the tensor spectral index $n_{T}$, for the combination of CMB B-modes experiments and laser interferometers (LiteBIRD+LISA and LiteBIRD+BBO), using a Monte Carlo Markov Chain exploration of the full cosmological parameters space.

We differentiate our work from the previous literature in three ways. First, we provide frequency-integrated error bars from the binned sensitivity curves for all the detectors. Second, we include astrophysical foregrounds for all experiments. Finally, we use the latest and realistic CMB sensitivity curves for the LiteBIRD mission, including state-of-the-art simulations for the CMB foregrounds.  

The paper is organized as follows.
In Section \ref{sec:tensor_ps} we describe the two theoretical tensor power spectrum models for which we will provide forecasts in the subsequent sections: the single-field slow-roll model and the spectator axion-SU(2) model. In Section \ref{sec:cmb} we discuss the experimental setup for the CMB B-mode experiment LiteBIRD, including the instrumental noise, the lensing contribution and the astrophysical foregrounds contamination. In Section \ref{sec:direct} we construct the instrumental sensitivity curves for the  interferometers and illustrate the effect of the astrophysical foregrounds on each  experiment as well as on the PTA. Section \ref{sec:results} is dedicated to the discussion of our results concerning forecasts on the sensitivity of all the experiments for the spectator axion-SU(2) and single-field slow-roll models. We also present the updated forecasts on the tensor spectral index $n_T$ exploiting the combination of CMB experiments and laser interferometers. We conclude in Section \ref{sec:conclusions} with future perspectives. 

\section{Theoretical Models for the Primordial Tensor Power Spectrum}\label{sec:tensor_ps}
In this Section we review the theoretical models of the primordial tensor power spectrum for which we will provide forecasts in the rest of the paper. We consider two possibilities in this respect: one is the nearly scale-invariant tensor power spectrum predicted in the context of the single field-slow roll inflation, while the other is the one produced by the spectator axion-SU(2) model \cite{Dimastrogiovanni_fasiello_fujita_2016}.

\subsection{Single-Field Slow-Roll Model}\label{sec:standard_ps}
In the single-field slow-roll inflationary scenario \cite{linde:1982,albrecht/steinhardt:1982,starobinsky:1980},  cosmological scalar \cite{mukhanov/chibisov:1981,hawking:1982,guth/pi:1982,starobinsky:1982} and tensor \cite{starobinsky:1979,abbott/wise:1984} perturbations are produced by the quantum vacuum fluctuations. The power spectrum for the scalar perturbations is parametrized by a power-law
${\mathcal P}^{vac}_{{\mathcal 
		R}}(k)= A_{S}
\left(k / k_0\right)^{n_{{S}}-1}\,$,
where $A_{S}$ is the amplitude of the scalar perturbations, $n_S$ the scalar spectral index, $k$ the wavenumber of the perturbation, $k_{0}=0.05\,\SI{}{\per\mega\parsec}$ the pivot-scale and the superscript $vac$ indicates that it is produced by the  quantum vacuum fluctuations. The same applies to the tensor power spectrum  
\begin{equation}
{\mathcal P}^{vac}_{T}(k)= A_{T}
	\left(\frac{k}{k_0}\right)^{n_T +\frac{1}{2}\alpha_{T}\ln(k/k_0)},
\end{equation}
 where $A_{T}$ is the amplitude of the tensor perturbations, $n_T$ the tensor spectral index and $\alpha_{T}=d n_{T}/d\ln k$ its running. We then define the tensor-to-scalar ratio $r$ as
$r=A_{T} / A_{S}$. We also enforce the inflationary consistency relation in single-field slow-roll inflation \cite{Lyth:1998xn}, connecting the spectral index and the amplitude of the tensor spectrum as $n_{T}=-r/8$, while its running satisfies the relation $\alpha_{T} = (r/8) \left[(n_{S}-1)+ r/8 \right]$.  

Currently no detection of $r$ exists and there are only upper limits available. The best limits come from CMB experiments, $r<0.06$ at $95\%$ CL, from the combination of the B-mode polarization data of the BICEP2/Keck \citep{bicep2_2018} and Planck 2018 data \citep{planck_2018}.

\subsection{Spectator Axion-SU(2) Model}\label{sec:axion_model}

Gauge fields are ubiquitous in physics and can affect the predictions of inflation (see \cite{Maleknejad_etal} for a review). In this paper we consider the SGWB produced in the spectator axion-SU(2) model \cite{Dimastrogiovanni_fasiello_fujita_2016} based on the ``chromo-natural'' inflation model \cite{adshead/wyman:2012}. This model has the Lagrangian 
\begin{equation}
\label{eq:action}
    \mathcal{L}=\mathcal{L}_{inflaton}+\frac{1}{2}\left(\partial_{\mu}\chi\right)^{2}-\mu^{4}\left[1+\cos\left(\frac{\chi}{f}\right)\right]-\frac{1}{4}F_{\mu\nu}^{a}F^{a\mu\nu}+\frac{\lambda}{4f}\chi F_{\mu\nu}\Tilde{F}^{a\mu\nu},
\end{equation}
where $\mathcal{L}_{inflaton}$ represents a generic inflaton sector generating the quasi-de Sitter expanding background and the curvature perturbations in agreement with the current CMB observations, $\chi$ is a pseudo-scalar axion field with a cosine-type potential, $\mu$ and $f$ are dimensionful parameters and $\lambda$ is a dimensionless coupling constant for the axion and gauge fields. The field strength tensor of the SU(2) gauge field is given by $F_{\mu\nu}^{a}=\partial_{\mu}A^{a}_{\nu}-\partial_{\nu}A^{a}_{\mu}-g \epsilon^{abc}A^{b}_{\mu}A^{c}_{\nu}$ with $g$ being the gauge field self-coupling constant, and $\Tilde{F}^{a\mu\nu}$ is its dual. We ignore the effect of the gravitational Chern-Simons term $R\tilde{R}$ because its effect on the SGWB is sub-dominant compared to the $F\tilde{F}$ term \cite{mirzagholi/etal:2020}.

During inflation the SU(2) gauge field establishes a homogeneous and isotropic vacuum expectation value, 
$\bar A_i^b=a(t)Q(t)\delta^b_i$ \cite{maleknejad/sheikh-jabbari:2011,maleknejad/sheikh-jabbari:2013}, which is an attractor solution \cite{maleknejad/erfani:2014,domcke/etal:2019,wolfson/maleknejad/komatsu:2020}. The perturbation around this value contains scalar, vector, and tensor modes \cite{maleknejad/sheikh-jabbari:2011,maleknejad/sheikh-jabbari:2013}, and the tensor mode linearly mixes with gravitons to produce the SGWB. In particular, the gauge field produces a chiral SGBW with either left- or right-handed circular polarization, depending on which circular polarization mode experiences a transient growth near horizon crossing
\cite{Adshead_etal,adshead/etal:2013,Dimastrogiovanni_etal,Maleknejad_etal}. 

Assuming that only left-handed polarized GWs are produced, we can write the sourced contribution to the tensor spectrum as \citep{Thorne}
   \begin{align}
        \mathcal{P}_{T}^{L,\, Sourced} (k) &= r_{*} \mathcal{P}_{\mathcal{R}}(k) \exp\left[-\frac{1}{2\sigma^{2}} \ln^{2}\left(\frac{k}{k_{p}}\right) \right], \\
        \mathcal{P}_{T}^{R,\, Sourced} (k) &\simeq 0,
    \end{align}
where  $\mathcal{P}_{\mathcal{R}}$ is the scalar curvature perturbation power spectrum, the parameter $r_{*}$, which is the tensor-to-scalar ratio at the peak scale $k=k_{p}$, controls the amplitude of the tensor power spectrum, and the parameter $\sigma$ controls the width of the Gaussian-shaped feature produced in the spectrum by this model. These parameters are related to the model parameters given in  Eq.~\ref{eq:action} (see below).  This form of the tensor power spectrum is valid for the cosine potential given in Eq.~\ref{eq:action} as well as for axion potentials with an inflection point \cite{fujita/sfakianakis/shiraishi:2019}.

The total tensor spectrum will be the sum of the sourced and the vacuum contributions
      \begin{align}
       \mathcal{P}_{T}(k, k_{p}, r_{*}, \sigma) &= \mathcal{P}_{T}^{vac}(k) + \mathcal{P}_{T}^{Sourced}(k, k_{p}, r_{*}, \sigma),\\
    \mathcal{P}_{T}^{Sourced}(k, k_{p}, r_{*}, \sigma) &= \mathcal{P}_{T}^{L,\, Sourced} (k) + \mathcal{P}_{T}^{R,\, Sourced} (k),
   \end{align}
while the contribution of the axion and SU(2) gauge fields to $\mathcal{P}_{\mathcal{R}}$ is negligible with respect to the vacuum one for an appropriate choice of the model parameters\footnote{We do not include the non-linear scalar curvature perturbation induced by the gauge field. This can be very large in the original chromo-natural model in which the axion plays the role of inflaton \cite{papageorgiou/peloso/unal:2018,domcke/etal:2019}. This contribution in the spectator axion-SU(2) model is smaller \cite{Agrawal_etal}, but may still affect the allowed parameter space in which the sourced GW is comparable to or larger than the vacuum contribution \cite{papageorgiou/peloso/unal:2019}. There is also a possibility of having a non-negligible contribution to the scalar sector for a very large $\sigma$ parameter choice, if the energy fraction of the axion grows after inflation and the axion decays faster than the inflaton \citep[see][ and references therein]{Thorne}.}, i.e., $m_{Q}\equiv gQ/H\geq \sqrt{2}$ where $H$ is the Hubble expansion rate during inflation \citep{Dimastrogiovanni_etal,Dimastrogiovanni_fasiello_fujita_2016}; thus, $\mathcal{P}_{\mathcal{R}}(k)= \mathcal{P}^{vac}_{\mathcal{R}}(k)$.  The parameters $\left\{r_{*}, k_{p}, \sigma \right\}$ can be connected to the physical parameters in the model Lagrangian  $\left\{g, \lambda, \mu, f \right\}$ \cite{Thorne,fujita/sfakianakis/shiraishi:2019}.
The peak wavenumber $k_p$ corresponds to the time $t_*$ at which $\chi$ is at the inflection point of the potential, $\chi(t_*)=\pi f/2$. The value of $m_Q$ is given by $m_*\equiv m_Q(t_*)=(g^2\mu^4/3\lambda H^4)^{1/3}$. The other relevant dimensionless variable is $\xi_*\equiv \lambda\dot\chi(t_*)/(2fH)\approx m_*+m_*^{-1}$.  With these variables, we can write 
$k/k_p=e^{H(t-t_*)}$, $\sigma^2=(\lambda/2\xi_*)^2/[2{\cal G}(m_*)]$, and ${\cal G}(m_*)\approx 0.666+0.81m_*-0.0145m_*^2-0.0064m_*^3$. The effective tensor-to-scalar ratio at the peak scale $r_{*}$ can also be related to the model parameters, but in principle can assume any positive value, while the width of the Gaussian feature $\sigma$ is bounded by the peak scale choice $k_{p}$ because of the attractor behaviour of the background axion field coupled to the SU(2) gauge fields. 

In the rest of this paper we will consider three sets of parameters:
\begin{equation}
\label{eq:params}
\left\{r_{*}, k_{p}, \sigma \right\} = \left\{400, 10^{15}\,\SI{}{\per\mega\parsec}, 9.1\right\},\quad \left\{0.15, 10^{11}\,\SI{}{\per\mega\parsec}, 8\right\}, \quad \left\{50, 10^{6}\,\SI{}{\per\mega\parsec},4.8\right\}\,,
\end{equation}
and we will refer to them as \textit{AX1}, \textit{AX2} and \textit{AX3} models, respectively. 
For all cases we will assume the vacuum contribution to the tensor-to-scalar ratio of $r_{vac}=10^{-5}$ \citep{Thorne}, although this choice might be subject to backreaction of particle production of the gauge field \cite{maleknejad/komatsu:2019,papageorgiou/peloso/unal:2019}. To avoid this we can simply assume a larger value for $r_{vac}$, which would add the scale-invariant component to all the figures we show in this paper.

 We chose the parameters given in Eq.~\ref{eq:params} to provide representative examples for our analysis. The first set of parameters represents a tensor spectrum model that is simultaneously detectable by both CMB and laser interferometers, while still satisfying the upper bound provided by the BICEP2/Keck/Planck analysis (see the end of Section \ref{sec:standard_ps}). The second set produces instead a spectrum that is just outside the reach of LiteBIRD and at the same time comfortably detectable by the advanced interferometers $\mu$Ares, DECIGO and BBO, thanks to the large bump feature produced at  $k_{p}=10^{11}\SI{}{\per\mega\parsec}$. The third parameter set produces a spectrum that peaks in the PTA experiments frequency range while still being compatible with the BICEP2/Keck/Planck upper limit in the CMB range. Due to the relationship between $\sigma$ and $k_{p}$, which tends to flatten out the spectrum, we could get an SGWB detectable by SKA only in the case without foreground contamination (Section \ref{sec:results}).
 
\begin{figure}
\centering
        \includegraphics[scale=1.2]{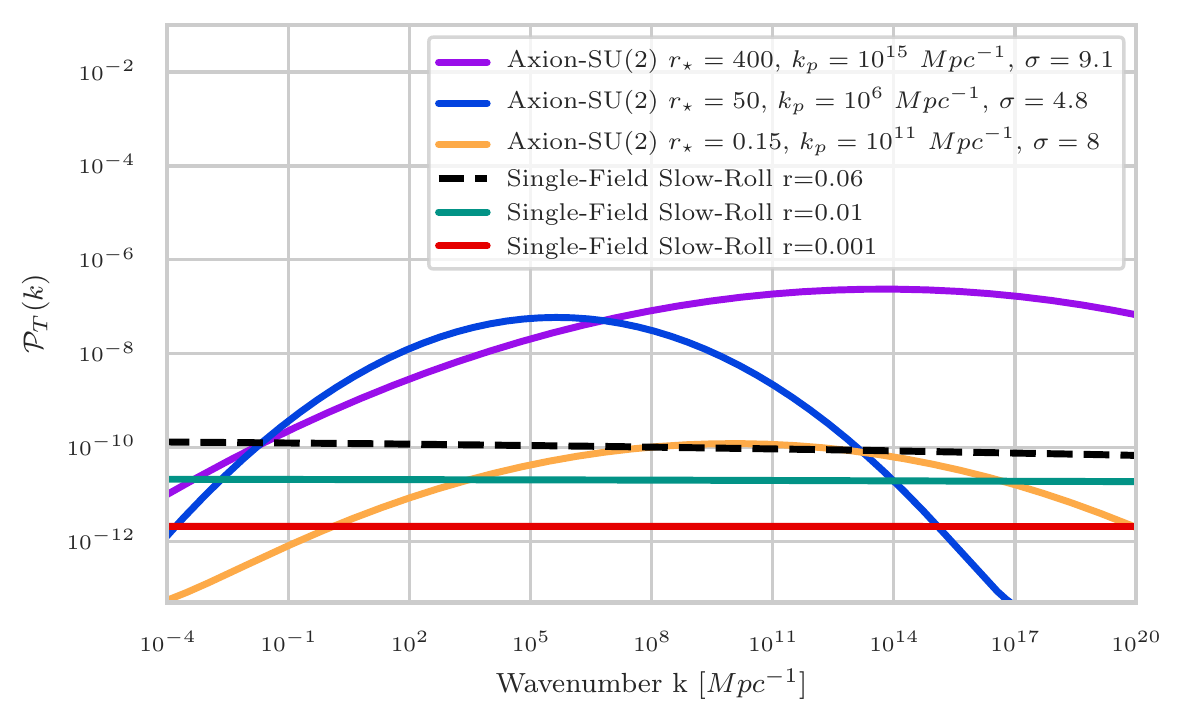}
	\caption{Tensor power spectra as a function of the wavenumber $k$ for the five representative cases considered in this paper, plus a standard $r=0.06$ model representing the CMB upper bound from the BICEP2/Keck/Planck combined analysis (black dashed curve).}\label{fig:spectra}
\end{figure}

 In Figure \ref{fig:spectra} we show the tensor power spectra $\mathcal{P}_{T}$ as a function of the wavenumber $k$ for the five cases considered in this paper.
 We have checked that all models are consistent with the current CMB shortwave and second-order back-reaction \citep{Clarke_2020}, indirect upper bounds \citep{Cabass_2015}, PTA limits \citep{NANOGrav_2018} and ground-based interferometers LIGO/Virgo \citep{LIGO_SGWB_upper_limit_2019} limits.
 
\subsection{Gravitational Wave Energy Density}
A quantity commonly used in the literature to show the sensitivities of GW observatories is the fractional energy density in GWs at the present (conformal) time
 $\tau_{0}$ \cite{kolb/turner:1990}\footnote{Throughout this paper we adopt the notation $c=1$ unless stated otherwise.} 
\begin{equation}\label{eq:omega}
				    \Omega_{GW}(k, \tau_{0}) = \frac{1}{\rho_{c}(\tau_0)}\frac{\partial \rho_{GW}(k,\tau_0)}{\partial\ln k}=\frac{\mathcal{P}_{T}(k)}{12 H_{0}^{2}}\cdot \left[\mathcal{T}^{'}(k, \tau_0)\right]^{2}.
\end{equation}
In the equation above $\rho_{c}$ is the critical energy density of the Universe and $\rho_{GW}$ the energy density of GWs, given by 
 $   \rho_{GW} = \langle h'_{ab} h'^{ab} \rangle/(32\pi a^{2}G)$,
where the tensor $h_{ab}$ represents the GW metric perturbation and the $'$ indicates the conformal time derivative.  
The second equality in Eq.~\ref{eq:omega} can be obtained from the definition of the tensor power spectrum and by expressing the time evolution of the primordial GW amplitude -- solution of the linearized Einstein equation --  in terms of the GW transfer function $\mathcal{T}(k,\tau)$ \citep[see][and references therein]{Watanabe_Komatsu}.
In the rest of this paper, we will use approximate analytical expressions for $\Omega_{GW}$, derived in    
\citep{Watanabe_Komatsu}, and valid for two different regimes: 
\begin{equation}\label{eq:analytical_omega_gw}
    \Omega_{GW}(k, \tau_{0}) = \frac{\mathcal{P}_{T}(k)}{12 H^{2}_{0}} k^{2}\cdot
    \begin{cases}
    \begin{aligned}
      &\frac{\tau^{2}_{*}}{\tau_{0}^{2}}\left[A(k)j_{2}(k\tau_{0})+B(k)y_{2}(k\tau_{0})\right]^{2},  \quad \text{if}\ k>k_{*}, \\[1ex]
      &\left[\frac{3j_{2}(k\tau_{0})}{k\tau_{0}}\right]^{2},  \quad \text{if}\  k<k_{*},\\
       \end{aligned}
    \end{cases}
 \end{equation}
where $\tau_{*}$ is the conformal time at the epoch of the matter-radiation equality and $k_{*}\approx 1/\tau_{*} $ is the comoving wavenumber of the modes that entered the horizon at that time, $j_{n}$, and $y_{n}$ with integer $n$ are the spherical Bessel functions of first and second kind, respectively, and the functions $A(k)$ and $B(k)$ are given by
\begin{align}
    A(k) &= \frac{3}{2k\tau_{*}}-\frac{\cos(2k\tau_{*})}{2k\tau_{*}}+ \frac{\sin(2k\tau_{*})}{(2k\tau_{*})^{2}},\\
    B(k) &= -1 + \frac{1}{(k\tau_{*})^{2}}-\frac{\cos(2k\tau_{*})}{(k\tau_{*})^{2}}-\frac{\sin(2k\tau_{*})}{2k\tau_{*}}.
\end{align}
As noted in Ref.~\citep{caprini_figueroa}, the analytical solution \ref{eq:analytical_omega_gw} reproduces well the fully numerical solution of the Einstein equation if we choose $\tau_{*}$ as the time  at which the scale factors for a matter-dominated Universe $a_{m} = H_{0}^{2}\Omega_{m} a_{0}^{3}\tau^{2}/4$ and the one for a radiation-dominated one $a_{r} =  H_{0}\sqrt{\Omega_{r}} a_{0}^{2}\tau$ cross. Setting $a_{m}(\tau_{*}) = a_{r}(\tau_{*})$ gives $\tau_{*} \simeq \SI{420}{\mega\parsec}$, corresponding to $k_{*}\approx 1/\tau_{*} \approx 2.38\times 10^{-3} \SI{}{\per\mega\parsec}$.

Following Refs.~\citep{Watanabe_Komatsu,caprini_figueroa}, we can approximately quantify the suppression effect on $\Omega_{GW}(k,\tau_{0})$ due to the change in effective relativistic degrees of freedom, by multiplying Eq.~\ref{eq:analytical_omega_gw} with the factor $\left(g_{*}/g_{*0}\right)\left(g_{s*}/g_{s*0}\right)^{-4/3}$, where $g_{*}$ and $g_{s*}$ are the effective number of relativistic particle species contributing to the energy density and the entropy, respectively. Furthermore, free-streaming neutrinos damp $\Omega_{GW}$ in the range $\approx 10^{-16} - 10^{-10}\,\SI{}{\hertz}$ \citep{Watanabe_Komatsu}: this effect is included in the angular power spectra used for our CMB forecasts (Section \ref{sec:fisher}), while it can be safely neglected  for all other experiments, since they probe frequencies $\gtrsim 10^{-9}\,\SI{}{\hertz}$. Finally, we checked that our resulting analytical formula for $\Omega_{GW}(k,\tau_0)$ approximates well (within $\sim 6 \%$) the fully numerical solution to the Einstein equation at PTA and interferometers scales given in \citep{saikawa_shirai_2018}.

 In the following we will often pass from the GW wavenumber $k$ to the frequency $f$ of the GW today, which are related to each other via $k=2\pi f/c$ (here we reinstate the factor $c$), 
 \begin{equation}
     \frac{k}{\SI{}{\per\mega\parsec}}= 6.5\times10^{14} \frac{f}{\SI{}{\hertz}},
 \end{equation} 
 making explicit the units of measure.

\section{CMB B-mode Experiments}\label{sec:cmb}
CMB experiments are at the forefront of the search for a primordial SGWB. As we discussed in Section \ref{sec:standard_ps}, the current best observational bounds on the SGWB come from the CMB. Furthermore, as it will be shown in Section \ref{sec:results}, they represent our best opportunity to detect a SGWB if the correct model for its production is the single-field slow-roll inflation with $r\lesssim 0.001$. 

The current generation of operating CMB experiments includes BICEP2/Keck, POLARBEAR, ACT, SPT and CLASS while the next generation of experiments, planned for this decade, will comprise the Simons Array, SO, SPO and CMB-S4 on the ground-based side, and the LiteBIRD mission observing from space. In this paper, we will focus on making forecasts for the LiteBIRD, which is expected to be -- together with CMB-S4 -- the most sensitive among the planned missions, capable of detecting a tensor-to-scalar ratio $r \lesssim 0.001$. 

The  signature of the primordial SGWB in the B-mode polarization has two main contributions: one at very large scales (around $k\sim 6\times10^{-4}\,\SI{}{\per\mega\parsec}$) where the CMB photons are re-scattered by the free electrons made available by cosmic reionization \cite{zaldarriaga:1996}, producing the so-called \textit{reionization bump}, and the other at intermediate scales ($k\sim 6\times 10^{-3}\,\SI{}{\per\mega\parsec}$) corresponding to the \textit{recombination bump} \citep{hiramatsu_etal_2018}. This primordial signal, however, is fainter than the contaminating signals of the secondary origin: smaller scales are dominated by the \textit{gravitational lensing} due to the cosmological large-scale structure, which converts the E-mode polarization of the CMB into a secondary B-mode \cite{zaldarriaga/seljak:1998}, while larger scales are contaminated by the presence of the \textit{diffuse Galactic foregrounds}. 

In this Section we first review the formalism of CMB power spectra (Section \ref{sec:CMB_ps}). We then describe the relevant noise sources for CMB experiments, including the instrumental, the lensing and the astrophysical foreground contributions (Section \ref{sec:CMB_noise}). Finally, we review the Fisher matrix approach for computing the binned uncertainties on the tensor power spectrum for a CMB experiment (Section \ref{sec:fisher}).

\subsection{CMB Angular Power Spectra}\label{sec:CMB_ps}

CMB experiments do not observe directly the scalar or tensor power spectra described in Sections \ref{sec:tensor_ps}, but rather
their effects on the CMB temperature and polarization
angular power spectra $C_{\ell}^{XX'}$, defined by the correlation function $\langle a_{\ell m}^{X*} a_{\ell' m'}^{X'} \rangle = \delta_{\ell\ell'}\delta_{mm'} 
C_\ell^{XX'}\,$,
where the indices $X,X' = \{T,E, B\}$ label the total intensity (T), gradient (E) and curl (B) modes of the CMB polarization \citep{kamionkowski_kosowski_stebbins_1997,seljak_zaldarriaga_1997} and the $a_{\ell m}^{X}$ are the coefficients of the spherical harmonic expansion of the total intensity and polarization. 

Assuming that vector modes get diluted by the expansion of the Universe,  each angular power spectrum will have contributions only from scalar and tensor modes, so that
$C_{\ell}^{XX',\,prim}= C_{\ell,s}^{XX'}+ C_{\ell,t}^{XX'}$.
We can now connect the observable angular power spectra to the primordial scalar and tensor ones through the scalar or tensor transfer functions $T^{X}_{\ell, y}$
\begin{equation}\label{eq:cmb_transfer}
C_{\ell, x}^{XX'} = \frac{2 \pi}{\ell(\ell+1)} \int d \ln k\,\mathcal{P}_{y}\left(k\right) 
T^{X}_{\ell, y}\left(k\right) T^{X'}_{\ell, y}\left(k\right)\,,
\end{equation}
with indices $X,X'=\{T,E\}$, $x=\{s\}$ and $y=\{\mathcal{R}\}$ for the scalar case and indices $X,X'=\{T,E, B\}$, $x=\{t\}$ and $y=\{T\}$ for the tensor one . The transfer functions depend on the cosmological parameters, for which we assume the Planck 2018 values \citep{planck_2018}, and can be computed from a Boltzmann solver such as \texttt{CAMB} \citep{camb} or \texttt{CLASS} \citep{CLASS_code_paper}. 

To conclude this Section, we specialize Eq.~\ref{eq:cmb_transfer} to the axion-SU(2) sourced contribution to the tensor spectrum, defined in Section \ref{sec:axion_model} 
\begin{equation}
   C_{\ell, t}^{XX',\,Sourced} = \frac{2 \pi}{\ell(\ell+1)} \int d \ln k\,\left[\mathcal{P}_{T}^{L,\,Sourced}(k)+\mathcal{P}_{T}^{R,\,Sourced}(k)\right]\left(k\right) 
T^{X}_{\ell, T}\left(k\right) T^{X'}_{\ell, T}\left(k\right)\,,
\end{equation}
with $XX'=\{TT,EE,TE,BB\}$.
Note that the chiral tensor spectrum produced in the axion-SU(2) model also yields non-zero parity-odd cross-spectra such as $TB$ and $EB$ spectra, which could be used as an observational marker to distinguish it from the standard SGWB from the vacuum fluctuations \citep{lue/wang/kamionkowski:1999,Saito_2007,contaldi/magueijo/smolin:2008}.
However, these cross-power spectra are difficult to detect unless $r\gtrsim 0.05$ \cite{Thorne}; thus, in this paper we will be concerned only by the intensity of the SGWB rather than by its circular polarization, and consider only the $BB$ spectrum in our analysis.

\subsection{Noise and Foregrounds for CMB Experiments}\label{sec:CMB_noise}
In this paper we will consider the LiteBIRD satellite and its constraining power on the SGWB.  For our purpose we can characterize this instrument using the following parameters: the polarization sensitivity (in $\mu$K-arcmin units) at each frequency channel, the Full Width at Half Maximum (FWHM) for the instrument beams, the observed sky fraction $f_{sky}$ and the multipole range of the measurement. For LiteBIRD we adopt a multipole range from $\ell_{min}=2$ to $\ell_{max}=200$. We report all the other specifications in Table \ref{table:litebird}. 

\begin{table}
	\centering
	\begin{tabular}{|c|c|c|c|}
		\hline \hline
		Experiment
		& Frequency
		& Sensitivity 
		& FWHM 
		\\
		
		& [GHz]
		& [$\mu$K-arcmin]
		&  [arcmin]
		\\
		\hline	\hline
		
		&	40 & 59.29 & 60 \\
		&	50 & 32.78 & 56  \\
		&	60 & 25.76 & 48 \\
		&	68 & 15.91 & 43 \\
		&	78 & 13.10 & 39  \\
		&	89 & 11.25  & 35  \\
	\textbf{LiteBIRD}	&	100 & 7.74 & 29 \\
	($f_{sky}=0.6$)	&	119 & 5.37 & 25  \\
	&	140 &   5.65    &  23  \\
		&	166 &   5.81    &    21     \\
		&	195 &    6.48   &     20      \\
		&	235 &   15.16    &   19   \\
		&	280 &    17.98   &     24      \\
		&	337 &    24.99   &    20        \\
		&	402 &     49.90  &    17      \\
		\hline\hline
	\end{tabular}
	\caption{Instrumental specifications adopted for the LiteBIRD CMB experiment (LiteBIRD collaboration, private communication).}
	\label{table:litebird}
\end{table}

As we already mentioned above, there are three relevant noise sources which contribute to the total observed CMB B-mode spectrum $C_{\ell}^{BB}$:
\begin{equation}\label{eq:total_signal}
C_{\ell}^{BB} = C_{\ell}^{BB,\,prim} + C_{\ell}^{BB,\,noise} +  C_{\ell}^{BB,\,lens} + C_{\ell}^{BB,\,fgs}\,,
\end{equation}
where $C_{\ell}^{BB,\,prim}$ is the primordial signal, $C_{\ell}^{BB,\,lens}$ is the gravitational lensing B-mode, $C_{\ell}^{BB,\,fgs}$ the residual contamination due to polarized diffuse foregrounds, and 
$C_{\ell}^{BB,\,noise}$  the post component separation noise. 
We model the instrumental noise \citep{stompor_etal_2016} at each frequency channel $\nu$ as  
\begin{equation}\label{eq:noise}
N_{\ell, \nu}^{BB}=\left[w_{B,\nu}^{-1}\exp\left(\ell(\ell+1)\frac{\theta^{2}_{FWHM,\nu}}{8\ln 2}\right)\right],
\end{equation}
where $w_{B,\nu}^{-1/2}$ is the white noise level (or sensitivity) in each frequency channel in $\mu K$-rad and $\theta_{FWHM,\nu}$ is the beam size in radians. 
 
The lensing represents a contaminant of the unknown amplitude when searching for
a primordial signal and affects especially the smaller angular scales of the CMB B-modes. We compute $C_{\ell}^{BB,\,lens}$ using the \texttt{CAMB} code. Note that for LiteBIRD we conservatively do not consider any cleaning from the lensing contamination, i.e., a procedure called ``delensing'' \citep{delensing1,delensing2, delensing3, delensing4}, but we stress that high resolution ground-based experiments such as CMB-S4 can be exploited to delens LiteBIRD data to enhance its capability in reconstructing the SGWB.

On the other hand, the dominant source of noise on large scale B-mode polarization is the diffuse Galactic foregrounds \citep[see, e.g.,][ and references therein]{planck_2018_diffuse_component_separation}. In particular, in this paper we will consider the two main sources of foregrounds for B-mode experiments: the thermal emission of \textit{dust} grains and the \textit{synchrotron} radiation emitted by cosmic-ray electrons spiraling in the Galactic magnetic field \citep[see][and references therein]{Dickinson}. 
We generate simulated sky maps of the polarized Galactic foreground emission using the ``d1s1'' sky model in the Python Sky Model (\texttt{PySM}) code \citep{pysm_ref}, and degrade them to a \texttt{HEALPIX} \citep{healpix} resolution $N_{side}=128$. We add to the simulated maps an instrumental white noise realization generated by the model in Eq.~\ref{eq:noise}. We perform component separation for three possible spectral energy distributions (SEDs): the CMB SED, for which we assume no free parameters; the thermal dust SED, for which we take the one-component modified black-body 
\begin{equation}\label{eq:power_law_dust}
	\textbf{A}_{dust} (\nu) = \left(\frac{\nu}{\nu_{d}}\right)^{\beta_{d}+1} \frac{e^{\frac{h\nu_{d}}{kT_{d}}}-1}{e^{\frac{h\nu}{kT_{d}}}-1},
\end{equation} 
with the spectral index $\beta_{d}$ and the temperature $T_d$ as free parameters and the reference frequency $\nu_d$ fixed to $\SI{353}{\giga\hertz}$ ; and the synchrotron SED,
for which we assume the curved power-law 
\begin{equation}\label{eq:power_law_sync}
	\textbf{A}_{sync}(\nu) = \left(\frac{\nu}{\nu_{s}}\right)^{\beta_{s}+C_{s} \ln(\nu/\nu_{s})},
\end{equation}
with the spectral index $\beta_{s}$ and the curvature $C_{s}$ as free parameters and $\nu_{s} = \SI{70}{\giga\hertz}$.

We compute the contributions of residual foregrounds  $C_{\ell}^{BB,\,fgs}$ and post component separation noise $C_{\ell}^{BB,\,noise}$ to the observed spectrum using the parametric maximum likelihood approach \citep{errard_stivoli_stompor_2011, cmb4cast, stompor_etal_2016, stompor_leach_stivoli_baccigalupi} implemented in the publicly available ForeGroundBuster (\texttt{FGBuster}) code\footnote{See \url{https://github.com/fgbuster/fgbuster} and reference therein.}. This code allows for several different choices of cleaning techniques, among which we choose the \textit{Multi-Resolution} procedure, an evolution of the \textit{Multi-patch} technique presented in \citep{errard_stompor_2018}.
While in the Multi-Patch approach we fit all the spectral parameters over independent sky patches equal to \texttt{HEALPIX} pixels with the same resolution parameter $N_{side}$, in the Multi-resolution approach, each of the free spectral parameters is fitted on a different \texttt{HEALPIX} grid with different resolution. The patches resolution for each parameter are gathered in the Multi-resolution vector $\mathbf{N}_{sides}$, for which we adopt the choice $\mathbf{N}_{sides} = [\beta_{d}, T_{d}, \beta_{s},  C_{s}] = [64,8,8,0]$, obtained by prioritizing the characterization of dust SED over synchrotron SED and by requiring that systematic residuals are much smaller than the statistical ones (J. Errard 2019, private communication). This selection of parameters provides appropriate residuals for the current foreground modeling in LiteBIRD.  

\begin{figure}
\centering
        \includegraphics[scale=0.9]{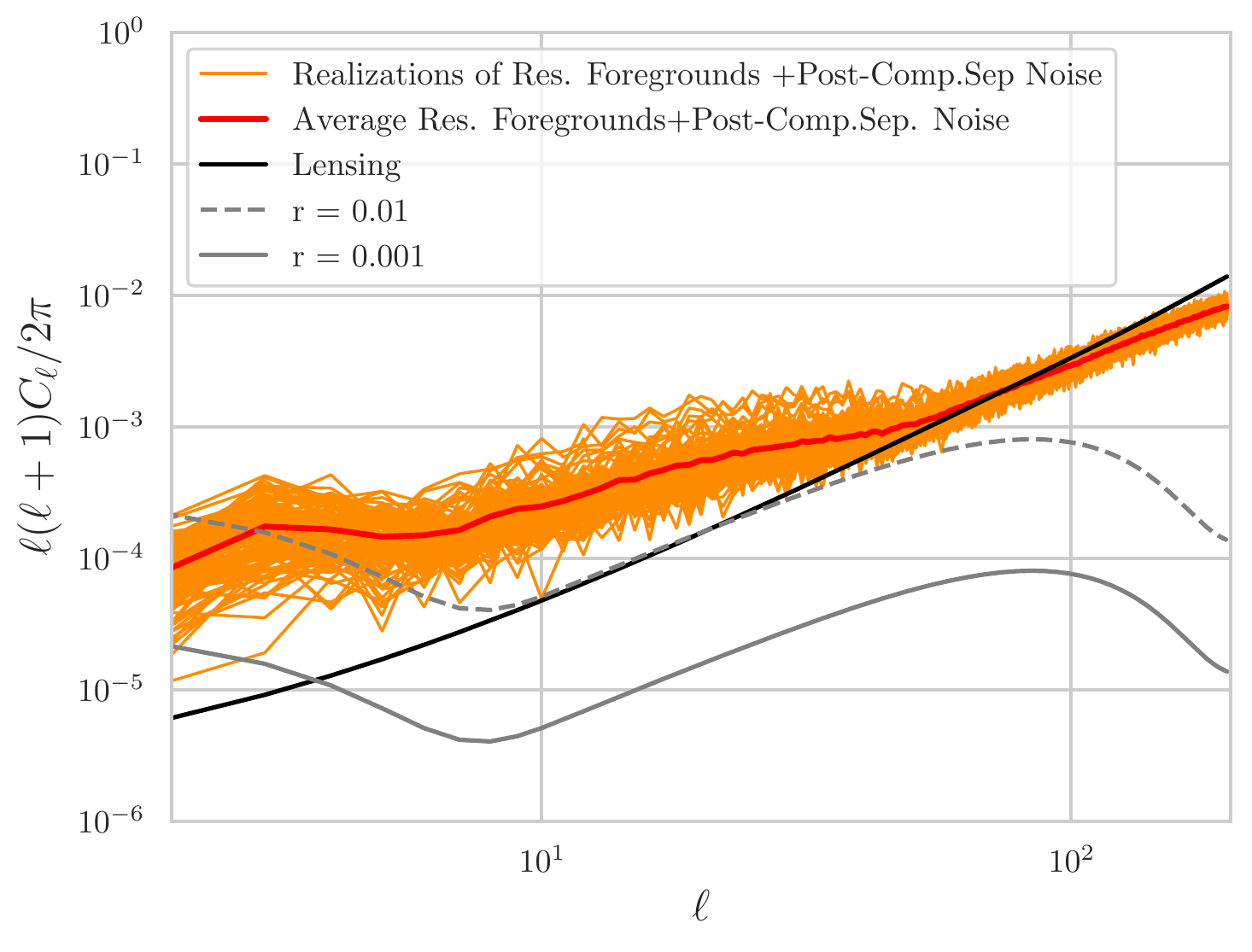}
        \caption{The sum of the residual foregrounds and post-component separation noise for 100 noise realizations (in orange) and their average (in red). We also show the lensing power spectrum $C_{\ell}^{BB,\,lens}$ (black solid line) and the primordial B-mode signals for $r=0.01$ (dashed grey) and $r=0.001$ (solid grey).   }\label{fig:cl_spectra}
    \end{figure}

We average the resulting residual foregrounds plus post-component separation noise spectra over 100 noise realizations, obtaining the final spectrum in Figure \ref{fig:cl_spectra} (red curve). This spectrum is roughly composed by two parts. In the angular domain, the diffuse Galactic  foregrounds are usually characterized by a decaying power law with the angular multipole. 
Therefore, at high $\ell$, the foreground contamination is less relevant, and the component separation noise is the co-addition of sensitivity in multi-frequency channels corresponding to the CMB solution. On the other hand, at low and intermediate multipoles, the structure is dominated by the component separation residuals from the large scale pattern of foregrounds.

\subsection{Fisher Matrix for the Tensor Power Spectrum}\label{sec:fisher}
To compute the binned uncertainties on the tensor power spectrum for LiteBIRD, we use a Fisher matrix approach similar to the one described in Refs.~\cite{Campeti,hiramatsu_etal_2018}. We report here the main ingredients of the method.
The tensor power spectrum $\mathcal{P}_{T}$ can be discretized as
\begin{equation}\label{eq:expansion}
\mathcal{P}_{T}(k) = A_{S} \sum_i p_i W_i(\ln k)\,,
\end{equation}
where 
$W_i$ is the discretization window function, which we choose to be equal to 1 inside the $i$-th of the $N$ power spectrum bins and 0 outside
\begin{equation}\label{eq:window}
W_i(\ln k) =  \begin{cases} 
      1 & \,\text{for}\, k_{i-1} \leq k <k_{i} \, \text{with}\, 1\leq i\leq N\\
      0 & \, \text{for}\, k< k_{i-1} \,\text{and}\, k>k_{i}
   \end{cases},
\end{equation}
and $\Delta \ln k= (\ln k_{i+1} -\ln k_i)$ 
is the width of the $i$-th bin. The discretization process allows  to write the derivative of the $C_{\ell}^{XX'}$ with respect to the power spectrum parameters $p_i$ in a simple way:
\begin{equation}\label{eq:derivative}
D_{\ell i}^{BB} = \frac{\partial C_\ell^{BB}}{ \partial p_i}\bigg|_{\rm fid}
	=\frac{2 \pi}{ \ell(\ell+1)} A_{S} \int d\ln k\,
	T^{B}_\ell(k)\, T^{B}_\ell(k)\,  W_i(\ln k)\,.
\end{equation}
We choose the $k$ range to be  $10^{-4}\,\SI{}{\per\mega\parsec}< k < \SI{1}{\per\mega\parsec}$ such that it contains the whole sensitivity curve of the LiteBIRD experiment. 
To obtain the error bar on each power spectrum wavenumber bin, we first compute the Fisher information matrix  \citep[see, e.g.,][]{tegmark_1996}
\begin{equation}
\label{eq:fisher}
F_{ij} = f_{sky}
\sum_{\ell=2}^{\ell_{\rm max}}
\frac{2 \ell +1}{2} 
{\rm Tr} \left[ D^{BB}_{\ell i} 
 \left(C_{\ell}^{BB}\right)^{-1}
 D_{j \ell}  \left(C_{\ell}^{BB}\right)^{-1}\right]\,,
\end{equation}
where the factor $f_{sky}$ takes into account the loss of modes by a partial sky coverage. We then take the diagonal of its inverse to obtain
    \begin{equation}\label{eq:diag}
        \sigma^{2}_{PS}(k_{i})=(F^{-1})_{ii}.
    \end{equation}
The desired binned uncertainty on $\Omega_{GW}$ is then easily obtained from Eq.~\ref{eq:omega} 
\begin{equation}\label{eq:cmb_binning}
    \sigma_{\Omega_{GW}}(k_{i})= \sigma_{PS}(k_{i}) \cdot \frac{A_{S}}{12 H^{2}_0} \left[\mathcal{T}^{'}(k, \tau_0)\right]^{2}.
\end{equation}

\section{Interferometers and PTA}\label{sec:direct}
The landscape of the current and future interferometers and PTA experiments is vast, characterized by their complementarity in probing the GW spectrum across a wide range of frequencies.  The frequency window between $\sim 10^{-7}$ and $\sim 10$ Hz is expected to be observed from space through a host of funded and proposed laser interferometers, ranging from $\mu$Ares \citep{Ares_paper} in the micro-Hertz band, to LISA \citep{LISAxcosmo} and AMIGO \cite{Baibhav:2019rsa} in the milli-Hertz band, to BBO \citep{Crowder_Cornish,Smith_and_Caldwell_2016}, DECIGO \citep{Seto:2001qf} and DO \citep{DO_paper} in the deci-Hertz bands. In this work we also include the recently proposed space-based atom interferometer AEDGE \citep{AEDGE_paper}, which will observe the deci-Hertz band as well. Going higher in the GW frequency ($\sim 10 - 10^{3}$ Hz), the next-generation ground-based detectors (CE \citep{CE_paper} and ET \citep{ET_paper}), also exploiting laser interferometry, will complement the previous observations in the high-frequency part of the GW spectrum.

We summarize in Table \ref{table:experiments} the main instrumental characteristics and capabilities of GW observatories treated in this paper, including the experiment type, the arm lenght ($L$) for traditional interferometers, the total observation length ($T_{obs}$), the observational efficiency $\epsilon$ to compute the actual observation time $T_{eff}=\epsilon T_{obs}$, the frequency range at which the experiment is operating, and the minimum of the binned sensitivity curve with and without foregrounds.

\begin{table}
	\centering
	\footnotesize{
	\begin{tabular}{|c|c|c|c|c|c|c|c|c|}
		\hline \hline
		Experiment
		& Type
		& L 
		& $T_{obs}$ 
		& $\epsilon$
		& Freq.Range
		& $h^2 \Omega^{min}_{GW}$
		& $h^2 \Omega^{min}_{GW}$
		& References
		\\
		
		&  
		& [m]
		&  [$\SI{}{\year}$]
		& 
		& [$\SI{}{\hertz}$]
		& w/o Fgs
		& w/ Fgs
		&
		\\
		\hline	\hline
		 & &  &  & & & & &\\
LISA &  Space   & $2.5\times 10^{9}$ & 4 & 75\% & $10^{-4}-10^{-1}$ & $5.9\times 10^{-14}$ & $9.9\times 10^{-14}$ & \citep{LISAxcosmo} \\
 &	M.I. &  &  & &  & & &\\
 \hline
 		 &	 &  &  & & & & &\\
DO  &  Space  & $10^{8}$ & 4 & 75\% & $10^{-3}-10^{1}$ & $3.7\times 10^{-15}$ & $2.1\times 10^{-14}$ & \citep{DO_paper}\\
Cons. & M.I.	 &  &  & && & &\\
 \hline
 		 &	 &  &  & && & &\\
DO  &  Space  & $10^{8}$ & 4 & 75\% & $10^{-3}-10^{1}$ & $7.1\times 10^{-16}$ & $3.7\times 10^{-15}$ & \citep{DO_paper}\\
 Opt. & M.I.	 &  &  & & && &\\

 \hline
 		 &	 &  &  & & & & &\\
$\mu$Ares &  Space & $430\times10^{9}$ & 10 & 100\% & $10^{-6}-10^{-2}$ & $4.7\times 10^{-18}$& $8.4\times 10^{-17}$ & \citep{Ares_paper}\\
 & M.I.	 &  &  & && & &\\
 \hline
 		 &	 &  &  & & & & &\\
DECIGO &  Space  & $10^{6}$ & 10 & 100\% & $10^{-4}-10^{1}$& $2\times 10^{-17}$& $9.8\times 10^{-17}$ & \citep{DECIGO_paper}\\
 &	F.P.I. &  &  & & & &  &\\
 \hline
 		 &	 &  &  & && & &\\
BBO &  Space & $5\times10^{7}$ & 10 & 100\% & $10^{-4}-10^{1}$ & $1.8\times 10^{-18}$& $1.8\times 10^{-18}$ & \citep{BBO_paper}\\
 & M.I.	 &  &  & & & & &\\
 \hline
 		 & 	 &  &  & & & & &\\
AEDGE &  Space & $4.4\times 10^{7}$ &  5    & 60\% & $10^{-2}-1$ &$4.2\times 10^{-16}$ & $2.6\times 10^{-15}$ & \citep{AEDGE_paper}\\
 &	A.I. &  &  & & & & &\\

\hline
 		 &	 &  &  & & & & &\\
ET &  Ground  & $1\times10^{4}$ & 1 & 100\% & $1-10^{3}$& $4.5\times 10^{-14}$& $2.8\times 10^{-13}$ & \citep{ET_paper}\\
 & M.I.	 &  &  & & & & &\\
\hline
 		 &	 &  &  & & & & &\\
SKA & PTA & - & 10 & 100\% & $10^{-9}-10^{-7}$ & $3.8\times 10^{-14}$& $2.7\times 10^{-13}$ & \citep{SKA_paper,Mingarelli:2019mvk},\\
 &	 &  &  & & & & & \\

		\hline\hline
	\end{tabular}}
	\caption{Summary of the instrumental specifications for interferometers and PTA considered in this work. ``M.I.'' stands for Michelson Interferometer, ``F.P.I.'' stands for Fabry-P\'erot Interferometer and ``A.I.'' stands for Atomic Interferometer. The binning used to compute the values of $h^2 \Omega^{min}_{GW}$ is $\Delta\ln{k}=1.2$.} 
	\label{table:experiments}
\end{table}

Going lower in the frequency, PTAs will probe GWs in the $\sim 10^{-9}-10^{-7}$ Hz region. There are several planned and ongoing PTA experiments (NANOGrav \citep{Arzoumanian:2015liz,NANOGrav_2018}, EPTA \citep{EPTA_paper}, PPTA \citep{PPTA_paper1, PPTA_paper2}, IPTA \citep{IPTA_paper}).
In this paper we show the expected constraints for the most ambitious experiment of this kind, i.e., the SKA \citep{SKA_paper}. 

All of the experiments listed above will target several GW sources, both stochastic and deterministic, but in the following we will be interested only in the stochastic ones, and in particular in the possibility of detecting a SGWB of the \textit{primordial} origin. Therefore, we will consider other SGWB sources, such as unresolved Galactic and extra-Galactic compact binaries for instance, as a \textit{foreground} or \textit{confusion noise} to our sought-after primordial signal. 

In this Section, we first describe the formalism required to compute the sensitivity curves for  interferometers (Section \ref{sec:sensitivity}). We then describe in detail our choices concerning the astrophysical foreground contamination (Section \ref{sec:laser_fgs}) and how it affects the sensitivity curve for each experiment. In Section \ref{sec:SKA} we describe how we calculate the sensitivity curve for the SKA. To supplement these sections, in Appendix we describe the construction of the interferometers response functions (Appendix \ref{sec:response}) and the noise properties of each  interferometer (Appendix \ref{sec:interf_noise}).

\subsection{Instrumental Sensitivity Curves}\label{sec:sensitivity}

In this Section we derive the equation for the sensitivity curve of a GW laser interferometer to an homogeneous and isotropic SGWB. Three of the experiments considered in this work ($\mu$Ares, DECIGO, BBO) are designed as two independent triangular interferometers, with consequently uncorrelated instrumental noises. The target of these experiments is to measure the \textit{cross-correlation} of the outputs of the two independent detectors. Therefore, in the following we will provide formulae for both the sensitivity of a single detector (suited for LISA, DO, ET ) and for the cross-correlation of two independent detectors. Our discussion follows Refs.~\citep{Smith_and_Caldwell_2016,Romano_Cornish,Kai}, and we refer to those papers for a more complete and detailed derivation. For a derivation of the  sensitivity curve of a PTA experiment, which will not be reproduced here, we refer the reader to Refs.~\citep{Romano_Cornish,Hazboun_2019}.

A SGWB can be expanded in plane waves as
\begin{equation}\label{eq:plane_wave}
    h_{ab}(t, \Vec{x}) =  \int_{-\infty}^{+\infty}df\int d^{2}\hat{n} \sum_{P=+,\times}  \Tilde{h}_{P}(f, \hat{n}) e_{ab}^{P}(\hat{n}) e^{i2\pi f(t-\hat{n}\cdot \Vec{x})},
\end{equation}
where $\Tilde{h}_{P}$ is the amplitude of a sinusoidal plane GW, $P=+,\times$ is the linear polarization state of GW, $\hat{n}$ the GW propagation direction and $e_{ab}^{P}$ the polarization tensor.
In time domain, the data $d_{I}$ of a detector $I$ can be written as the sum of the signal $s_{I}$ and noise $n_{I}$ 
\begin{equation}
    d_{I}(t) = s_{I}(t)+n_{I}(t).
\end{equation} 
Moving to Fourier space, the noise spectrum for a single detector is determined by
\begin{equation}\label{eq:noise_psd}
    \big\langle \Tilde{n}_{I}(f) \Tilde{n}^{*}_{I}(f')\big\rangle = \frac{1}{2} \delta(f-f') \mathcal{S}_{n}^{I}(f).
\end{equation}
Similarly, we define the GW signal strain power spectrum $S_{s}$ through the correlation of the GW Fourier modes defined in Eq.~\ref{eq:plane_wave}:\footnote{More generally, the covariance matrix of $\tilde h_P$ can be written in terms of the ``GW Stokes parameters'' in analogy to the electromagnetic waves \cite{seto:2006}
\begin{equation}
\big\langle \tilde{h}_{P}(f,\hat{n}) \tilde{h}^{*}_{P'}(f',\hat{n}')\big\rangle
= \frac{1}{2} \delta(f-f')\frac{\delta^{(2)}(\hat{n}-\hat{n}')}{4\pi}
\left(\begin{array}{cc}I+Q&U-iV\\U+iV&I-Q\end{array}\right).
\end{equation}
Here, $I$ is the Stokes $I$  and should not be confused with the index for the detector used in the main text. Circular polarization from chiral GW due to the SU(2) gauge field would appear as the Stokes $V$ \cite{Thorne}. In this paper we are concerned only with the total intensity of the SGWB and ignore $Q$, $U$, or $V$, hence $\delta^{PP'}$ in Eq.~\ref{eq:signal_psd}.}
\begin{equation}
\label{eq:signal_psd}
    \big\langle \Tilde{h}_{P}(f, \hat{n}) \Tilde{h}^{*}_{P'}(f',\hat{n}')\big\rangle = \frac{1}{2} \delta(f-f') \frac{\delta^{(2)}(\hat{n}-\hat{n}')}{4\pi}\delta^{PP'} \mathcal{S}_{s}(f).
\end{equation}
We can now introduce the \textit{response function} $T^{P}_{I}(f, \hat{n})$ to describe the signal response of a detector $I$ to a sinusoidal plane GW, which will be computed in Appendix \ref{sec:response} for several different detector configurations. Using this we write the signal response $\Tilde{s}_{I}$ of a detector $I$ in Fourier space as
\begin{equation}\label{eq:s_i_fourier}
    \Tilde{s}_{I}(f) = \int d^{2}\hat{n} \sum_{P=+,\times} T^{P}_{I}(f, \hat{n}) \Tilde{h}_{P}(f, \hat{n}),
\end{equation}
with $T^{P}_{I}(f, \hat{n}) = T^{ab}_{I}(f, \hat{n})\, e_{ab}^{P}(\hat{n}) e^{-i2\pi f\hat{n}\cdot\Vec{x}}$.

For a network of detectors $I,J=1,2,...$, we write
\begin{equation}
     \big\langle \Tilde{s}_{I}(f) \Tilde{s}^{*}_{J}(f')\big\rangle = \frac{1}{2} \delta(f-f') \Tilde{C}_{IJ}(f) = \frac{1}{2} \delta(f-f') \mathcal{R}_{IJ}(f) \mathcal{S}_{s}(f),
\end{equation}
where $\tilde{C}_{IJ}$ is the covariance matrix of the signal response defined by
\begin{equation}
    \Tilde{C}_{IJ} = \big\langle \Tilde{s}_{I} \Tilde{s}_{J}\big\rangle - \big\langle \Tilde{s}_{I} \big\rangle \big\langle \Tilde{s}_{J} \big\rangle,
\end{equation}
and $\mathcal{R}_{IJ}(f)$ is the so-called \textit{overlap reduction function} for the detector pair $IJ$ \cite{Flanagan:1993ix} (see also discussion in Appendix \ref{sec:response})
\begin{equation}\label{eq:overlap}
    \mathcal{R}_{IJ}(f) = \int  \frac{d^{2}\hat{n}}{4\pi} \sum_{P=+,\times} T^{P}_{I}(f, \hat{n}) T^{P *}_{J}(f, \hat{n}).
\end{equation}
It can be shown that the optimal signal-to-noise ratio (hereafter SNR) for a cross-correlation measurement of a SGWB using a network of detectors $I,J=1,2,...$, takes the form \citep{Romano_Cornish, Allen_Romano_1999,Kai}
\begin{equation}\label{eq:SNR_nofgs}
    SNR = \left[n_{det} T \int^{f_{max}}_{f_{min}} \sum_{J>I} \frac{\mathcal{R}_{IJ}^{2}(f) \mathcal{S}_{s}^{2}(f)}{\mathcal{S}_{n}^{I}(f)\mathcal{S}_{n}^{J}(f)} df\right]^{1/2},
\end{equation}
where $n_{det}$ is the number of detectors in the network, $T$ is the mission observation time and $\left[f_{min}, f_{max} \right]$ is the detector pair bandwidth. 

Since the GW strain power spectrum density can be related to the fractional energy density spectrum in GW as \citep{Romano_Cornish} 
\begin{equation}
 S_{s}(f) = \frac{3H_{0}^{2}}{4\pi^{2}f^{3}}\Omega_{GW}(f),  
\end{equation}
we can write the sensitivity curve  in terms of the minimum detectable gravitational wave energy density $\Omega_{GW}$ with the desired SNR in a frequency bin $\Delta f$ as \citep{Smith_and_Caldwell_2016}
\begin{equation}\label{eq:binning}
    \Omega_{GW}^{min}(f_{i}) = \left[n_{det} T \int^{f_i+\Delta f/2}_{f_i-\Delta f/2}  \left(\frac{3H_{0}^{2}}{4\pi^{2}}\right)^{2} \sum_{J>I}\frac{\mathcal{R}_{IJ}^{2}(f)}{f^{6}\mathcal{S}_{n}^{I}(f)\mathcal{S}_{n}^{J}(f)} df\right]^{-1/2}.
\end{equation}

\begin{figure}
\centering
        \includegraphics[scale=1.0]{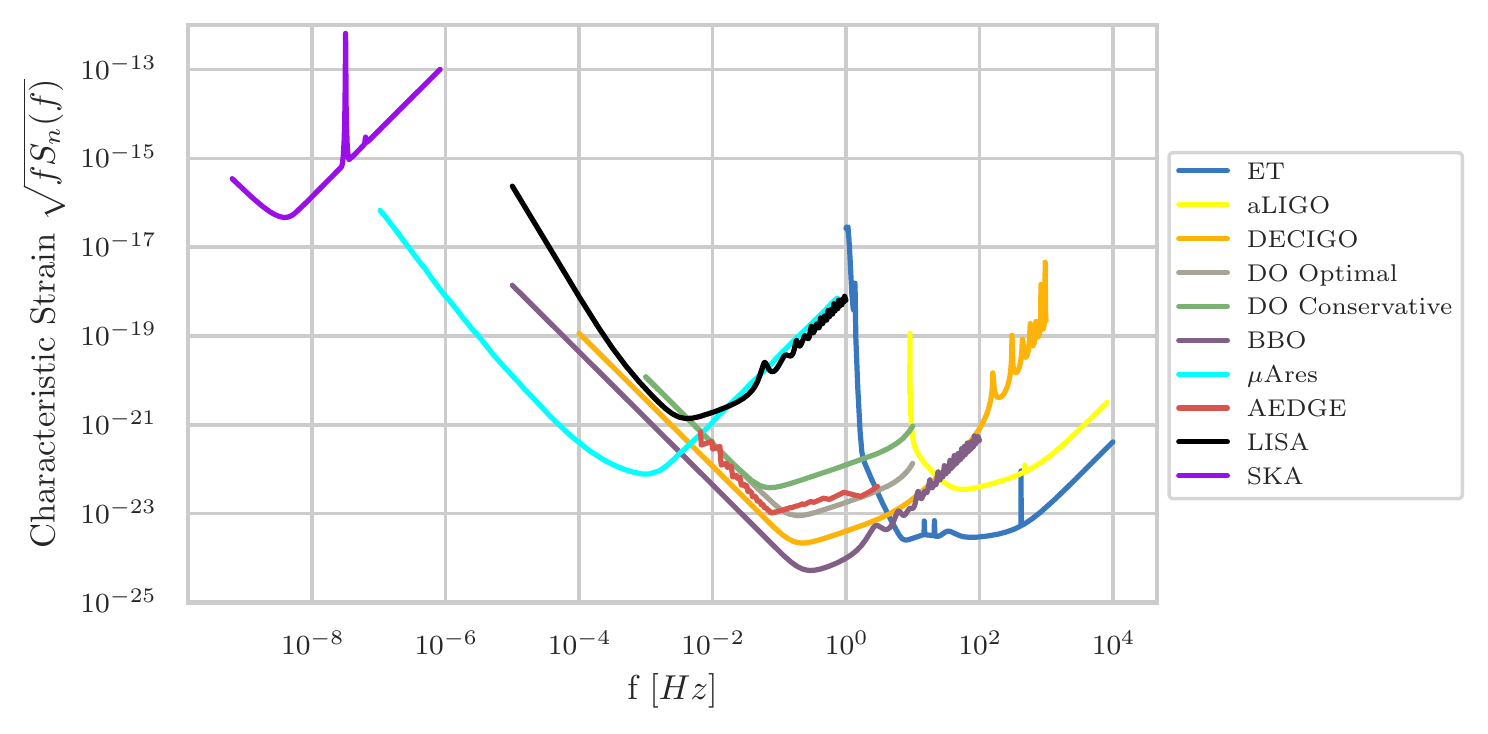}
        \caption{Strain sensitivity curves (without the contribution of astrophysical foregrounds discussed in Section \ref{sec:laser_fgs}) for all the  interferometers and PTA experiments considered in this paper. We also plot for reference the strain curve for the Advanced LIGO (aLIGO) experiment.}\label{fig:all_strains}
\end{figure}

Another useful quantity, which is common in the literature, is the strain spectral sensitivity $\mathcal{S}_{h}$ for the detector network, defined as
\begin{equation}\label{eq:strain}
\mathcal{S}_{h} = \left(\sum_{J>I}\frac{\mathcal{R}_{IJ}^{2}(f)}{\mathcal{S}_{n}^{I}(f)\mathcal{S}_{n}^{J}(f)}\right)^{-1/2}.
\end{equation}
In Appendices \ref{sec:response} and \ref{sec:interf_noise}, we give details on our computations for the overlap reduction function $\mathcal{R}_{IJ}(f)$ and the noise spectrum $\mathcal{S}_{n}(f)$ for each experiment, respectively. 
In Figure \ref{fig:all_strains} we show the strain sensitivity curves for all the interferometers and PTA experiments considered in this paper.

\subsection{Astrophysical Foregrounds for interferometers and PTA}\label{sec:laser_fgs}
We proceed now to describe the main sources of astrophysical foregrounds considered in this work: we first provide approximate analytical fitting formulas for each component (Subsection \ref{sec:fgs_fit}); then we introduce, in Subsection \ref{sec:new_filter}, a filter function capable of maximizing the SNR expression including foregrounds (possibly mitigating them with external information) and finally we describe our foreground cleaning strategy on a case-by-case basis (Subsection \ref{sec:fgs_strategy}).

\subsubsection{Approximate analytical fits for the foreground sources}\label{sec:fgs_fit}

The main sources of astrophysical foregrounds for laser and atomic interferometers are represented by the unresolved populations of \textit{Galactic} and \textit{Extragalactic White Dwarfs} (hereafter GWD and EGWD, respectively), as well as unresolved \textit{stellar mass} \textit{Black Hole}, \textit{Neutron Star} and \textit{Black Hole - Neutron Star Binaries} (hereafter BBH, BNS and BH-NS, respectively). At PTA frequencies the dominant foreground is expected to be the one due to the coalescence of \textit{Massive Black Hole Binaries} (MBHB). This latter foreground could be of importance also for the $\mu$Ares interferometer, at frequencies below $\sim 10^{-5}$ Hz. 

We now describe the model adopted for each of the foreground contributions mentioned above, starting with the GWD binaries confusion noise. Following Refs.~\citep{Cornish_2017,Cornish_2018}, we parametrize it as
\begin{equation}\label{eq:galactic}
 \mathcal{S}_{fg}^{gal}(f) = A f^{-7/3}e^{-f^{\alpha}+ \beta f \sin\kappa f }[1+\tanh (\gamma(f_{k}-f))] \,\SI{}{\per\hertz},
\end{equation}
where $A=9\times 10^{-45}$ and the parameters $\alpha$, $\beta$, $\kappa$, $\gamma$ and $f_{k}$ are reported in Table 1 of Ref.~\citep{Cornish_2018}. These parameters vary according to the total mission observation time, hence the amount of cleaning that is possible to perform on the data. 
On the other hand, the contribution to the SGWB of the EGWD binaries can be analytically approximated as \citep{Nishizawa}
\begin{equation}\label{eq:extragalactic}
    \mathcal{S}_{fg}^{exgal}(f) = 4.2\times10^{-47} \left(\frac{f}{\SI{1}{\hertz}}\right)^{-7/3} \exp\left(-2\left(\frac{f}{5\times10^{-2}\SI{}{\hertz}}\right)^{2}\right) \SI{}{\per\hertz}.
\end{equation}

The contribution of unresolved BBH and BNS can be approximately expressed at interferometers frequencies by a power-law:  
\begin{equation}\label{eq:NS}
    \mathcal{S}_{fg}^{BBH+BNS}(f) = \frac{3H_0^2}{4\pi^2 f^3}\Omega_{*}\left( \frac{f}{f_{*}}\right)^{2/3}, 
\end{equation}
where we assume for the amplitude of the BBH+BNS foreground a value of $\Omega_{*}= 8.9\times 10^{-10}$ at $f_{*} = \SI{25}{\hertz}$, which is the best estimate according to the
current measured merging rates of these compact objects \citep{LIGO_SGWB_upper_limit_2019}.

As for the unresolved MBHB foreground, we use the analytical model given in \citep{Sesana_2008}
\begin{equation}\label{eq:MBHB_foregrounds_strain}
    \mathcal{S}_{fg}^{MBHB}(f) = \frac{h_{0}^{2}}{f} \left(\frac{f}{f_{0}}\right)^{-4/3} \left(1 + \frac{f}{f_{0}}\right)^{2\gamma}, 
\end{equation}
where the parameters $h_0$, $f_{0}$ and $\gamma$ are determined by the particular astrophysical model assumed for the MBHB system. The shape and amplitude of the MBHB foreground can vary greatly according to the theoretical model considered and, in particular, to the eccentricity of the binary system. However, just for the purpose of showing an indicative level for this foreground, we adopt the VHMhopk model \citep{Lodato}, with parameters $h_{0}=0.69\times10^{-15}$, $f_{0}=4.27\times 10^{-8}$ Hz and $\gamma=-1.08$, which are consistent with the current upper limits from the 11-year NANOGrav data set \citep{NANOGrav_2018}. Note also that the fit \ref{eq:MBHB_foregrounds_strain} can be well approximated by a power-law $\mathcal{S}_{fg}^{MBHB}\propto f^{-7/3}$ in the PTA range, while in the $\mu$Hz range by another power-law  $\mathcal{S}_{fg}^{MBHB}\propto f^{-4/3+2\gamma}$. 

\subsubsection{A filter for foreground mitigation}\label{sec:new_filter}

In this Subsection, we will describe a possible strategy for the mitigation of astrophysical foreground, using in particular the filter $Q(f)$ introduced in Poletti D. 2020 (\textit{in preparation}), which maximizes the SNR including foregrounds, taking into account the possibility of mitigation by using external information provided by other experiments. 

Specifically, we consider the possibility of a \textit{multi-band} cleaning strategy for the BBH+BNS foreground in the space interferometers bands: it has been shown in Refs.~\citep{PCA_fgs,Pan_Yang_2019} that it is indeed possible to use the information on the BBH+BNS populations gathered by ground-based experiments -- such as Advanced LIGO/Virgo or the future ET and CE detectors -- to remove to a certain degree this foreground contamination from the band of space-borne interferometers like LISA.

In the following, we will assume to \textit{know the foreground} $\mathcal{S}_{fg}^{BBH+BNS}$ \textit{up to a fractional uncertainty on its amplitude} $\sigma_{fg}$. In this case, the cross-correlation estimator \citep{Allen_Romano_1999} 
\begin{equation}\label{eq:cross_estimator}
 \hat{X} = \int_{-\infty}^{+\infty} df \int_{-\infty}^{+\infty} df' \delta_{T}(f-f') d_{I}^{*}(f) d_{J}(f') Q(f),  
 \end{equation}
where $Q(f)$ is a filter function and $\delta_T$ is a finite-time approximation to the Dirac delta function,
will contain also the foreground contribution -- which we denote by $\mathcal{S}_{fg}$ in the following, dropping the superscript $BBH+BNS$ for the sake of notational simplicity -- besides the primordial one ($S_s$):
\begin{equation}
\langle \hat{X} \rangle = T \int_{0}^{\infty} [\mathcal{S}_s(f) + \mathcal{S}_{fg}(f)] \mathcal{R}_{IJ}(f) Q(f) df.
\end{equation}
 Thus, we define the following cross-correlation estimator of the primordial signal $S_{s}$ 
\begin{equation}
\hat{Y} = \hat{X} -  T \int_{0}^{\infty} \mathcal{S}_{fg}(f) \mathcal{R}_{IJ}(f) Q(f) df, 
\end{equation}
and we can write the associated SNR as \citep{Pan_Yang_2019} 
    \begin{equation}\label{eq:SNR_with_fgs}
    SNR_{\hat{Y}} = \frac{\mu }{\sqrt{\sigma_{instr}^{2} + \sigma_{sys}^{2}} }  
    \end{equation}
    where $\mu$ is the mean value of the primordial signal $S_{s}(f)$,
    \begin{equation}
   \mu  =  T \int_{0}^{\infty} \mathcal{S}_s(f) \mathcal{R}_{IJ}(f) Q(f) df,
    \end{equation}
    $\sigma_{instr}$ is the statistical uncertainty due to detector noise
    \begin{equation}
    \sigma_{instr}^{2} \approx \frac{T}{2} \int_{0}^{\infty} \mathcal{S}_{n}^{I}(f) \mathcal{S}_{n}^{J}(f) |Q(f)|^{2} df,
    \end{equation}
    and $\sigma_{sys}$ is the systematic bias due to the limited accuracy of the foreground measurement
    \begin{equation}
   \sigma_{sys}^{2} = \sigma^{2}_{fg} \left( T \int_{0}^{\infty} \mathcal{S}_{fg}(f) \mathcal{R}_{IJ}(f) Q(f) df \right)^{2}.
    \end{equation}
It can be shown (Poletti D. 2020 \textit{in preparation}) -- in a similar way to the fact that the filter $Q(f)=\mathcal{S}_{s}(f)\mathcal{R}^{*}_{IJ}(f)/(\mathcal{S}_{n}^{I}(f) \mathcal{S}_{n}^{J}(f))$ maximizes the foreground-less SNR in Eq.~\ref{eq:SNR_nofgs} \citep{Allen_Romano_1999} -- that the filter (neglecting the $f$ dependency for the sake of notational simplicity )
\begin{equation}\label{eq:Q_Davide}
Q(f) = \frac{\mathcal{S}_{s}R_{IJ}^{*}}{\mathcal{S}_{n}^{I} \mathcal{S}_{n}^{J}} - 2 T \frac{I_{s\times fg}}{\sigma_{fg}^{-2}+2T \,I_{fg\times fg}}\frac{\mathcal{S}_{fg}R_{IJ}^{*}}{\mathcal{S}_{n}^{I} \mathcal{S}_{n}^{J}},
\end{equation}
where
\begin{equation}
I_{s\times fg} = \int \frac{\mathcal{S}_{s} \mathcal{S}_{fg}}{\mathcal{S}_{n}^{I} \mathcal{S}_{n}^{J}}|\mathcal{R}_{IJ}|^2 df,
\end{equation}
and
\begin{equation}
I_{fg\times fg} = \int \frac{\mathcal{S}_{fg}^2}{\mathcal{S}_{n}^{I} \mathcal{S}_{n}^{J}}|\mathcal{R}_{IJ}|^2 df,
\end{equation}
maximizes the SNR in Eq.~\ref{eq:SNR_with_fgs}, taking into account the presence of foregrounds.

Note that, if we know the spectral shape of the foreground but we do not have any external constraint $\sigma_{fg}$ on its amplitude, we can still apply Eqs.~\ref{eq:SNR_with_fgs} and \ref{eq:Q_Davide} in the limit $\sigma_{fg} \rightarrow \infty$ (and $T>0$): in this case the filter has zero response to the foreground template and this corresponds to subtracting the foregrounds by only exploiting its spectral dependence. Moreover, the SNR \ref{eq:SNR_with_fgs} and the filter \ref{eq:Q_Davide} can obviously be applied to whatever foreground with known power-law spectral shape, not only to the BBH+BNS foreground, and can be extended also to multiple foreground components (Poletti D. 2020 \textit{in preparation}).
Finally, we note that an expression for the binned sensitivity curve in terms of the minimum detectable GW energy density can be easily obtained in an analogous way to Eq.~\ref{eq:binning} also in the case with foregrounds.

\subsubsection{Foreground cleaning strategy for interferometers}\label{sec:fgs_strategy}

We discuss now, on a case-by-case basis, our treatment of the contamination of the astrophysical foregrounds when attempting a detection of the primordial SGWB for each interferometer considered in this work, starting with LISA.

\begin{figure}
    \begin{subfigure}[b]{0.5\textwidth}
        \includegraphics[width=1.0\textwidth]{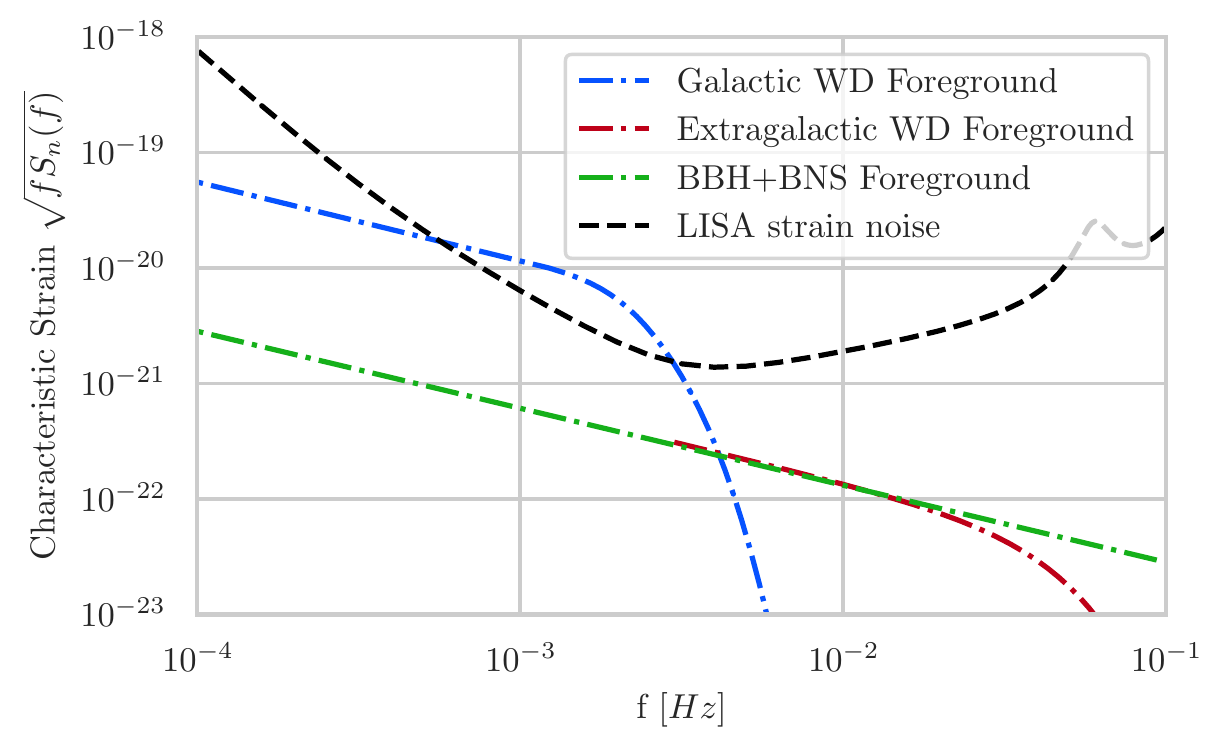} 
    \end{subfigure}
    \begin{subfigure}[b]{0.5\textwidth}
        \includegraphics[width=1.0\textwidth]{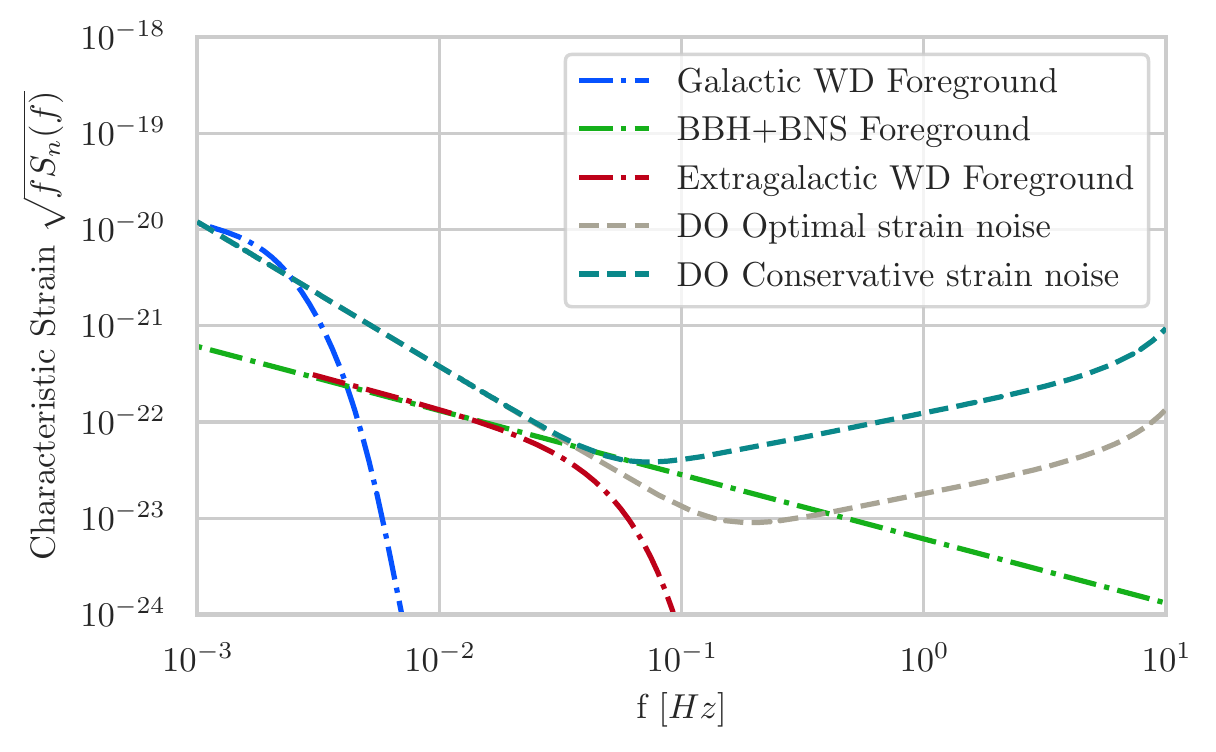} 
    \end{subfigure}
	\caption{Characteristic strain noise curve (dashed curves) for LISA (left panel) and DO (right panel). The foregrounds due to unresolved GWD (dot-dashed blue), EGWD (dot-dashed red) and BBH+BNS populations (dot-dashed green) are also shown.}\label{fig:LISA_DO_fgs}
\end{figure}

\paragraph{LISA ---}
As evident from the left panel of Figure \ref{fig:LISA_DO_fgs}, the WD binaries constitute one of the most relevant confusion noise source in the  LISA band, at frequencies $f\lesssim \SI{5}{\milli\hertz}$. However, it has been shown in \citep{adams_cornish} that this foreground can be subtracted almost completely by exploiting its anisotropy and its time-modulation due to the motion of LISA's constellation \citep{PCA_fgs}. In this work, therefore, we will optimistically assume that the GWD foreground can be perfectly subtracted.

The EGWD foreground could also be relevant in the LISA band, in particular between $f\sim \SI{5}{\milli\hertz}$ and $\sim \SI{0.2}{\hertz}$, beyond which it starts to deviate from a power-law behavior. Differently from the GWD one, this foreground is expected to be almost isotropic, with a hint of anisotropy due to the stronger signal by nearby galaxies which may be used to favor the subtraction. Moreover, its unique spectral shape could also help in separating and subtracting this contaminant from the primordial signal \citep{adams_cornish}. As shown in \citep{Pan_Yang_2019}, the impact of this foreground on the SNR for LISA is secondary with respect to the BBH+BNS foreground.
Therefore, we will neglect the EGWD foreground in our approximate treatment.

The main foreground in the LISA band is represented by the unresolved BBH+BNS populations. Differently from the GWD and EGWD foreground -- which cannot be subtracted using ground-based experiments, since WD binaries never enter their bands -- we exploit a multi-band cleaning technique for the BBH+BNS foreground. We adopt a value $\sigma_{fg}\sim 0.1$ for the fractional uncertainty on the foreground amplitude, justified by the analysis of \citep{PCA_fgs}, involving multi-band cleaning with Advanced LIGO and Virgo. A value $\sigma_{fg}\sim 1.3\times 10^{-2}$ could also be reached using a network of three CE detectors located in Australia, China and US \citep{Pan_Yang_2019}. Furthermore, the level $\sigma_{fg}\sim 10^{-3}$ could be reached using external information provided by ET (M. Pieroni, A. Ricciardone and E. Barausse \textit{in preparation}).

\begin{figure}
    \begin{subfigure}[b]{0.5\textwidth}
        \includegraphics[width=1.0\textwidth]{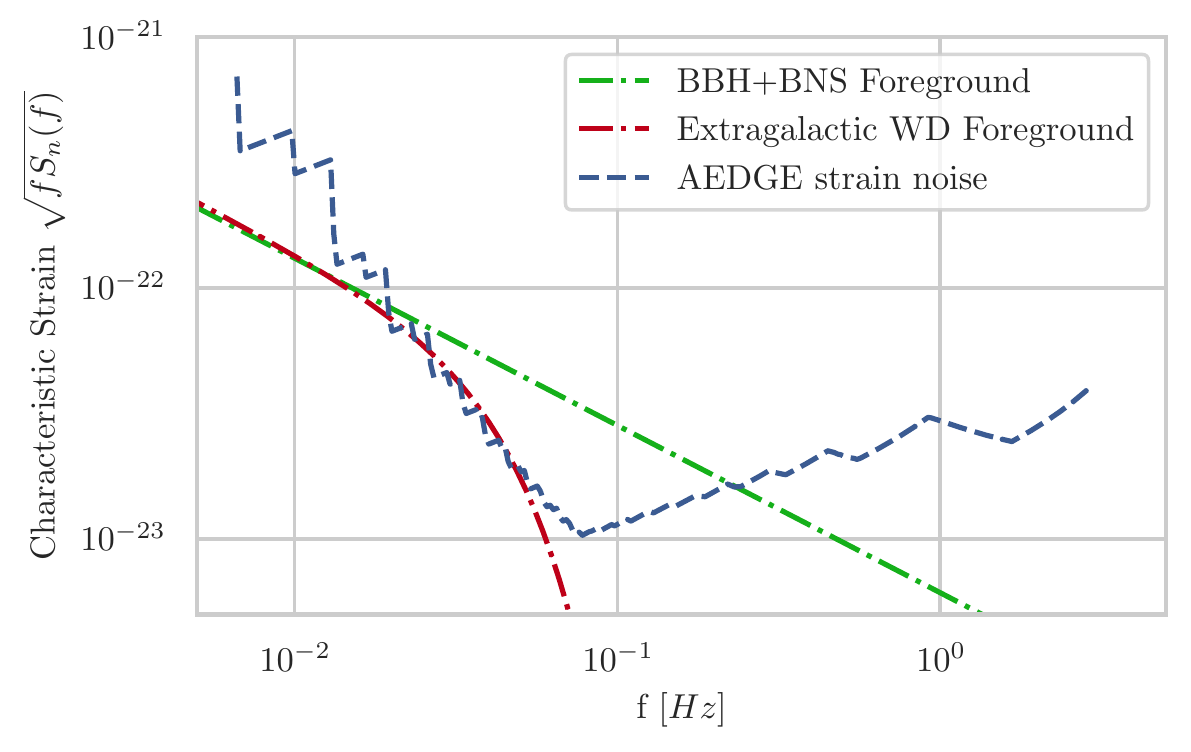}
    \end{subfigure}
    \begin{subfigure}[b]{0.5\textwidth}
        \includegraphics[width=1.0\textwidth]{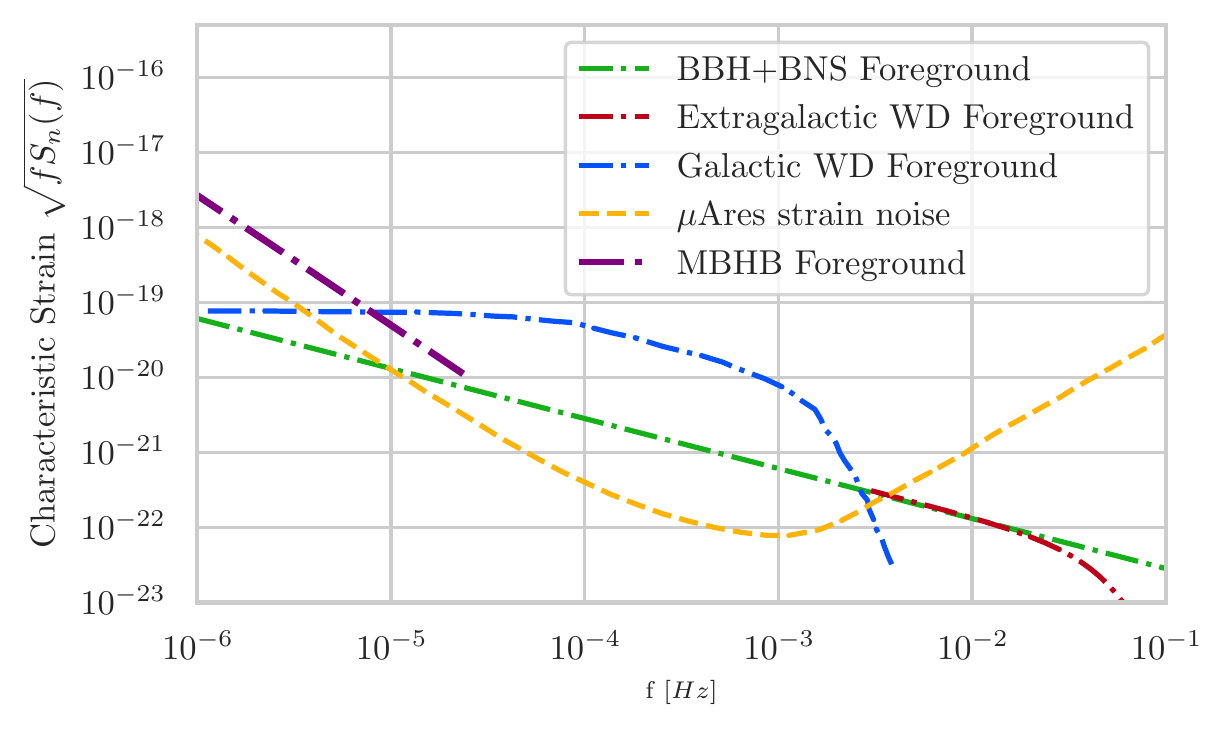} 
    \end{subfigure}
	\caption{Same as Figure~\ref{fig:LISA_DO_fgs} but for AEDGE (left panel) and $\mu$Ares (right panel). The relevant foregrounds are the BBH+BNS (dot-dashed green) and the EGWD binaries (dot-dashed red) for AEDGE and additionally the unresolved MBHB (dot-dashed purple) and GWD binaries (dot-dashed blue) for $\mu$Ares.}\label{fig:AEDGE_Ares}
\end{figure}

\paragraph{DO Optimal/Conservative and AEDGE ---}
The DO interferometer, both in its Optimal and Conservative incarnations, suffers mainly from the presence of the BBH+BNS foreground (right panel of Figure \ref{fig:LISA_DO_fgs}). The contribution from GWD is almost irrelevant in both designs, as it affects only the very low-frequency part of both sensitivity curves. The same holds for the AEDGE experiment (left panel of Figure \ref{fig:AEDGE_Ares}), since it has similar sensitivity and frequency range to DO Optimal (Figure \ref{fig:all_strains}). For DO (Optimal and Conservative) and AEDGE, we make the same foreground cleaning assumption described above for LISA.

\begin{figure}
    \begin{subfigure}[b]{0.5\textwidth}
        \includegraphics[width=1.0\textwidth]{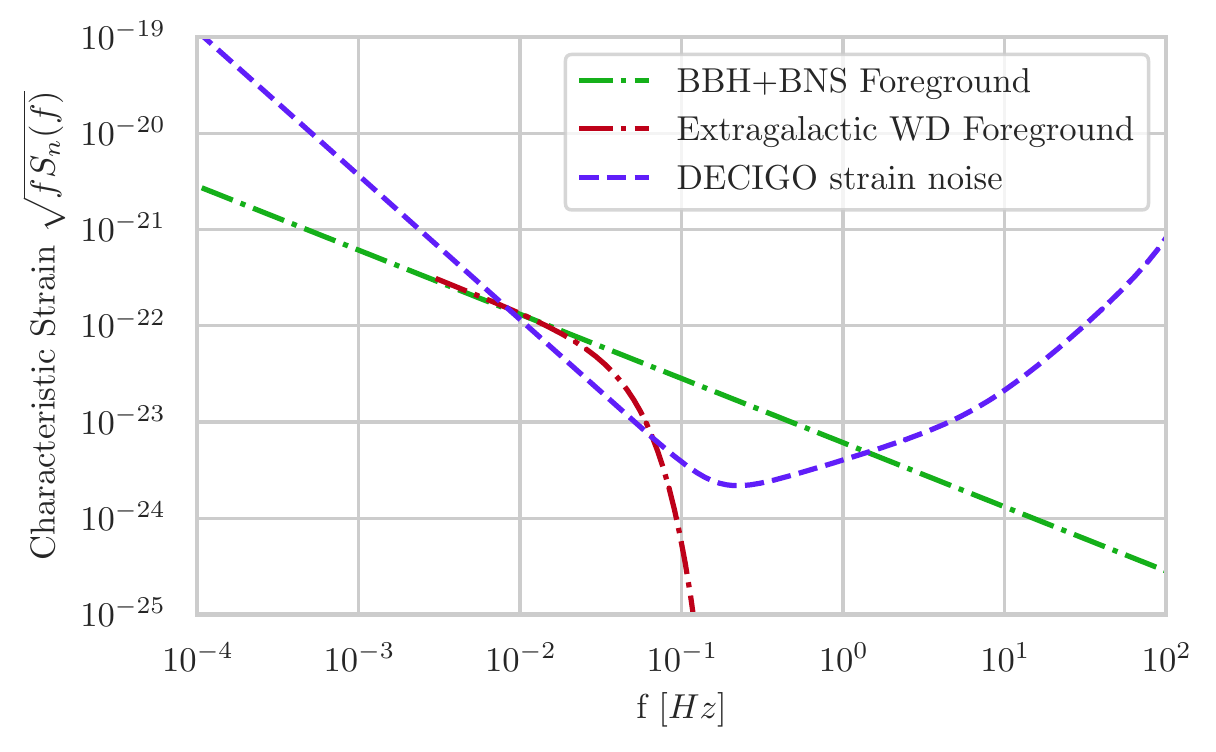} 
    \end{subfigure}
    \begin{subfigure}[b]{0.5\textwidth}
        \includegraphics[width=1.0\textwidth]{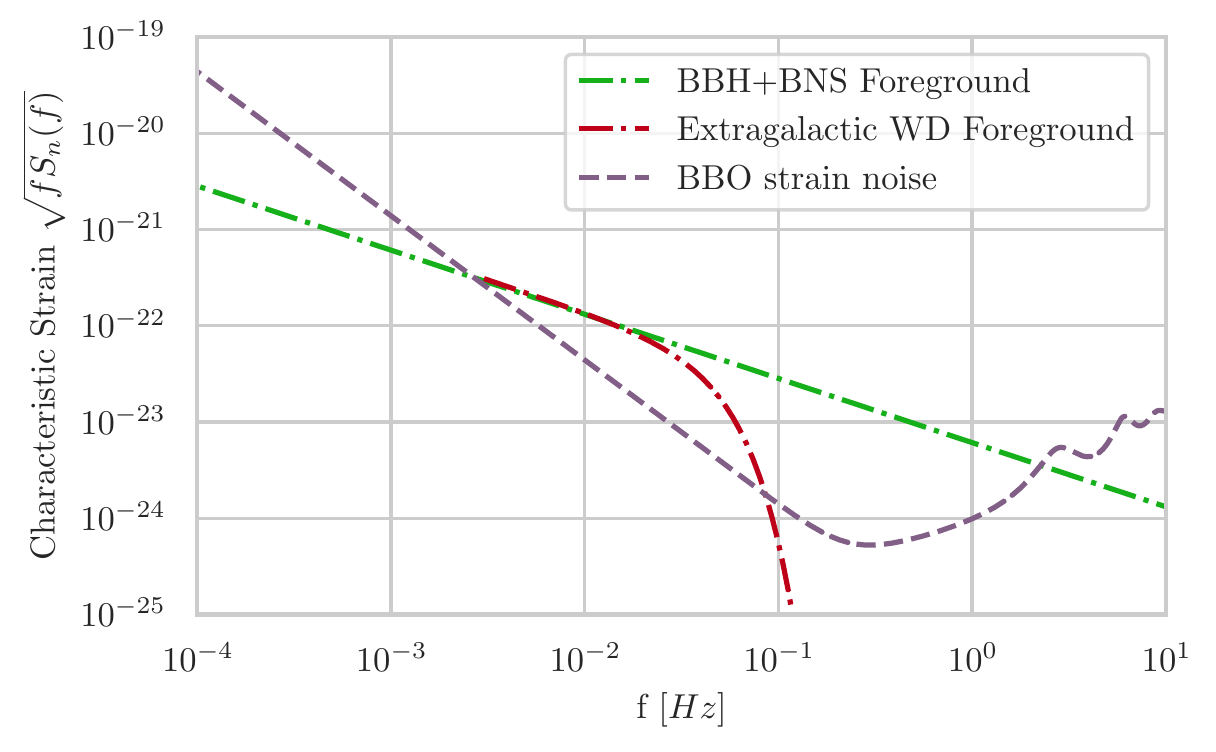} 
    \end{subfigure}
	\caption{Same as Figure~\ref{fig:LISA_DO_fgs} but for DECIGO (left panel) and BBO (right panel). The relevant foregrounds are the EGWD (dot-dashed red) and the BBH+BNS foreground (dot-dashed green).}\label{fig:DECIGO_BBO_fgs}
\end{figure}

\paragraph{DECIGO and BBO ---}
    As for DECIGO and BBO, the BBH+BNS foreground constitutes the main  contaminant in their bands (Figure \ref{fig:DECIGO_BBO_fgs}). BBH and BNS can be individually resolved and subtracted by these two ultra-sensitive interferometers, leaving a residual  foreground with amplitude a factor $\sim 5\times 10^{-3}$ smaller than the original one in the DECIGO case \citep{Nishizawa,Yagi:2011wg}, while BBO should be capable instead of resolving all compact binaries in its band, thanks to its deeper sensitivity, which allows to fully subtract them \citep{cutler_harms,Nishizawa}. 
    
    Despite the subtraction of resolved sources, the BBH+BNS foreground remains a strong limiting factor for DECIGO: it has been suggested in the past that an improvement of a factor $\sim 2$ or $\sim 3$ in sensitivity should be enough to fully resolve and subtract this foreground \citep{Nishizawa}. In alternative to this sensitivity boost, we also try to consider the possibility  of a multi-band cleaning for DECIGO using ET, corresponding to $\sigma_{fg}\sim 10^{-3}$; however, in this case the sensitivity does not improve significantly with respect to the case in which we only use the spectral dependence to subtract the foreground. We conclude that in order to fully restore the SGWB detection power of DECIGO, we need at least a value $\sigma_{fg}\sim 10^{-6}$, which seems to be at the moment outside of the reach of external ground-based experiments.    
    
    The EGWD confusion noise is expected to contribute mostly in DECIGO's and BBO's bands in the range  $10^{-3}-10^{-1}$ Hz. As we discussed in LISA's case, potentially this foreground could  be subtracted in a very efficient way using its unique spectral dependence, and moreover its contribution is expected to drop very steeply beyond $f\sim \SI{0.2}{\hertz}$. For these reasons, we discuss results for DECIGO and BBO for both an optimistic case, neglecting this foreground, and a pessimistic case without any subtraction, cutting off frequencies $\lesssim \SI{0.2}{\hertz}$ \citep{Yagi:2011wg, Farmer_Phinney_2003}.

\paragraph{$\mu$Ares ---}
    The sensitivity curve of $\mu$Ares appears to be strongly affected by the Galactic WD foreground in almost all the experiment bandwidth (right panel of Figure \ref{fig:AEDGE_Ares}). Although a treatment similar to LISA's one \citep{adams_cornish} does not exist at the moment for $\mu$Ares, we assume that a subtraction strategy, exploiting the anisotropy and time-modulation of this foreground can be applied, perfectly subtracting this contaminant. The EGWD foreground, on the other hand, should be of secondary importance for $\mu$Ares, and we neglect it in the following.
    
   The BBH+BNS foreground is also very relevant in the $\mu$Ares band, therefore we adopt a multi-band cleaning approach as in LISA's case: for the AX1 model, a value $\sigma_{fg} \sim 0.1$ is enough to obtain multiple detections over $\mu$Ares band, while for the other SGWB models we show results for $\sigma_{fg}\sim 10^{-3}$ (in conjunction with third-generation detector ET). However, even the latter value for $\sigma_{fg}$ strongly limits the prospects for detection for an inflationary SGWB (see Figure \ref{fig:flat_Ares}): we conclude that to restore the full ``sensitivity bucket'' of $\mu$Ares, we would need a value $\sigma_{fg}\lesssim 10^{-5}$, which is not foreseen at the moment using for proposed third-generation ground-based detectors. 
   
     In addition to the three components we described above,  the coalescence of MBHB could produce an unresolved foreground between $\sim 10^{-7}$ and $\sim 10^{-5}$ Hz (dot-dashed purple curve). 
     We approximate this foreground as a power-law in the $\mu$Ares band (see Section \ref{sec:fgs_fit}) and we subtract it by means of its spectral shape. Therefore, in the $\mu$Ares case we will use a version of the filter \ref{eq:Q_Davide} extended to multiple foregrounds, adopting for the BBH+BNS component a multi-band approach with finite $\sigma_{fg}$ and the limit $\sigma_{fg}\rightarrow \infty$ for the MBHB one.

\paragraph{ET ---}
Concerning the ground-based ET, the dominant source of confusion noise in this frequency band is represented by BBH and BNS (left panel of Figure \ref{fig:SKA_fgs}). The BBH contribution can be remarkably reduced by individually resolving sources, leaving a residual with amplitude a factor $\sim 200$ smaller than the initial foreground, while the BNS and BH-NS contributions are more strenuous and can be reduced only by a factor $\sim 2$  and $\sim 10$, respectively \citep{Zhu:2012xw}. The total BBH+BNS foreground can be reduced through individual subtraction by an overall factor of $\sim 3$ \citep{Zhu:2012xw}, thus still constituting an important limiting factor when attempting a detection of the primordial SGWB.   We adopt in this case the $\sigma_{fg}\rightarrow \infty$ limit in our filter for ET, corresponding to subtraction using only the spectral dependence of this foreground.


\subsection{SKA}\label{sec:SKA}
 For SKA we optimistically include only the white noise component in the noise budget; however, we note that the so-called ``red noise'' component due to pulsar timing noise \citep{Hazboun_2019} could be present in the data, raising considerably the noise level in the lower frequency part of the PTA sensitivity curves. We use the the codes \texttt{hasasia}\footnote{\url{https://hasasia.readthedocs.io/en/latest/index.html}} \citep{Hazboun_2019} and \texttt{gwent}\footnote{\url{https://gwent.readthedocs.io/en/latest/index.html}} to compute the sensitivity to the SGWB, choosing for the pulsars an rms timing residual of  $\sigma_{t}=50$ ns, an observing time  $T=10$ yr, a number of pulsars $N_{p}=200$ and an average observation cadence of 1 per week \citep{Mingarelli:2019mvk}. 
 
 \begin{figure}
     \begin{subfigure}[b]{0.5\textwidth}
        \includegraphics[width=1.0\textwidth]{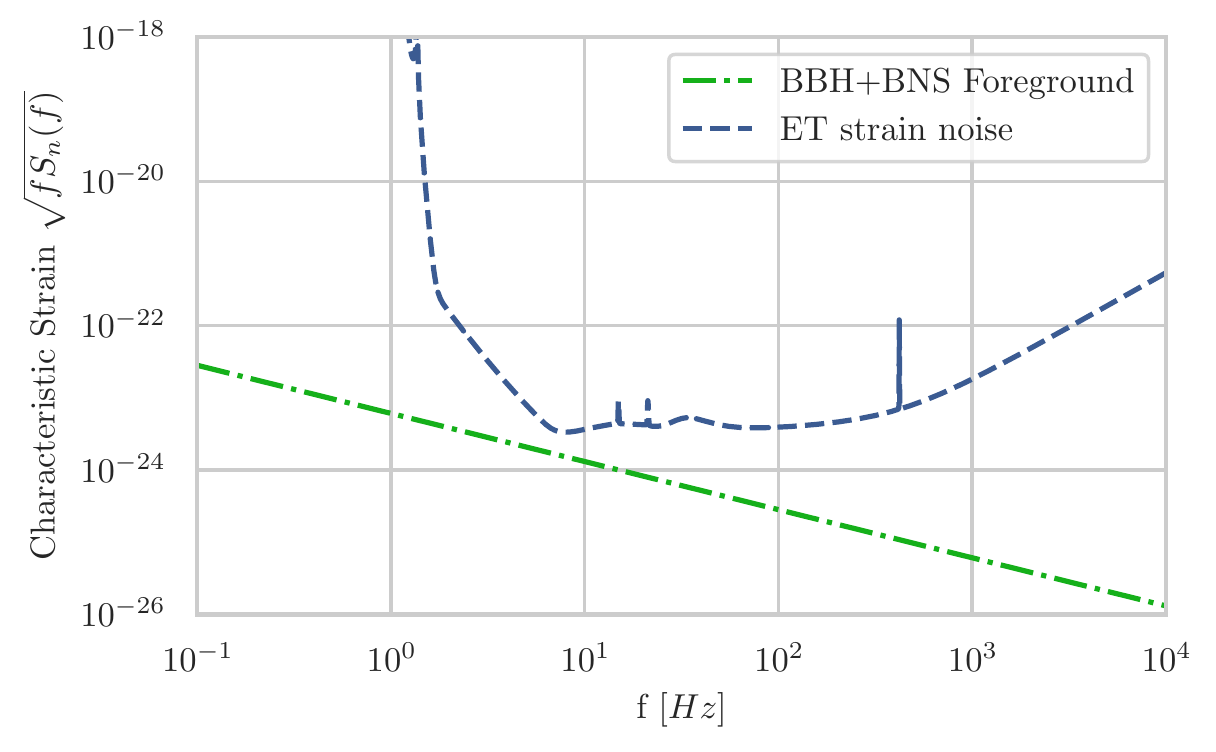} 
    \end{subfigure}
    \begin{subfigure}[b]{0.5\textwidth}
        \includegraphics[width=1.0\textwidth]{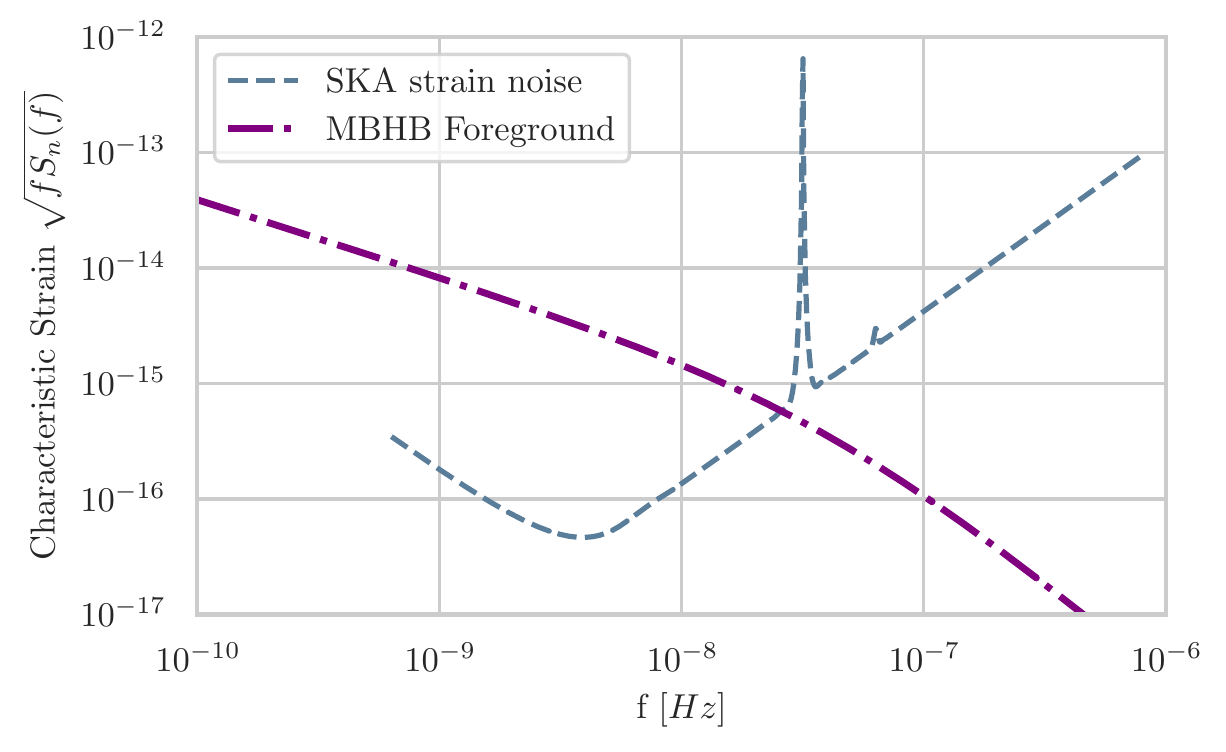} 
    \end{subfigure}
    \caption{Same as Figure~\ref{fig:LISA_DO_fgs} but for ET (left panel) and SKA (right panel). The relevant foreground for this experiments, namely the unresolved BBH+BNS populations for ET and the MBHB foreground for SKA, are shown in the dot-dashed green and purple lines, respectively.}\label{fig:SKA_fgs}
\end{figure}

 As for the foreground, Figure \ref{fig:SKA_fgs} shows that the most sensitive part of the SKA bandwidth will be limited by the presence of the MBHB astrophysical foreground. We approximate this component as a power-law in the SKA band (Section \ref{sec:fgs_fit}) and we adopt a subtraction strategy based on the spectral shape of this contaminant, using  the limit $\sigma_{fg}\rightarrow \infty$ in the SNR and filter expressions in Eqs.~\ref{eq:SNR_with_fgs} and \ref{eq:Q_Davide}, respectively.
 
\section{Results}\label{sec:results}
In this Section we present the forecasts for CMB, PTA, and interferometers, described respectively in Sections \ref{sec:cmb} and \ref{sec:direct}. We will first show the binned sensitivity curves obtained for LiteBIRD, SKA, and all the direct detection experiments (Section \ref{sec:binned_curves}). Then, in Section \ref{sec:errorbar_results} we will proceed to show the error bars for each experiment and each of the five example tensor power spectrum models described in Section \ref{sec:tensor_ps}.

\subsection{Binned $\Omega_{GW}$ Sensitivity Curves}\label{sec:binned_curves}
\begin{figure}
\centering
        \includegraphics[scale=0.9]{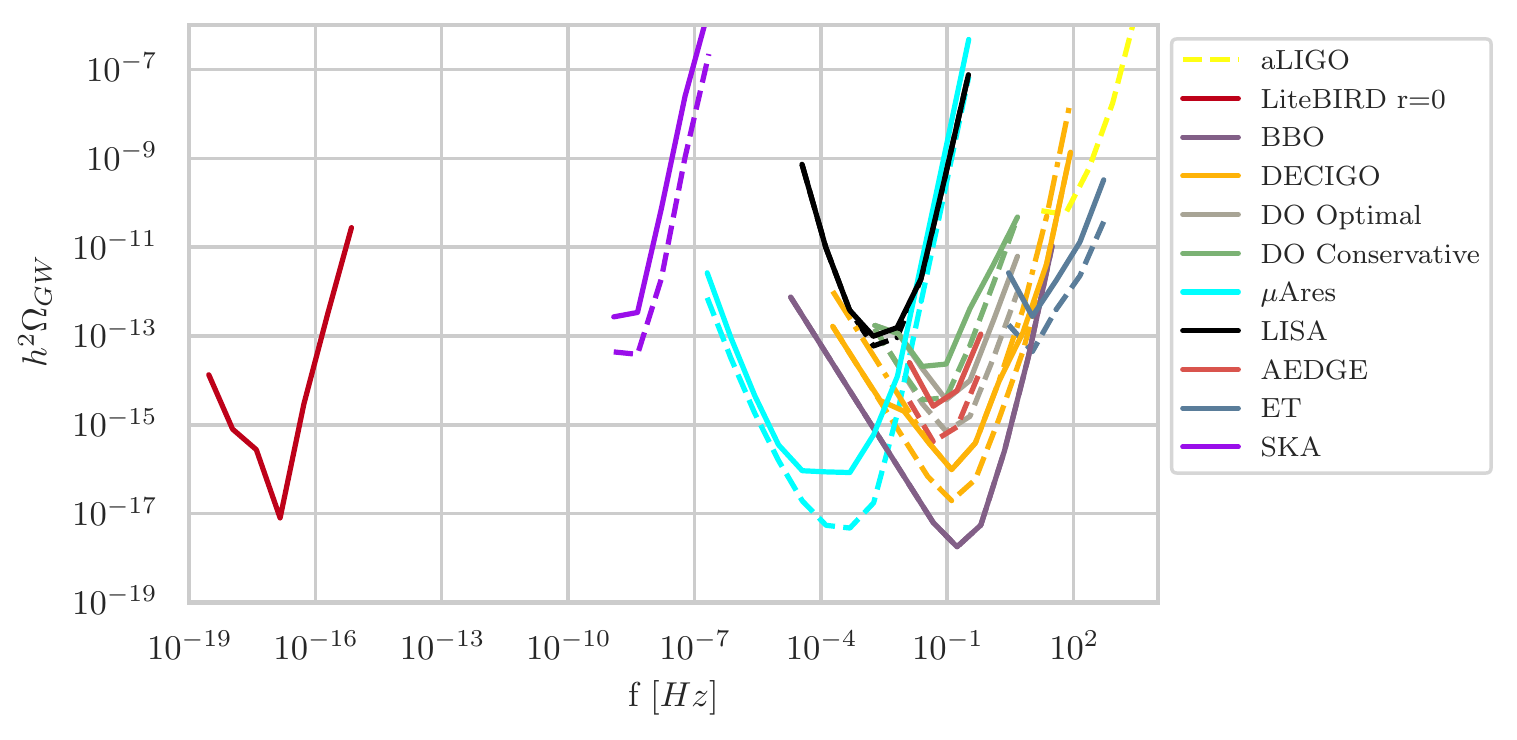}
        \caption{Sensitivity curves on the energy density of gravitational waves $\Omega_{GW}$ with (solid and  dot-dashed lines) and without (dashed lines) the contribution of the astrophysical foregrounds for all the experiments considered in this work, obtained with a logarithmic binning in wavenumber with $\Delta\ln k = 1.2$. Note that we assume for the BBH+BNS cleaning a value $\sigma_{fg} \sim 0.1$ for LISA, DO Optimal and  DO Conservative for the case with foregrounds, while for $\mu$Ares we choose $\sigma_{fg} \sim 10^{-3}$. For DECIGO we show two curves with foregrounds: the dot-dashed one assumes subtraction using only the spectral dependence of the foreground, while the dashed one assumes multi-band cleaning with $\sigma_{fg}\sim 10^{-3}$. We also plot for reference the sensitivity curve for the aLIGO experiment without the astrophysical foregrounds.}\label{fig:binned_curves}
\end{figure}

We calculate the binned sensitivity curves to the gravitational wave energy density $\Omega_{GW}$ using Eq.~\ref{eq:cmb_binning} for the LiteBIRD CMB experiment, Eq.~\ref{eq:binning} for all the interferometers and PTA experiments (and its equivalent when taking the foregrounds into account, as described Section \ref{sec:laser_fgs}). We plot them in Figure \ref{fig:binned_curves}, choosing $\Delta\ln k = 1.2$ as the power spectrum discretization scale. 
The solid and dashed lines show the sensitivities obtained with and without the foregrounds, respectively. 
The sensitivity of a CMB experiment to $\Omega_{GW}$, as computed in Section \ref{sec:fisher}, depends on the fiducial tensor power spectrum  used to compute the $C_{\ell}$ in the Fisher matrix given in  Eq.~\ref{eq:fisher}; thus, the sensitivity curve of LiteBIRD (in red) is computed for  $r=0$  -- that is without including cosmic variance -- for consistency with the interferometers.  Note also that the LiteBIRD sensitivity always includes the foregrounds.

We find  that the best sensitivity of LiteBIRD (including foregrounds) at frequencies $f\sim 10^{-17}\,\SI{}{\hertz}$ is similar to those of the most advanced among the interferometers, namely $\mu$Ares at $f\sim 10^{-3}\,\SI{}{\hertz}$ and DECIGO and BBO at $f\sim 10^{-1}\,\SI{}{\hertz}$. However, when plotting error bars on the  model predictions in the next sub-Section, we find that the shape of the GW spectrum is very different for CMB and interferometer frequencies. It has a rising spectrum towards the CMB frequency after the transition between the matter and radiation dominated eras, while for the single-field slow-roll model it rapidly flattens out at higher frequencies, making a detection challenging for interferometers. The situation changes dramatically for some parameter choices of the axion-SU(2) model, which can produce a strongly blue-tilted signal easily detectable at interferometer frequencies \citep{Thorne}.

Figure~\ref{fig:binned_curves} also highlights the fact that the frequency window between $\sim10^{-16}-10^{-9}\,\SI{}{\hertz}$ is devoid of any experiment. The constraints on the SGWB intensity in this range come only from indirect limits, such as the BBN, second-order back-reaction and CMB shortwave calculations \citep{Clarke_2020}. 

Concerning the effect of the foregrounds, we find significant impacts in the frequency range $10^{-3}-10^{2}\,\SI{}{\hertz}$,  due to unresolved BBH and BNS, mainly limiting the sensitivity of $\mu$Ares, DO, AEDGE, DECIGO and ET. In particular, the DO Optimal and Conservative (grey and green curves in Figure \ref{fig:binned_curves}  respectively) and AEDGE (in light red) experiments seem to be significantly affected by this component for a fractional uncertainty $\sigma_{fg}\sim 0.1$  on the amplitude of the BBH+BNS foreground \citep{PCA_fgs}. However, this does not prevent detection of the AX1 model at high significance in multiple bins for these three experiments, as we will see in the next Section. Moreover, it will possible to further decrease the $\sigma_{fg}$ for these experiments to the $\sim 1.3\times 10^{-2}$ level using external information from third generation ground-based experiments such as CE \citep{Pan_Yang_2019} (Section \ref{sec:laser_fgs}), or even to the $\sim 10^{-3}$ level using ET (M. Pieroni, A. Ricciardone e E. Barausse in preparation). On the other hand, DECIGO (in orange) and $\mu$Ares (in cyan) are appreciably limited by the BBH+BNS foreground even for values $\sigma_{fg}\sim 10^{-3}$. Moreover, $\mu$Ares is also affected -- especially at lower frequencies -- by the presence of the MBHB foreground, which limits the sensitivity of SKA (light purple curve) as well. 

\subsection{Error bars for the spectator Axion-SU(2) models}\label{sec:errorbar_results}
Next, we calculate the $1\sigma$ error bars on $\Omega_{GW}$ for five models of the primordial tensor spectrum. Of these, three are the AX1, AX2 and AX3 models defined in Section \ref{sec:axion_model} (see Eq.~\ref{eq:params}), while two are single-field slow-roll ones with $r=0.01$ and $r=0.001$ with tensor tilt and running satisfying the inflationary consistency relation. In this Section we discuss the results for the former models, while in the next Section~\ref{sec:results_single_field} we discuss the latter. Note that we include cosmic variance in the error bars for all experiments \citep[see][for cosmic variance in interferometers cross-correlation]{Cornish_2002,saikawa_shirai_2018}. We also report here a caveat on inferring the global SNR from the binned error bars, considering two possible regimes for the filter \ref{eq:Q_Davide}: if filter is dominated by the external information ($\sigma_{fg}^{-2}\gg 2T I_{fg\times fg}$), the information is correlated among the bins and therefore simply coadding the SNR of all bins would be too optimistic; in the regime $\sigma_{fg}^{-2}\ll 2T I_{fg\times fg}$ instead, the simple coaddition of each bin's SNR leads to pessimistic estimates of the global SNR, as the information lost due to the binning increases with the number of components removed, which is proportional to the number of bins. We checked that both these effects are irrelevant for our results in the selection of cases presented in this work.

\begin{figure}
\centering
        \includegraphics[scale=1.1]{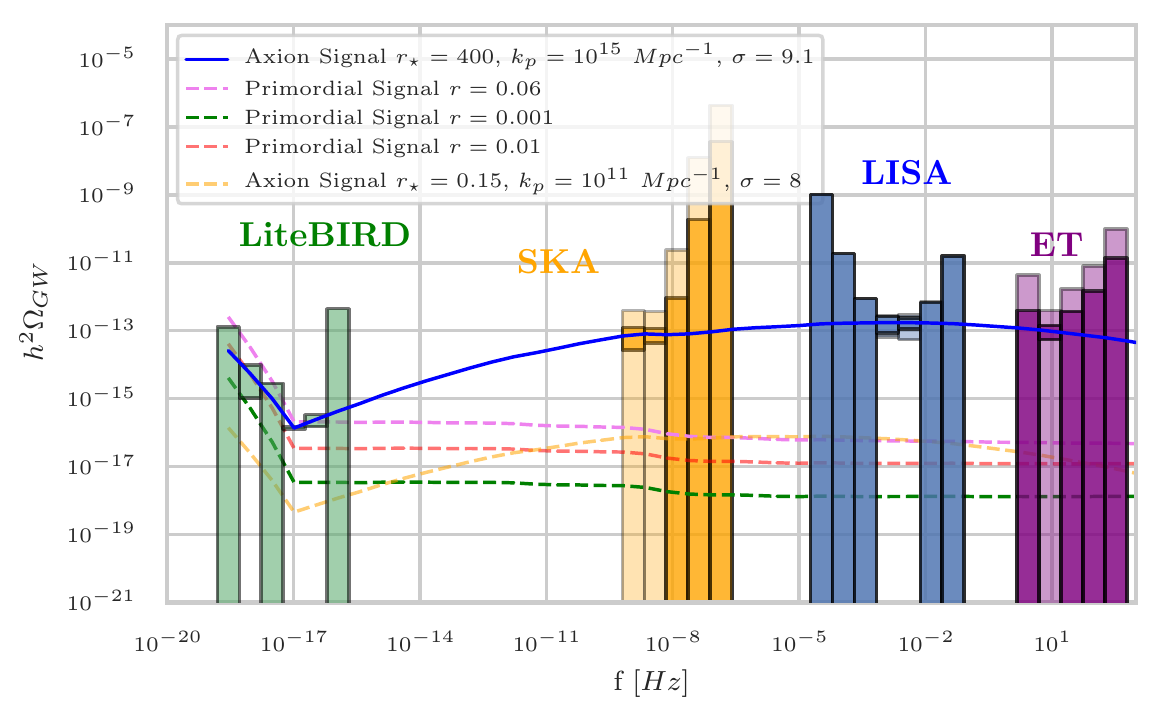}
        \caption{Expected $1\sigma$ error bars on $\Omega_{GW}$ for the AX1 model (the solid blue line) for the LiteBIRD (green), SKA (orange),  LISA (blue), and ET (purple). We show the constraints with and without the astrophysical foregrounds in the light and dark shared areas, respectively.  We use the logarithmic binning in wavenumber with $\Delta\ln k = 1.2$. We also show for comparison the other tensor spectrum models adopted in this paper (dashed lines), including the BICEP2/Keck/Planck upper bound $r=0.06$. 
        }\label{fig:axion1_LISA}
\end{figure}

In Figure~\ref{fig:axion1_LISA}, we show the results for LiteBIRD, SKA, LISA and ET. 
The light and dark shaded areas show the error bars for the AX1 model with and without the astrophysical foregrounds included in our calculation. We always take the foregrounds into account for the LiteBIRD CMB satellite, as explained in Section \ref{sec:CMB_noise}. 

For what concerns the AX1 model, we tuned its parameter set to have simultaneous detections in both the CMB and the interferometers ranges, while still being consistent with the BICEP2/Keck/Planck upper bound at CMB scales (the dashed pink curve in Figure \ref{fig:axion1_LISA}). As can be seen from the plots, this model can be detected in the PTA and ground-based range,  only when neglecting the foreground contamination. 

By observing closely the CMB part of the spectrum, the LiteBIRD error bars clearly show two peaks of sensitivity corresponding to the reionization bump (second bin from the left) and the recombination bump (fourth and fifth bins from the left), as we anticipated in Section \ref{sec:cmb}. Both these bumps corresponds to detections of this model (green error bars in Figure \ref{fig:axion1_LISA}).

For the space-borne interferometer LISA, we adopt a multi-band cleaning of the BBH+BNS foreground, exploiting external information from Advanced LIGO/Virgo, which provides $\sigma_{fg}\sim 0.1$, as we discussed in Section \ref{sec:laser_fgs}: this allows detection in two bins either with or without foregrounds (blue error bars in Figure \ref{fig:axion1_LISA}).

The ground-based ET  shows detections only in the absence of foregrounds (purple error bars in Figure \ref{fig:axion1_LISA}. We tried to tune the axion-SU(2) parameter set to have detections also from ground-based interferometers in the presence of the foreground, but were not successful due to the attractor behaviour of the theory and the CMB upper bounds, as explained in Section \ref{sec:axion_model}. 

Similarly, SKA shows two detections in the foreground-less case but none in the case with foregrounds for this model. This confirms PTA surveys as a useful instrument to characterize exotic SGWB models with bump- or peak-like features in their frequency range \citep{Garcia-Bellido_2016}, highlighting however the fact that they could be limited in practice by the presence of the MBHB astrophysical foreground.

\begin{figure}
\centering
        \includegraphics[scale=1.1]{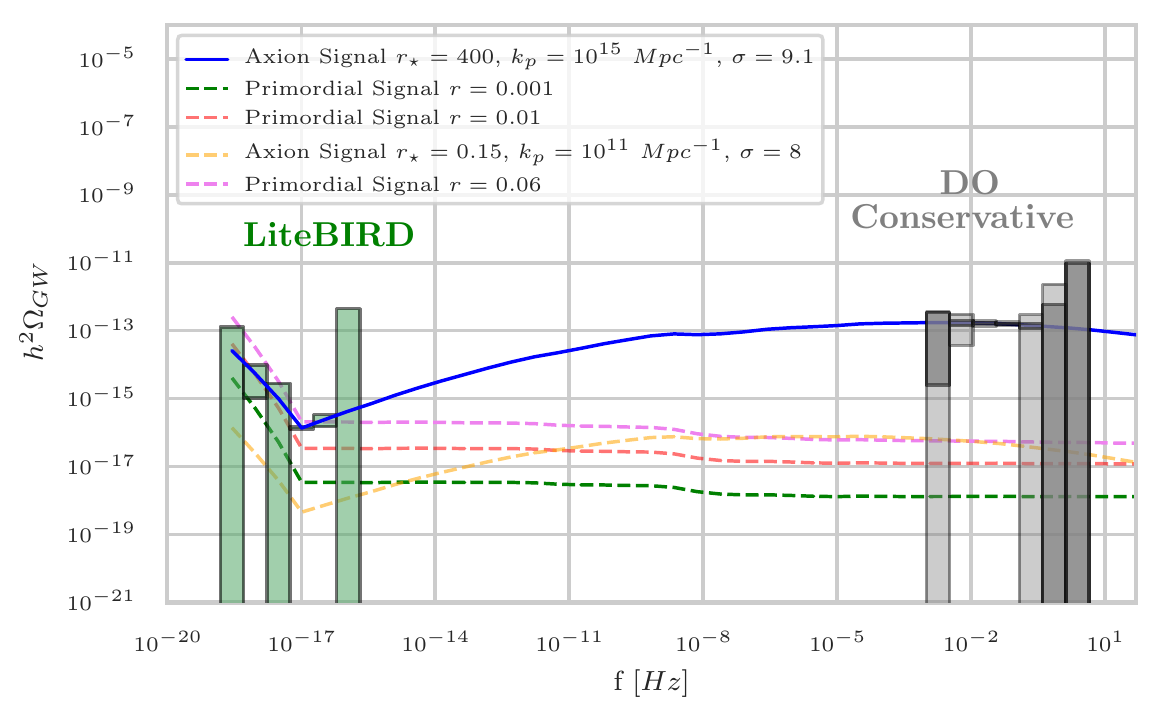}
        \caption{Same as Figure~\ref{fig:axion1_LISA} but for the DO Conservative.
        }\label{fig:axion1_DO_cons}
\end{figure}

\begin{figure}
\centering
        \includegraphics[scale=1.1]{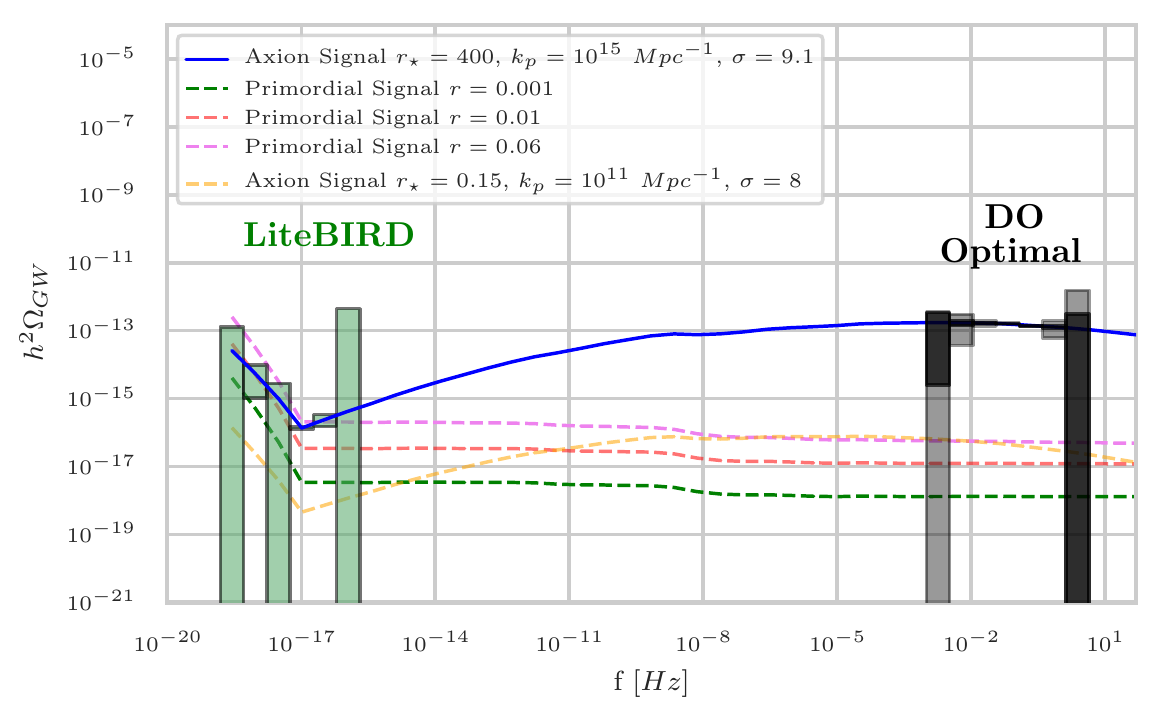}
        \caption{Same as Figure~\ref{fig:axion1_LISA} but for the DO Optimal.
        }\label{fig:axion1_DO}
\end{figure}

\begin{figure}
\centering
        \includegraphics[scale=1.1]{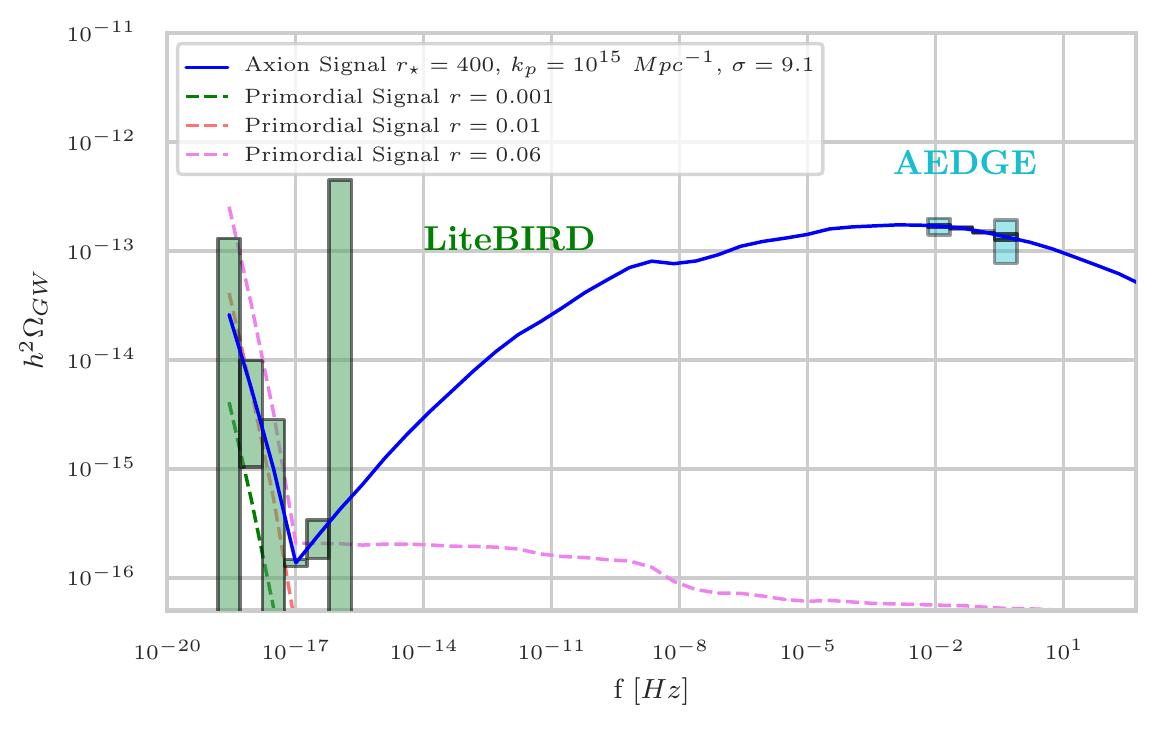}
        \caption{Same as Figure~\ref{fig:axion1_LISA} but for the AEDGE. Note a different scale for the vertical axis.
        }\label{fig:axion1_AEDGE}
\end{figure}

In Figures~\ref{fig:axion1_DO_cons}--\ref{fig:axion1_AEDGE}, we show the expected error bars for the AX1 model for the other interferometers. We show the error bars only for the experiments that can give a detection (without the foregrounds contamination) in at least one bin. Therefore, we show DO Conservative, DO optimal and AEDGE only for the AX1 model, which has the strongest signal in the frequency range favorable to them. We do not show them for the other models because they would not be able to have a detection in at least one bin. However, we make an exception for DECIGO and BBO and do not show their error bars for the AX1 and AX3 models despite excellent prospects for the detection, since these experiments are so sensitive that the error bars would be invisible, similarly to what happens for the $\mu$Ares experiment.


Figures \ref{fig:axion1_DO_cons} and \ref{fig:axion1_DO} show that the error bars for the DO Conservative and Optimal designs are similar for this particular model, with the less-sensitive Conservative setup having one detection missing with respect to the Optimal case in the next to last bin. In both cases, the foreground contamination appears to  have a small impact.  
Figure \ref{fig:axion1_AEDGE} shows the error bars for the AEDGE atomic interferometer: this detector shows a similar sensitivity to the DO Optimal design, with the latter being slightly less sensitive while covering a wider frequency range.

\begin{figure}
\centering
        \includegraphics[scale=1.1]{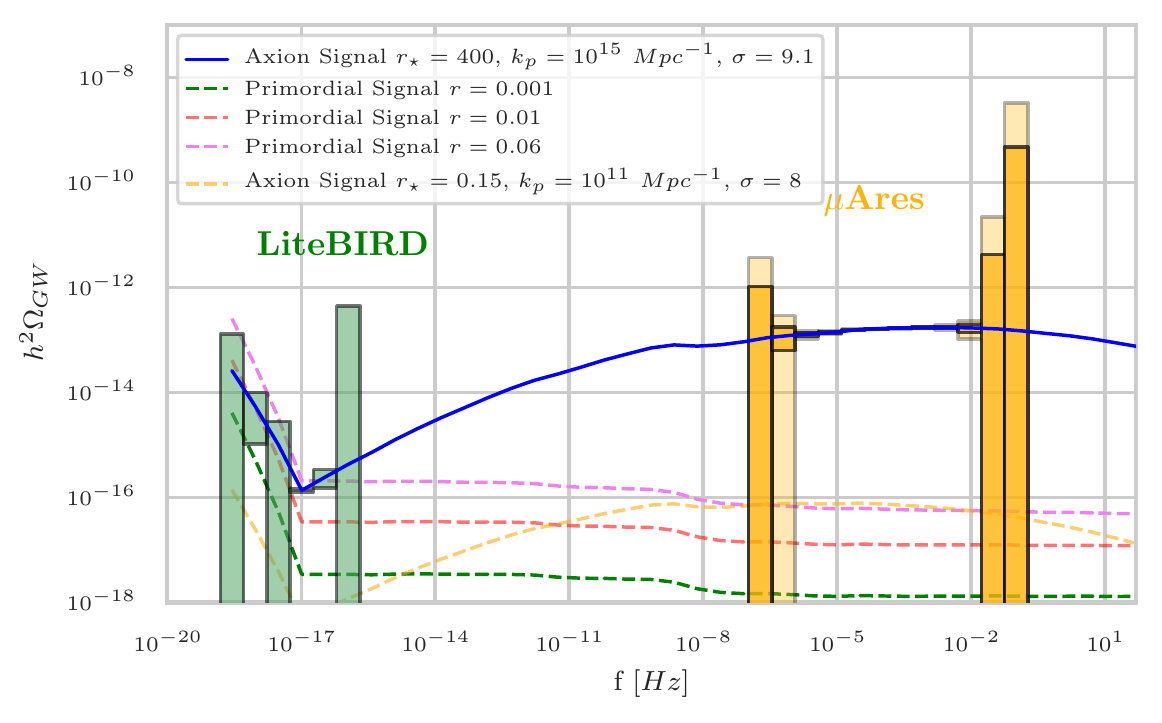}
        \caption{Same as Figure~\ref{fig:axion1_LISA} but for the $\mu$Ares. Note a different scale for the vertical axis.
        }\label{fig:axion1_Ares}
\end{figure}

The error bars on the AX1 model for the $\mu$Ares mission are shown in Figure \ref{fig:axion1_Ares}. The foreground contamination plays a minor role in this very high SNR case, and the $\mu$Ares is capable of detecting this model across an impressive range of frequencies $\sim 10^{-6}-10^{-2}\,\SI{}{\hertz}$, even for a value as high as $\sigma_{fg}=0.1$ for the BBH+BNS multi-band foreground cleaning, as provided by Advanced LIGO/Virgo.

\begin{figure}
\centering
        \includegraphics[scale=1.1]{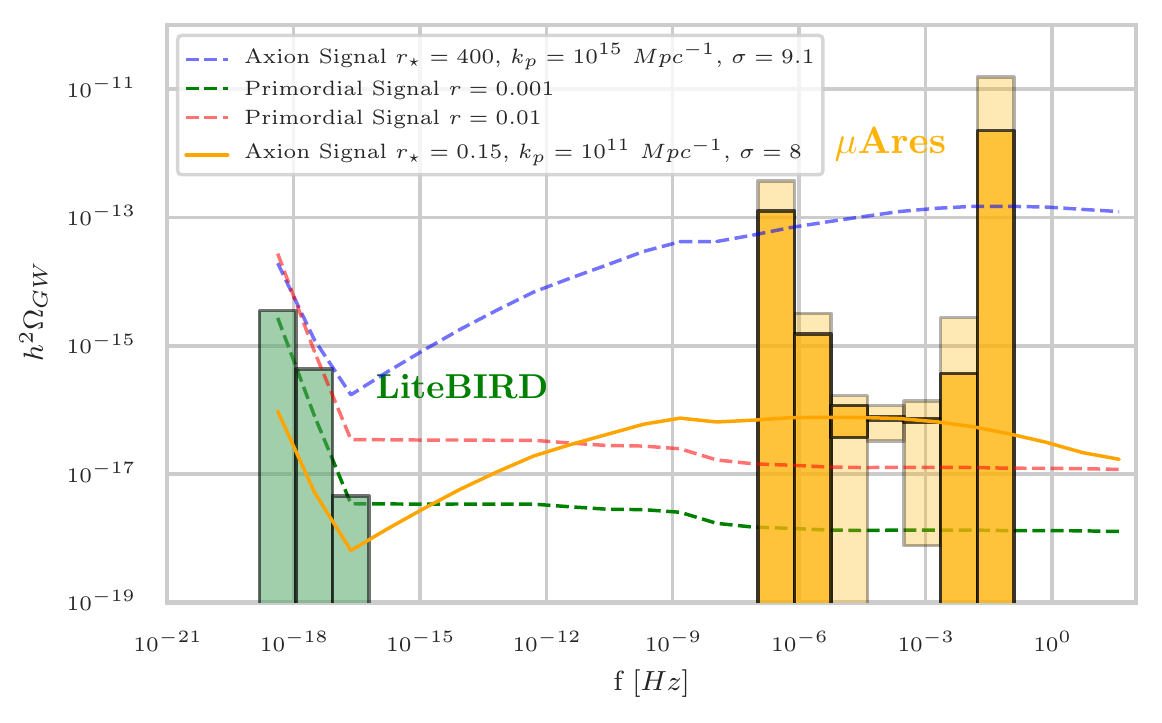}
        \caption{Expected $1\sigma$ error bars on $\Omega_{GW}$ for the AX2 model (the solid orange line) for the LiteBIRD (green) and 
        $\mu$Ares (orange). We show the constraints with and without the astrophysical foregrounds in the light and dark shared areas, respectively.  We use the logarithmic binning in wavenumber with $\Delta\ln k = 2.0$. Note that the error bars with foregrounds are computed assuming a multi-band cleaning with $\sigma_{fg} \sim 10^{-3}$. We also show for comparison the other tensor spectrum models adopted in this paper (dashed lines).
        }\label{fig:axion2_Ares}
\end{figure}

\begin{figure}
\centering
        \includegraphics[scale=1.1]{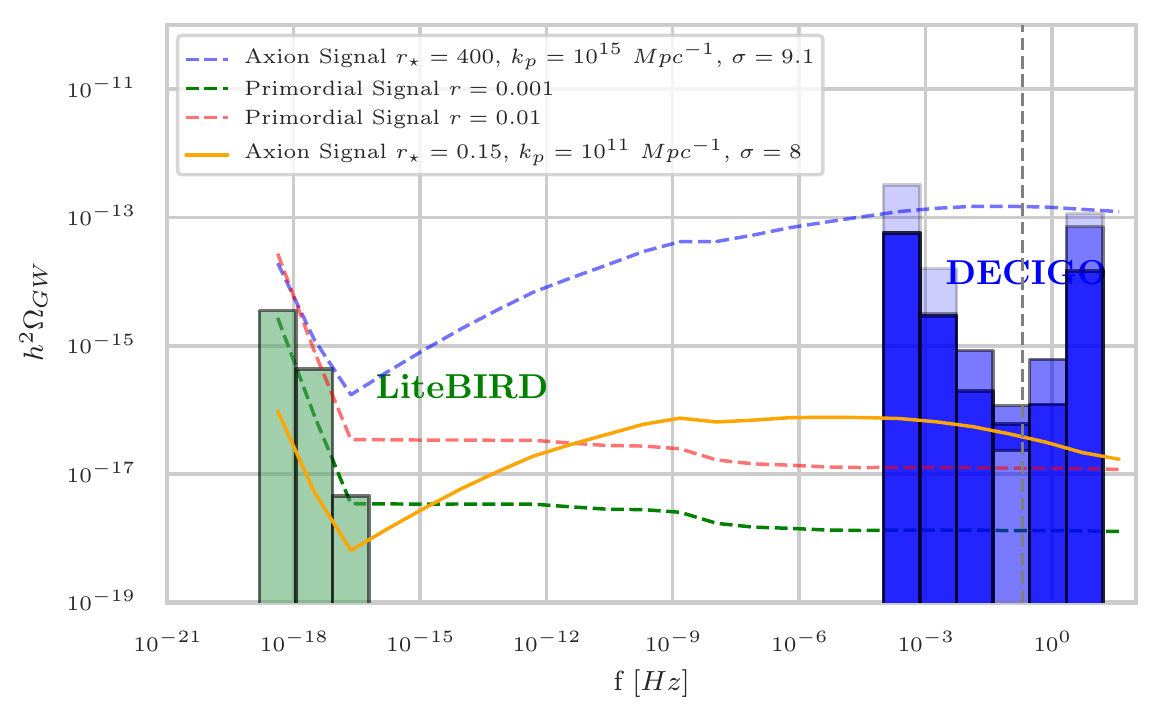}
        \caption{Same as Figure~\ref{fig:axion2_Ares} but for the DECIGO. We show two different cases for the error bars with foregrounds: the very light shaded areas correspond to the case in which we we only use the spectral dependence to subtract the foreground, while the light shaded error bars assume instead multi-band cleaning with $\sigma_{fg}\sim 10^{-3}$. The grey vertical line indicates the EGWD frequency cutoff at $\sim \SI{0.2}{\hertz}$, if we pessimistically assume no subtraction of this foreground is possible.
        }\label{fig:axion2_DECIGO}
\end{figure}

\begin{figure}
\centering
        \includegraphics[scale=1.1]{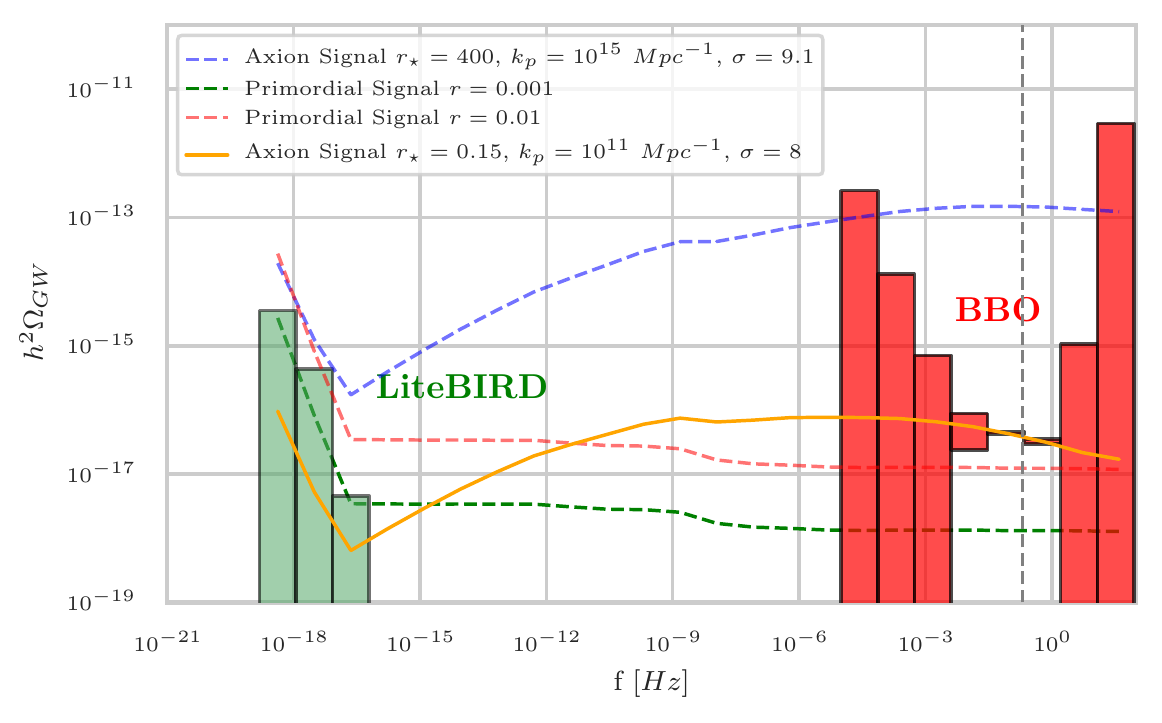}
        \caption{Same as Figure~\ref{fig:axion2_DECIGO} but for the BBO.
        }\label{fig:axion2_BBO}
\end{figure}

Next, we show the error bars for the AX2 model. This set was specifically tuned to show the capability of the axion-SU(2) to produce a signal out of the reach of LiteBIRD while being detectable in the interferometer bands.
For this case we use a larger bin size, $\Delta\ln k = 2.0$. 
In Figures \ref{fig:axion2_Ares}, \ref{fig:axion2_DECIGO} and \ref{fig:axion2_BBO}, we show the results for this model for $\mu$Ares, DECIGO and BBO, respectively. Concerning the BBH+BNS foreground cleaning, we adopt for $\mu$Ares  a value $\sigma_{fg} \sim 10^{-3}$, which is enough to have high-significance detections in three bins in the case with foregrounds, while the MBHB dforeground is subtracted exploiting its spectral shape (see Section \ref{sec:fgs_strategy}). For DECIGO, we show instead two different options for the foreground treatment: the very light shaded error bars represent the case in which we only use the spectral dependence to subtract the foreground, while the light shaded error bars assume instead multi-band cleaning with $\sigma_{fg}\sim 10^{-3}$. The dark shaded error bars show, as always, the case without foregrounds. In this case, we see that the addition of foregrounds for DECIGO does not allow any detection, while we have a detection at high significance in one bin in the case without foregrounds contamination. BBO, on the other hand, should be able to resolve and subtract all compact sources (see Section \ref{sec:fgs_strategy}): because of this, the error bars for BBO with and without foregrounds are the same, showing detections of the AX2 model at high-significance in three bins. We show in Figures \ref{fig:axion2_DECIGO} and \ref{fig:axion2_BBO} a grey vertical line indicates the EGWD frequency cutoff at $\sim \SI{0.2}{\hertz}$, if we pessimistically assume no subtraction of this foreground is possible: in this case DECIGO has no detections, even in the foreground-less case while BBO has detection in one bin.


\begin{figure}
\centering
        \includegraphics[scale=1.1]{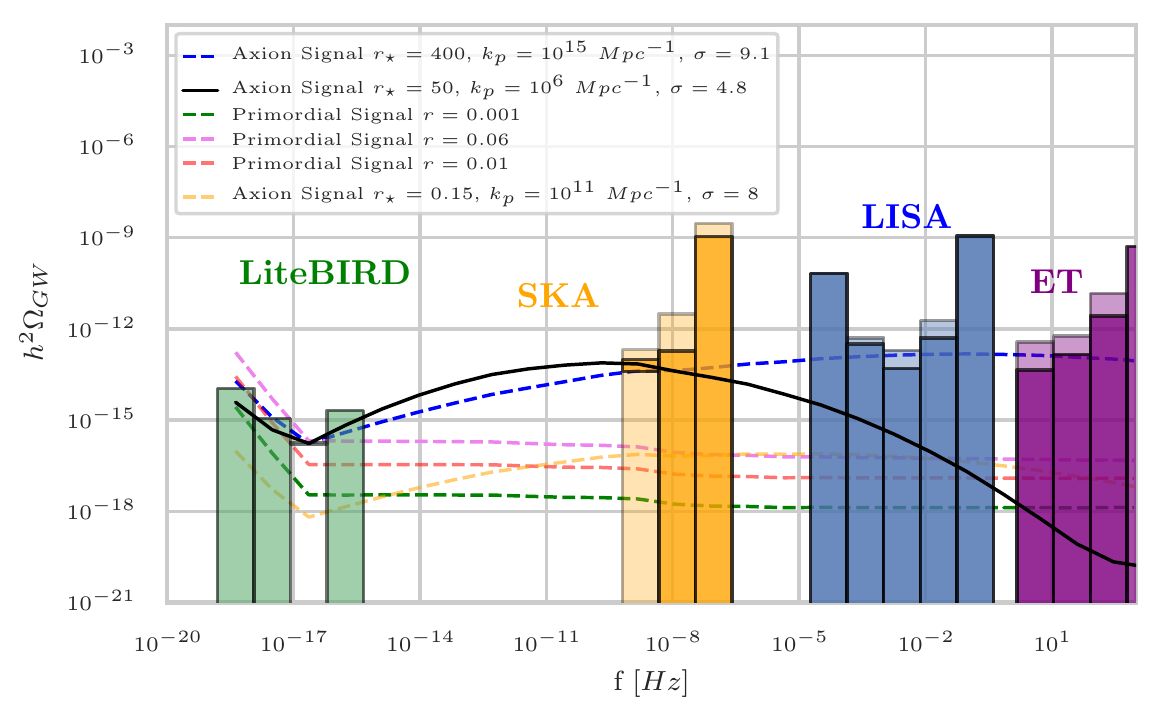}
        \caption{Same as Figure~\ref{fig:axion1_LISA} but for the AX3 model, with a logarithmic binning of $\Delta\ln k = 2.0$.
        }\label{fig:axion3_SKA}
\end{figure}

We have explored the possibility of having an axion-SU(2) tensor spectrum peaked in the PTA frequency range with the AX3 model, producing a signal detectable by SKA and LiteBIRD -- while still complying with the BICEP2/Keck/Planck upper bound on CMB scales  -- but outside the reach of LISA and ET (even without the foreground contamination, see Figure \ref{fig:axion3_SKA}) but we could not succeed in obtaining a signal detectable by SKA when accounting for the MBHB foreground contamination. This is due to the attractor nature of the axion-SU(2) model, which poses a minimum value for the Gaussian width of the spectrum bump $\sigma$ for a given peak scale $k_{p}$ (see Section \ref{sec:axion_model}). Although we do not show error bars for other experiments, we checked that the AX3 model is also not detectable by DO, AEDGE and DECIGO, while $\mu$Ares and BBO can detect it at high significance in three bins, even when accounting for foregrounds.

\subsection{Error bars on single-field slow-roll models and combined constraints on $n_{T}$}\label{sec:results_single_field}

\begin{figure}
\centering
        \includegraphics[scale=1.1]{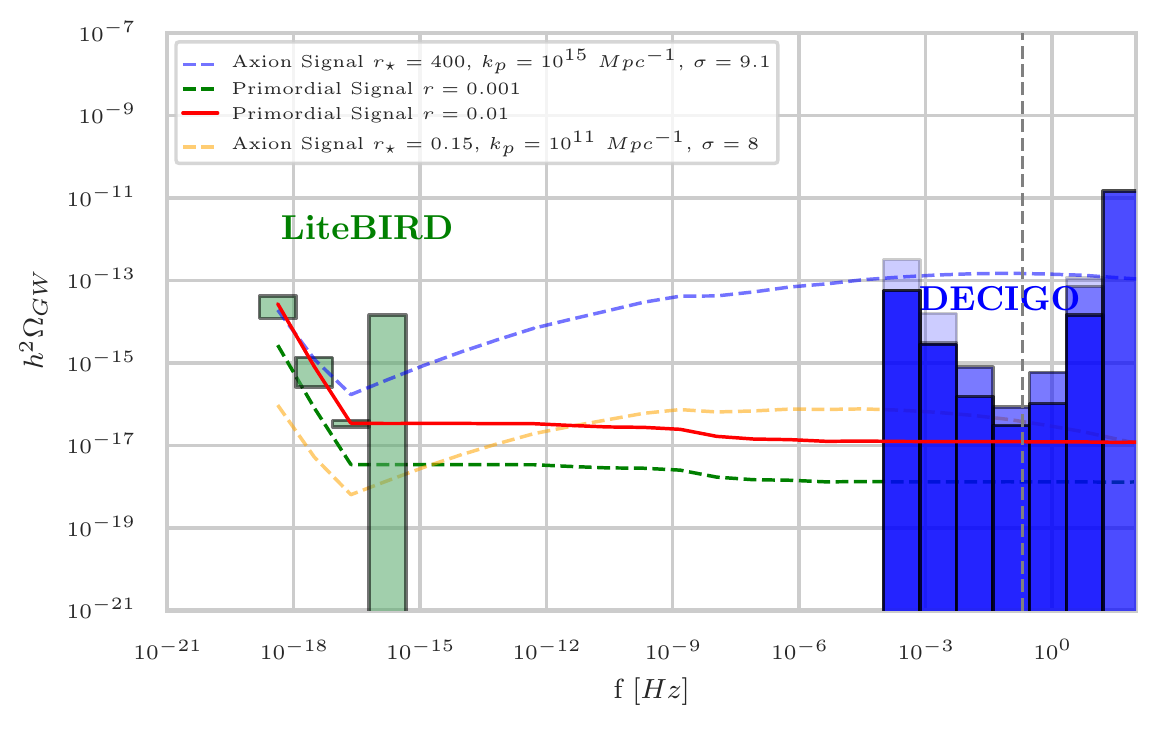}
        \caption{Expected $1\sigma$ error bars on $\Omega_{GW}$ for the single-field slow-roll model with $r=0.01$ (the solid red line) for the LiteBIRD (green) and 
        DECIGO (blue).  Dark shaded areas correspond to error bars without foregrounds. We show two different cases for the error bars with foregrounds: the very light shaded areas correspond to the case in which  we only use the spectral dependence to subtract the foreground, while the light shaded error bars assume instead multi-band cleaning with $\sigma_{fg}\sim 10^{-3}$. We use the logarithmic binning in wavenumber with $\Delta\ln k = 2.0$. The grey vertical line indicates the EGWD frequency cutoff at $\sim \SI{0.2}{\hertz}$, if we pessimistically assume no subtraction of this foreground is possible. We also show for comparison the other tensor spectrum models adopted in this paper (dashed lines).}
        \label{fig:flat_DECIGO}
\end{figure}

\begin{figure}
\centering
        \includegraphics[scale=1.1]{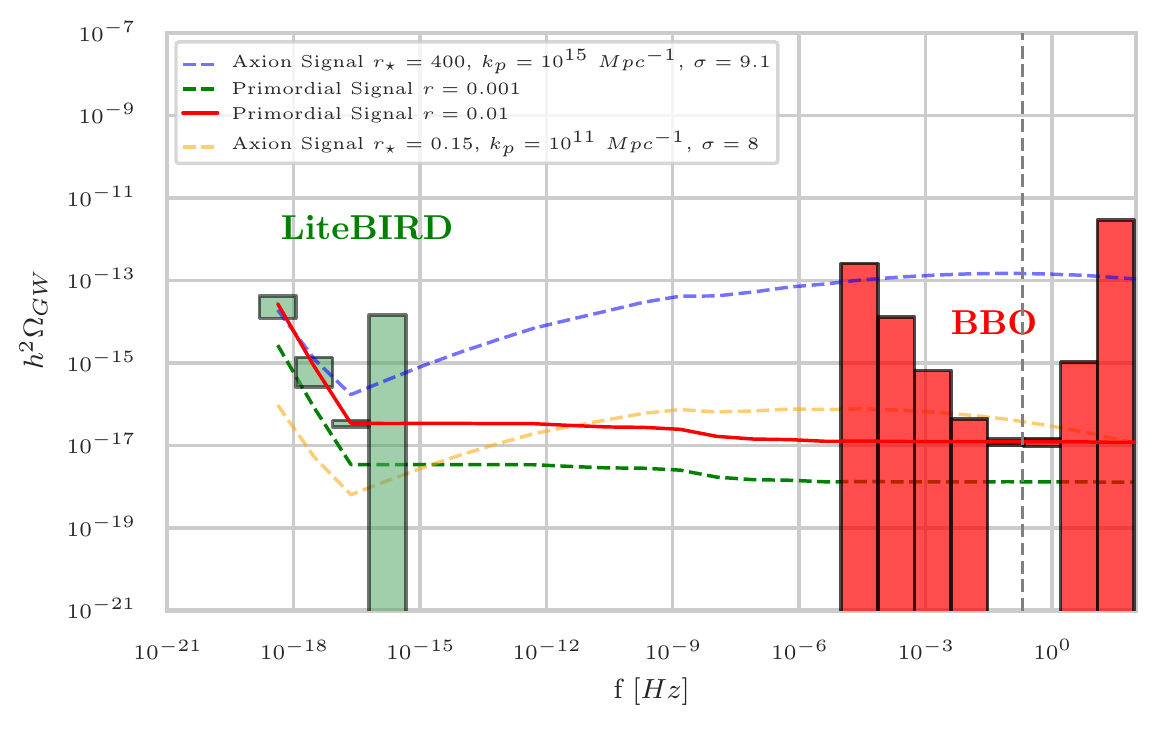}
        \caption{Same as Figure~\ref{fig:flat_DECIGO} but for the BBO.
        }\label{fig:flat_BBO}
\end{figure}

\begin{figure}
\centering
        \includegraphics[scale=1.1]{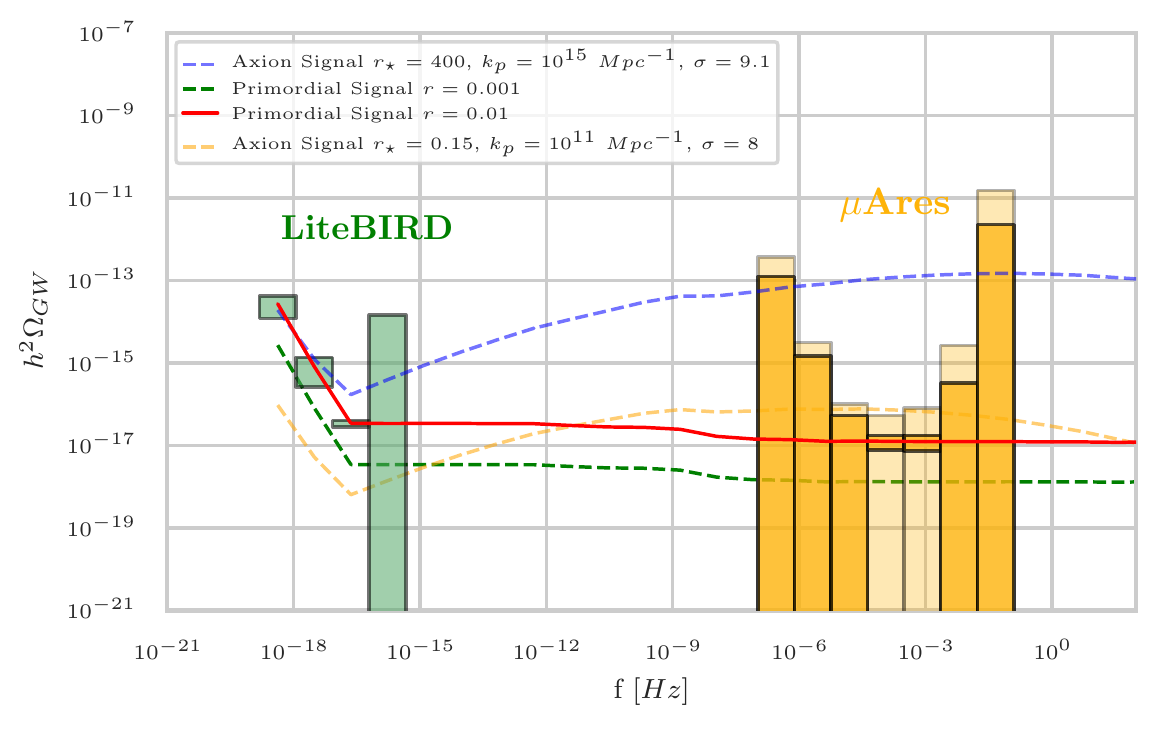}
        \caption{Same as Figure~\ref{fig:flat_DECIGO} but for the $\mu$Ares. Note that the error bars with foregrounds are computed assuming a multi-band cleaning with $\sigma_{fg} \sim 10^{-3}$.
        }\label{fig:flat_Ares}
\end{figure}

We consider now the two single-field slow-roll models with $r=0.01$ and $0.001$, with both $n_{T}$ and its running determined by the consistency relation. We choose a bin size of $\Delta\ln k = 2.0$. The model with the larger $r=0.01$ is easily detected by LiteBIRD in multiple bins. On interferometric scales, it can be detected in two bins -- or in one bin if pessimistically assuming no cleaning of the EGWD foreground -- by BBO (Figure~\ref{fig:flat_BBO}), while $\mu$Ares can detect this model only without foregrounds (Figure~\ref{fig:flat_Ares}), and DECIGO, in its standard design, cannot detect this model even without foregrounds (Figure \ref{fig:flat_DECIGO}). 


\begin{figure}
\centering
        \includegraphics[scale=1.1]{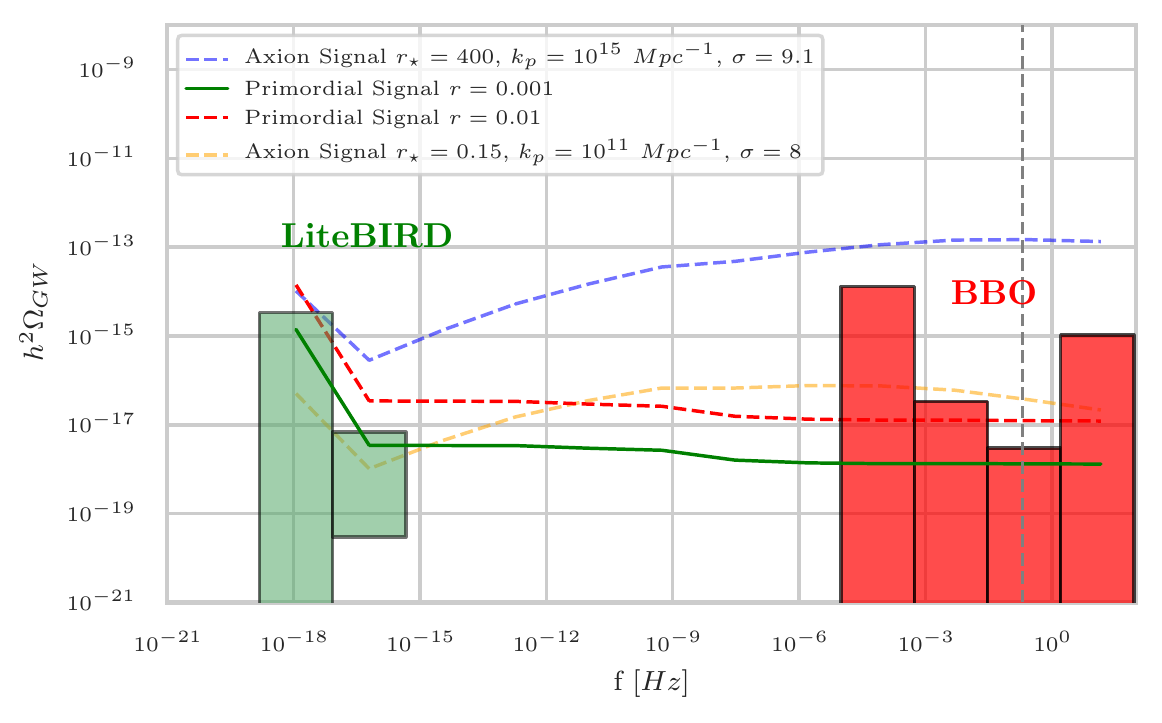}
        \caption{Same as Figure~\ref{fig:flat_BBO} but for $r=0.001$ with a logarithmic binning of $\Delta\ln k = 4.0$.
        }\label{fig:flat0001_BBO}
\end{figure}

To have a detection in at least one bin for the lower $r=0.001$, we increase  the binning scale to $\Delta\ln k = 4.0$. This model is detected by LiteBIRD on the CMB side, however  not even BBO, the most sensitive among the considered direct detection experiments, can detect it on the side of interferometers (Figure \ref{fig:flat0001_BBO}). The SNR remains smaller than 1 even considering a single bin enclosing the whole BBO band.

With the extremely high sensitivity of BBO at frequencies $\sim 16$ orders of magnitude larger than the CMB, ones creates a significant lever-arm, providing interesting constraints on the spectrum tilt $n_T$.
The path of multi-frequency measurements of the primordial tensor spectrum has been explored in the past, in the context of forecasts \citep[see for instance][for the combination of Planck/CMBPol and DECIGO/BBO\footnote{Note that in Ref. \citep{Smith_Peiris_Cooray_2006}, the specifications used for DECIGO reflect the so-called Ultimate DECIGO design -- an idealized instrument limited only by quantum noise \citep{Kudoh,Kuroyanagi:2014qaa} -- and are roughly a factor $\sim 5000$ more sensitive than the DECIGO design we adopt.}]{Smith_Peiris_Cooray_2006}, as well as of the analysis of available datasets, combining for instance the Planck, BICEP2/Keck, PPTA and LIGO data \citep{Meerburg_2015}, and adding SPTPol \citep{Lasky_2015} or COrE and other indirect constraints \citep{Cabass_2015} to the previous datasets. Notably, Ref.~\citep{Meerburg_2015} uses the original BICEP2 results, which have not been confirmed due to Galactic dust contamination, while the subsequent works \citep{Lasky_2015} and \citep{Cabass_2015} are based instead on the combined BICEP/Planck analysis, with the latter work more focused on smaller $r$ values detectable by future CMB missions.
     Furthermore, Ref.\citep{Liu:2015psa} explored the possibility of constraining early Universe physics -- in particular its equation of state, cosmic phase transitions and free-streaming neutrinos or other relativistic dark fluids -- focusing on the constraining power of PTA on significantly blue-tilted tensor spectra.  

Here, we update the forecasts on the tensor power spectrum amplitude $r$ and the tilt $n_{T}$ from the combination of CMB and laser interferometers, considering in particular, for the first time in the literature, the two configurations LiteBIRD+LISA and  LiteBIRD+BBO, focusing on testing the standard slow-roll scenario  and the inflationary consistency relation.
Differently from all the works cited above, we take into account foregrounds for all experiments. 
 We bin the LISA and BBO\footnote{We neglect the EGWD foreground for BBO in this Section, adopting the optimistic approach described in Section \ref{sec:fgs_strategy}.} sensitivity curves with $\Delta\ln k =2.0$ and $4.0$, respectively. We explore the full cosmological parameters space including \{$A_{S}$, $n_{S}$, $\tau$, $\Omega_{b}h^{2}$, $\Omega_{c}h^{2}$, $H_0$, $r$, $n_{T}$\} via the Monte Carlo Markov Chain (MCMC). We consider also another case in which we additionally fit for the running of the tensor spectral index $\alpha_{T}$, adding it to the list of explored parameters. We modify the \texttt{MontePython} MCMC package \citep{Audren:2012wb, Brinckmann:2018cvx} by adding a Gaussian likelihood for the interferometers \citep{Mandic_2012} 
\begin{equation}
    \mathcal{L}(\hat{\Omega}_{i}, \sigma_{i}|\Vec{\theta})\propto \exp\left[\frac{1}{2} \sum_{i}\frac{(\hat{\Omega}_{i}-\Omega_{M}(f_{i}; \Vec{\theta}))^{2}}{\sigma_{i}^{2}}\right],
\end{equation}
where $\Omega_{M}(f|\Vec{\theta})$ is the proposed model as a function of frequency $f$ and model parameters $\Vec{\theta}$, $\hat{\Omega}_{i}$ the fiducial model in the frequency bin $f_{i}$ and $\sigma_{i}^{2}$ its variance in the same bin. For the CMB, we adopt instead the standard Gaussian likelihood \citep{Perotto_2006}, with noise and foregrounds $C_{\ell}$ spectra determined from the LiteBIRD specifications (see Section \ref{sec:CMB_noise}).

\begin{figure}
\centering
        \includegraphics[scale=0.6]{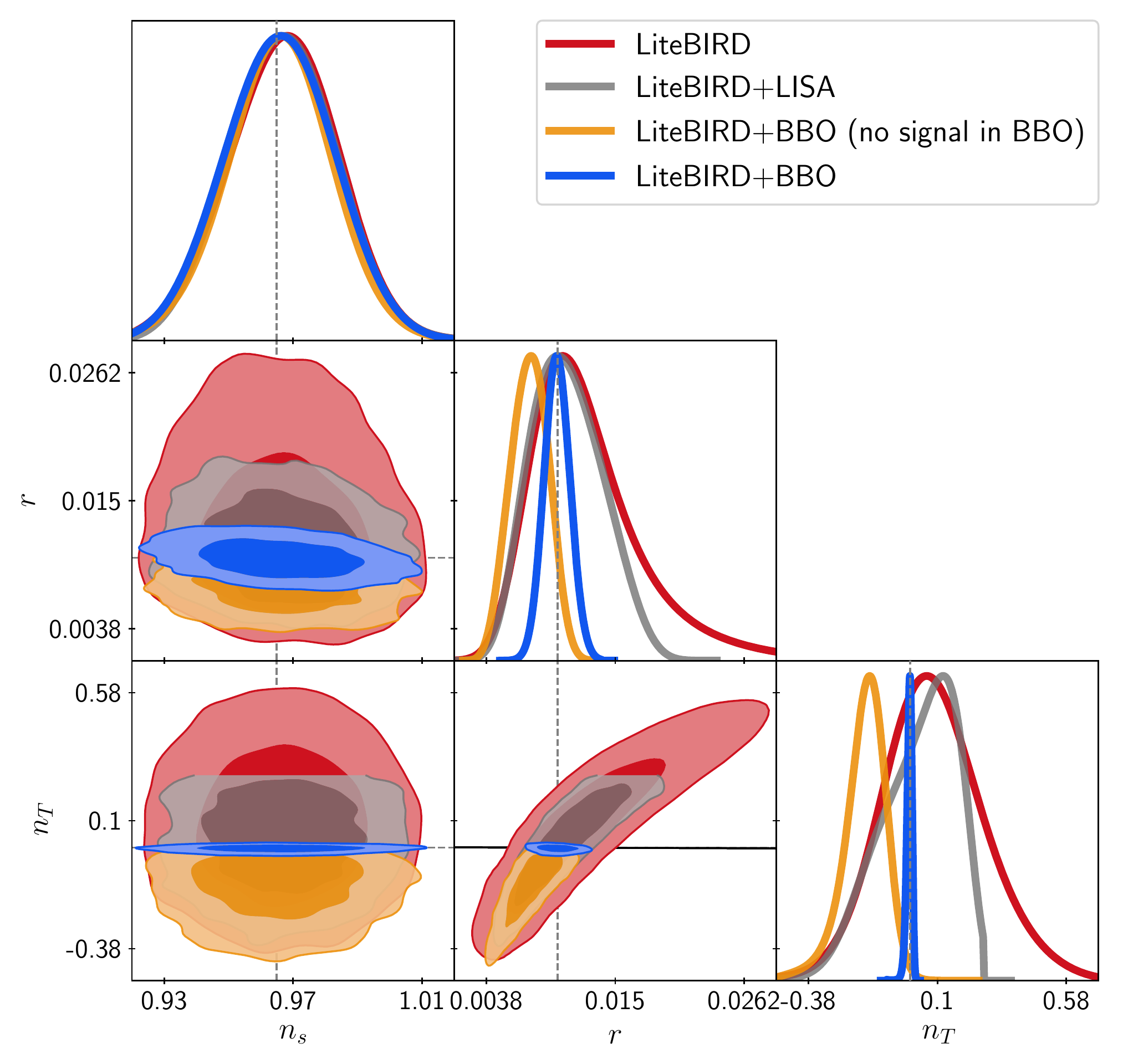}
        \caption{$1D$ and $2D$ marginal distributions of the $n_{S}$, $r$ and $n_{T}$ parameters for four possible observational configurations: LiteBIRD alone (red contours), LiteBIRD+LISA (grey), LiteBIRD+BBO (blue), and LiteBIRD+BBO assuming the fiducial signal in the LiteBIRD range but no signal in the BBO range (orange). The contours show the $68\%$ and $95\%$ CL, the dashed lines show the fiducial parameters, and the solid black line shows the consistency relation.}\label{fig:MCMC}
\end{figure}

We adopt for the fiducial model $r=0.01$, $n_T=-r/8$ and $\alpha_T = 0$ for the case without running; when fitting also for the running, we adopt as fiducial value $\alpha_{T} = (r/8) \left[(n_{S}-1)+ r/8 \right]$, given by the inflationary consistency relation. The values for the other cosmological parameters are taken from Ref.~\citep{planck_2018}. We show in Figure \ref{fig:MCMC} the $1D$ and $2D$ marginal distributions of the $n_{S}$, $r$ and $n_{T}$ parameters (without fitting for the running) for four possible observational configurations: $(i)$ constraints from LiteBIRD alone (red contours); $(ii)$ constraints from LiteBIRD and LISA (grey contours), $(iii)$ constraints from LiteBIRD and BBO (blue contours); and $(iv)$ constraints from LiteBIRD and BBO assuming the fiducial signal in the LiteBIRD range but no signal in the BBO range, that is $\hat{\Omega}_{i}=0$ in every bin $f_{i}$ (orange contours). This configuration is chosen to quantify possible deviations from the consistency relation in the eventuality of a detection at $\sim 5 \sigma$ by LiteBIRD, but no detection in BBO.

For $(i)$ we recover the following best-fitting parameters with  $1\sigma$ uncertainties:   $n_{S}=0.9665^{+0.017}_{-0.018}$,  $r= 0.013^{+0.003}_{-0.006}$ and $n_{T}= 0.09 ^{+0.18}_{-0.20}$; thus, a test of the consistency relation is out of discussion using the CMB alone: only extreme deviations from the consistency relation (e.g., axion-SU(2) models) can be detected in this case.

For $(ii)$ the addition of LISA impacts mainly the error on $n_T$ by limiting the range of allowed blue-tilted models, but this is still not enough to distinguish the consistency relation from the scale-invariant case. In this case the recovered parameters are  $n_{S}=0.9669 \pm 0.017$,  $r= 0.0109\pm 0.003$ and $n_{T}= 0.026 ^{+0.18}_{-0.10}$. The further inclusion of the ground-based interferometer ET  jointly with LISA does not improve significantly the constraints with respect to LISA alone because of the large foreground contamination affecting this experiment.

For $(iii)$ the effect of adding BBO is evident in Figure \ref{fig:MCMC}: the constraints on $r$ and $n_{T}$ become significantly tighter and also the maxima of the marginal distributions for the recovered parameters are very close to their fiducial values. Using the LiteBIRD+BBO configuration, we recover the following parameters:   $n_{S}=0.9649^{+0.016}_{-0.017}$,  $r= 0.0100\pm 0.0011$ and $n_{T}= -0.00125^{+0.011}_{-0.007}$. Also in this case, however, the error on the tensor spectral index, although remarkably smaller than the LiteBIRD only case, does not allow to distinguish the consistency relation from a scale-invariant case.

For $(iv)$ we recover   $n_{S}=0.96559^{+0.018}_{-0.016}$,  $r= 0.0076^{+0.0017}_{-0.0016}$ and $n_{T}= -0.15 ^{+0.12}_{-0.05}$. As it can be argued from Figure \ref{fig:MCMC}, the recovered tensor-to-scalar ratio shows a bias: this is because, to have an undetectable signal at interferometers scales, the spectrum must have a large red tilt, so large that it affects also the CMB scales. Therefore, even in the absence of a consistency relation detection, if we do not detect a signal in BBO, the red tilt in the power-law model of tensor power spectrum has to be so large that we can detect its departure from the single-field slow-roll consistency relation.

Finally, we try to additionally fit for the running $\alpha_T$ for the LiteBIRD+BBO configuration $(iii)$. In this case, we recover $n_{S}=0.9649^{+0.0184}_{-0.0168}$,  $r= 0.01^{+0.0021}_{-0.0054}$, $n_{T}= -0.00125 ^{+0.13}_{-0.19}$ and $\alpha_{T}= -0.00004 ^{+0.013}_{-0.008}$: the addition of this parameter degrades significantly the constraints on $r$ and $n_T$ because of the degeneracy introduced. Similarly, the weakening of the constraints due to the addition of the running does not allow the detection of a departure from the consistency relation as in configuration $(iv)$.

\section{Conclusions and prospects}\label{sec:conclusions}

We have calculated the sensitivities of CMB, PTA, and direct detection experiments for SGWB from the primordial GW across 21 decades in frequency. Not only do we provide the sensitivity curves for the GW energy density parameter $\Omega_{GW}$ (Figure~\ref{fig:binned_curves}) as commonly done in the literature, but also we provide the binned $1\sigma$ error bars on the model predictions for $\Omega_{GW}$ from two representative classes of sources of the primordial SGWB: the quantum vacuum fluctuation in the metric tensor (i.e., the homogeneous solution of Einstein's equation) from single-field slow-roll inflation models with $r=0.01$ and $r=0.001$ and the tensor tilt and its running given by the consistency relation, and the source-induced primordial GW from the spectator axion-SU(2) model (i.e., from the stress energy tensor in the right hand side of Einstein's equation). 

For CMB and PTA we considered the most ambitious future experiments LiteBIRD and SKA, respectively, while for direct detection experiments we considered a host of funded and proposed space (LISA, $\mu$Ares, DO, AEDGE, DECIGO, BBO) and ground-based (ET) GW observatories covering a wide range of frequencies from $10^{-7}$ to $10^3$~Hz. We took into account the instrumental noise, the response functions, and most importantly the contamination of the astrophysical foregrounds in the forecasts. We have presented all the details in our computation with homogeneous assumptions for all experiments in one place, which should provide convenient resources for the experiments in search of the primordial SGWB. 

We showed that it is possible to tune the axion-SU(2) model parameters to have detections with high significance in multiple frequency bins in both the CMB and space interferometers frequency ranges, even when accounting for the foreground contamination (Figures \ref{fig:axion1_LISA}-\ref{fig:axion1_Ares}), while remaining consistent with all current upper limits. We also showed that the parameters of the axion-SU(2) model can be chosen in such a way that the signal is out of reach for CMB experiments, while being detectable by the most sensitive space interferometers, i.e.,  $\mu$Ares, DECIGO (but only without the foregrounds) and BBO (Figures \ref{fig:axion2_Ares}-\ref{fig:axion2_BBO}).

On the other hand, the situation is different for future ground-based interferometers, for which the current estimates for the foreground contamination prevent detections of the axion-SU(2) model. 
It is also difficult to obtain a tensor spectrum detectable by the SKA experiment on PTA scales in presence of the foreground, while still complying with the BICEP2/Keck/Planck upper bound on CMB scales (Figure \ref{fig:axion3_SKA}). This is due to the attractor behaviour of the axion-SU(2) model, posing an upper limit on the width of the spectrum bump for a given peak scale $k_{p}$.

For what concerns the single-field slow-roll power spectrum, we showed that the $r=0.01$ model can be detected comfortably and simultaneously by LiteBIRD, by BBO (Figure \ref{fig:flat_BBO}), by $\mu$Ares (but only in the case without foregrounds, Figure \ref{fig:flat_Ares}), but not by DECIGO in its standard design (Figure \ref{fig:flat_DECIGO}).
We also found that the lower tensor-to-scalar ratio $r=0.001$ can be detected only by LiteBIRD, while not even the ultra-sensitive BBO can detect such a signal on the interferometers side (Figure \ref{fig:flat0001_BBO}).

Finally, we presented updated constraints on $r$ and $n_T$ combining LiteBIRD with LISA and LiteBIRD with BBO, to leverage on the scale dependence of the tensor spectrum. We conclude that distinguishing the single-field slow-roll consistency relation from the scale-invariant case remains out of reach even for LiteBIRD+BBO. However, if we detect tensors in the CMB but not in BBO, we would detect a significant deviation from the consistency relation in the context of the power-law primordial spectrum.

If the primordial SGWB is discovered during the next decade by ground-based CMB observatories or LiteBIRD, characterizing the power spectrum beyond the value of $r$ and testing chirality and Gaussianity would be of utmost importance for deciphering of the origin of the SGWB. If the discovered SGWB were found to be nearly scale-invariant, parity even and Gaussian, it would set a target for the $\mu$Ares, DECIGO and BBO to test the prediction of single-field slow-roll inflation models. On the other hand, if the SGWB were found to be blue-tilted, chiral or non-Gaussian, it would give excellent prospects for direct detection by LISA in the 2030s as well as by other proposed post-LISA direct detection experiments at any frequencies, opening up a new window to particle physics during inflation.


\acknowledgments
We thank E. Barausse for useful comments and discussion on the foregrounds treatment for interferometers.
We thank C. Berry, A. Sesana, and the AEDGE collaboration for providing us with the noise power spectra of DO, $\mu$Ares, and AEDGE, respectively. We also thank J. Errard for sharing the multi-resolution analysis of the foreground removal, the Joint Study Group of the LiteBIRD collaboration for useful discussions and the instrument specification given in Table~\ref{table:litebird}, and P. Adshead, E. Barausse, V. Domcke, O. Ozsoy, C. \"Unal, and I. Wolfson for comments on the manuscript.
PC thanks L. Boco, N. Krachmalnicoff, and T. Smith for useful discussions.
EK thanks SISSA and IFPU for hospitality, where this work was initiated, and A. Buonanno for useful discussion on the future direct detection mission proposals. This work has been supported by the network COSMOS by the Italian Space Agency (cosmosnet.net) and by the INDARK specific initiative of the National Institute of Nuclear Physics. The work of EK was supported in part by the Excellence Cluster ORIGINS which is funded by the Deutsche Forschungsgemeinschaft (DFG, German Research Foundation) under Germany's Excellence Strategy -- EXC - 2094 -- 390783311. We acknowledge the NERSC super-computing center in Berkeley and the Ulysses super-computer at SISSA for supporting numerical analyses in this work.

\appendix

\section{Interferometers Designs and Response Functions}\label{sec:response}
A necessary ingredient to compute the sensitivity curve of a GW direct SGWB experiment (Eq.~\ref{eq:binning}) is the \textit{overlap reduction function} $\mathcal{R}_{IJ}(f)$ of the detector pair $IJ$ (Eq.~\ref{eq:overlap}) \cite{Flanagan:1993ix}, which is computed  from the \textit{response function} $T_{I}(f,\hat{n})$ of each of the detector involved in the cross-correlation (Eq.~\ref{eq:s_i_fourier}).
We summarize here the formalism necessary to compute it, following Ref.~\citep{Smith_and_Caldwell_2016} to which we refer the reader for further details.

The overlap reduction function depends on the design of the detector and the combination of laser signals from the interferometer arms that we choose to form at the detector output. The response of space interferometers can also depend on time because of the orbital motion of the spacecrafts composing the detector; however, for simplicity we ignore this dependence. 

Let us start by considering the response of a single arm of the interferometer, from which we build the response of the full detector.  
The physical principle behind the detection of GWs in a laser interferometer is simple: the passage of GWs changes the proper distance between two freely moving test-masses at the opposite ends of an interferometer arm, causing phase-shifts in the laser beams which are traveling back-and-forth in each arm. It can be shown \citep{Smith_and_Caldwell_2016} that the phase change due to light traveling from the test-mass $i$ to the test-mass $j$ along a single interferometer arm is  
\begin{equation}
    \Delta\varphi_{ij}(t) = \int_{-\infty}^{+\infty}df\int d^{2}\hat{n} \sum_{P=+,\times} \Tilde{h}_{P}(f, \hat{n})e^{i2\pi f t_{i}} e_{ab}^{P}(\hat{n}) T^{ab}(\hat{l}_{ij}\cdot\hat{n},f),
\end{equation}
where  $L$ is the arm length, the test-masses $i$ and $j$ are located at $\Vec{x}_{i}$ and $\Vec{x}_{j}+L\hat{l}_{ij}$, respectively, $t_{i}$ is the time at which light left the mass $i$, $t$ is the time of arrival at the mass $j$ and $T^{ab}$ is the single-arm response function given by
\begin{align}
    T^{ab}(\hat{l}\cdot\hat{n},f) &= \hat{l}^{a}\hat{l}^{b}\, \mathcal{T} (\hat{l}\cdot\hat{n},f) \, e^{-i2\pi\hat{n}\cdot \Vec{x}_{i}}, \\
    \mathcal{T}(\hat{l}\cdot\hat{n},f) &=  \frac{1}{2} \sinc\left[\frac{f}{2f^{*}}(1-\hat{l}\cdot \hat{n})\right]e^{i\frac{f}{2f^{*}} \left(1-\hat{l}\cdot \hat{n}\right)},
\end{align}
where $f^{*}=1/(2\pi L)$. To measure the SGWB it is necessary to correlate the phase differences from different arms or paths around the interferometer. For example, we write the correlation between the $i\rightarrow j$ and the $k\rightarrow l$ paths as
\begin{equation}
    \langle \Delta\Tilde{\varphi}_{ij}(f) \Delta\Tilde{\varphi}^{*}_{kl}(f')\rangle = \frac{1}{2} \delta(f-f')\mathcal{R}_{ij,kl}(f)\mathcal{S}_{s}(f),
\end{equation}
where $\mathcal{R}_{ij,kl}$ is the overlap reduction function defined in Eq.~\ref{eq:overlap}, which we rewrite in this case as
\begin{equation}
    \mathcal{R}_{ij,kl}(f) = \int \frac{d^{2}\hat{n}}{4\pi} T^{ab}(\hat{l}_{ij}\cdot \hat{n}, f)\, T^{ab*}(\hat{l}_{kl}\cdot \hat{n}, f).
\end{equation}

\begin{figure}
\centering
        \includegraphics[scale=0.6]{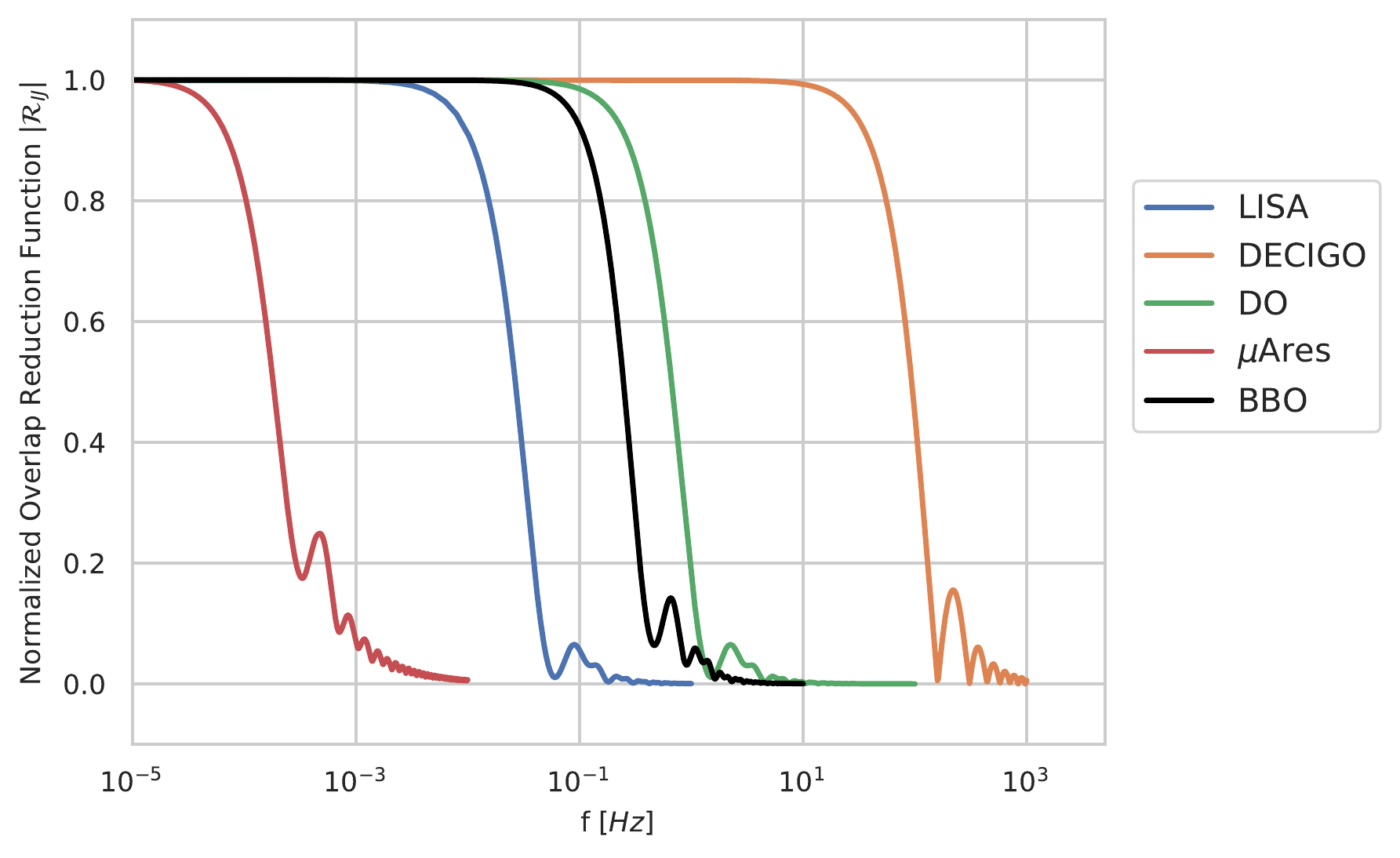}
        \caption{Absolute value of the overlap reduction functions $|\mathcal{R}_{IJ}|$ normalized to 1, computed for the interferometers LISA, DECIGO, DO, BBO, $\mu$Ares.}\label{fig:responses}
\end{figure}

To build the detector responses for the experiments we consider in this paper, we start from the simplest design adopted for the LISA mission.
The current proposal for LISA showcases three spacecrafts, each occupying a vertex $\Vec{x}_{i}$ with $i=A,B,C$ of an equilateral triangle $ABC$ of side $L=2.5\times10^{9}$ m; laser beams (six in total) travel back and forth along each of the triangle sides. 
We compute the response function for LISA using the standard Time-Delay Interferometry (TDI) signals. In this particular case \citep{LISAxcosmo}, the interferometer response function at the detector vertex $A$ reads 
\begin{align}\label{eq:michelson1}
    T^{ab}_{A\scalebox{.9}{$\scriptscriptstyle BC$}}(\hat{n},f) &= \frac{1}{2}e^{-i2\pi f \hat{n}\cdot\Vec{x}_{A}} \left[(\hat{l}_{AB}\otimes\hat{l}_{AB})\mathcal{T} (\hat{l}_{AB}\cdot\hat{n},f) - (\hat{l}_{AC}\otimes\hat{l}_{AC})\mathcal{T} (\hat{l}_{AC}\cdot\hat{n},f) \right],\\
    \mathcal{T}(\hat{l}\cdot\hat{n},f) &=  \frac{1}{2}W(f, f^{*}) \left( \sinc\left[\frac{f}{2f^{*}}(1-\hat{l}\cdot \hat{n})\right]e^{-i\frac{f}{2f^{*}}\left(3+\hat{l}\cdot \hat{n}\right)} \right. \nonumber\\ &\qquad\qquad\qquad\,\,\, +\left. \sinc\left[\frac{f}{2f^{*}}(1+\hat{l}\cdot \hat{n})\right]e^{-i\frac{f}{2f^{*}}\left(1+\hat{l}\cdot \hat{n}\right)} \right), \label{eq:michelson2}
\end{align}
where $W(f, f^{*})=1$ for the Michelson signals and $W(f, f^{*})=1-e^{-2if/f^{*}}$ for the TDI signals we are interested in. Specifically, the TDI $A$ and $E$ modes overlap reduction function\footnote{The three TDI signals are constructed by diagonalizing the signal covariance matrix and are named the $A$, $E$ and $T$ modes. Note that Eq.~\ref{eq:TDI_response} is valid only for the $A$ and $E$ TDI modes, which happen to be the most sensitive to the SGWB, while the $T$ mode is much less sensitive and is used instead to remove noise from the $A$ and $E$ modes \citep{LISAxcosmo}.}  for LISA (the blue curve in Figure \ref{fig:responses})  will be  
\begin{equation}\label{eq:TDI_response}
    \mathcal{R}_{A,E} = \mathcal{R}_{A\scalebox{.9}{$\scriptscriptstyle BC$},\, A\scalebox{.9}{$\scriptscriptstyle BC$}}-\mathcal{R}_{A\scalebox{.9}{$\scriptscriptstyle BC$},\, B\scalebox{.9}{$\scriptscriptstyle CA$}},
\end{equation}
where $\mathcal{R}_{A\scalebox{.9}{$\scriptscriptstyle BC$},\, A\scalebox{.9}{$\scriptscriptstyle BC$}}$ is the response for the auto-correlation at the vertex $A$ and $\mathcal{R}_{A\scalebox{.9}{$\scriptscriptstyle BC$},\, B\scalebox{.9}{$\scriptscriptstyle CA$}}$ is the one for the cross-correlation between the signals at the vertices $A$ and $B$ \citep{LISAxcosmo}. 

We use TDI signals to compute the overlap reduction function also for DO (green curve in Figure \ref{fig:responses}), which has been proposed as a LISA-like interferometer with shorter arms of lenght $L=10^{8}$ m. 

Differently from the LISA and DO detectors, BBO will feature six spacecrafts forming two independent triangular LISA-like interferometers $ABC$ and $A'B'C'$ with sides $L=5\times 10^{7}$ m. The two interferometers will be co-planar with one being rotated by $\SI{180}{\degree}$ with respect to the other, creating the so-called ``hexagram'' configuration.

In this case, it is convenient to introduce another signal combination that we can form from the Michelson signals $s_{mich,\scalebox{1.}{$\scriptscriptstyle A$}}(t)$ and $s_{mich,\scalebox{1.}{$\scriptscriptstyle C$}}(t)$ at the vertices $A$ and $C$ of one interferometer, respectively \citep{Smith_and_Caldwell_2016}
\begin{equation}
    s_{X}(t) = s_{mich,\scalebox{1.}{$\scriptscriptstyle A$}}(t) + 2s_{mich,\scalebox{1.}{$\scriptscriptstyle C$}}(t).
\end{equation}
The detector response function for the Michelson signal $s_{mich,\scalebox{1.}{$\scriptscriptstyle A$}}(t)$ at the vertex $A$ takes the form in Eq.~\ref{eq:michelson1}, while the one for the $s_{X}(t)$ signal combination is given by
\begin{equation}
       T^{ab}_{X}(\hat{n},f) = T^{ab}_{A\scalebox{.9}{$\scriptscriptstyle BC$}}(\hat{n},f) + 2T^{ab}_{C\scalebox{.9}{$\scriptscriptstyle AB$}}(\hat{n},f),
\end{equation}
and for both responses the transfer function $\mathcal{T}(\hat{l}\cdot\hat{n},f)$ is given by Eq.~\ref{eq:michelson2} with $W(f, f^{*})=1$. 

Now, to compute the overlap reduction function for the BBO hexagram  configuration, we cross-correlate the Michelson signal $s_{mich,\scalebox{1.}{$\scriptscriptstyle A$}}$ at the vertex $A$ on the interferometer $ABC$ and the combination $s_{X'}(t) = s_{mich,\scalebox{1.}{$\scriptscriptstyle A'$}}(t) + 2s_{mich,\scalebox{1.}{$\scriptscriptstyle C'$}}(t)$ on the other interferometer $A'B'C'$ \citep{Smith_and_Caldwell_2016} (the black curve in Figure \ref{fig:responses}). 
 As shown in \citep{Smith_and_Caldwell_2016}, it is convenient then to correlate the Michelson signal $s_{mich,\scalebox{1.}{$\scriptscriptstyle A$}}(t)$ with the signal combination $s_{X}^{'}(t)$, because the total noises for these two signals will be uncorrelated over the frequencies at which space-based interferometers are typically most sensitive.
 The final overlap reduction function for this signal combination \citep{Smith_and_Caldwell_2016} will be 
\begin{equation}\label{eq:do_resp}
    \mathcal{R}_{Hexagram} = \mathcal{R}_{A\scalebox{.9}{$\scriptscriptstyle BC$},\,A\scalebox{.9}{$\scriptscriptstyle BC$}} + \mathcal{R}_{X',X'} + 2\mathcal{R}_{A\scalebox{.9}{$\scriptscriptstyle BC$},\,X'}.
\end{equation}

The DECIGO design is similar to the BBO, with two independent triangular interferometers with arms $L=10^{6}$ m disposed in the hexagram configuration. Unlike BBO, however, the current DECIGO design envisages \textit{Fabry-P\'erot} (hereafter FP) interferometers; the response function at the vertex $A$ \citep{Kudoh} becomes therefore
\begin{equation}
    T^{ab}_{FP}(\hat{n},f) = \frac{1}{2}e^{-i2\pi f \hat{n}\cdot\Vec{x}_{A}} \left[(\hat{l}_{AB}\otimes\hat{l}_{AB}) - (\hat{l}_{AC}\otimes\hat{l}_{AC}) \right],
\end{equation}
and -- similarly to what we do for BBO -- we cross-correlate it with the response at the vertex $A'$ on the second interferometer, obtaining the overlap reduction function depicted in the orange curve in Figure \ref{fig:responses}.

The $\mu$Ares experiment will be composed, similarly to DECIGO and BBO, by two identical triangular LISA-like constellations with arms $L=430\times10^{9}$ m. However, in this case one of the two triangular interferometer would be trailing Mars orbit within the ecliptic plane while the other would be in the same orbit but $\SI{90}{\degree}$ tilted with respect to the ecliptic plane \citep{Ares_paper}. In order to compute the overlap reduction function for $\mu$Ares, we adopt again the same method employed for BBO, taking into account the design differences. We show the resulting curve in the purple line in Figure \ref{fig:responses}.

 Finally, we take into consideration the ET ground-based experiment. The current proposal consists of a network of three interferometers with arm opening of $60$ degrees, arranged in a such a way to form an equilateral triangle. For the ET experiment there is no need to compute the overlap reduction function, since the strain sensitivity curves (as defined in Eq.~\ref{eq:strain}) are publicly available\footnote{\url{http://www.et-gw.eu/index.php/etsensitivities}\label{footnote2}}.

\section{Interferometers Noise Models}\label{sec:interf_noise}
To compute the sensitivity curve in Eq.~\ref{eq:binning} we need not only the overlap reduction function, but also the \textit{noise power spectral density} $\mathcal{S}_{n}(f)$ for each detector (Eq.~\ref{eq:noise_psd}). Let us start from the LISA  mission. Following Ref.~\citep{LISAxcosmo}, we use the noise models reported in the LISA Science Requirements Document\footnote{\url{https://www.cosmos.esa.int/web/lisa/lisa-documents}}: the two main noise sources are acceleration noise and optical metrology noise, with spectra 
\begin{align}
    \mathcal{S}^{LISA}_{acc}(f) &= \frac{\left(\sqrt{(\delta a)^2}/L\right)^{2}}{(2\pi f)^{4}}\left(1+ (f_{1}/f)^{2}\right)\,\SI{}{\per\hertz}, \\
    \mathcal{S}^{LISA}_{opt} &= \left(\sqrt{(\delta x)^2}/L\right)^{2}\,\SI{}{\per\hertz},
\end{align}
where $\sqrt{(\delta a)^2}=3\times 10^{-15}\, \SI{}{\metre\per\square\second}$ and $\sqrt{(\delta x)^2}=1.5\times 10^{-11}\,\SI{}{\metre}$ are the rms amplitudes for acceleration  and optical metrology noise, respectively, and  $f_{1}=\SI{0.4}{\milli\hertz}$. The noise spectra for the TDI $A$ and $E$ signals that we used to compute the response function for LISA in Appendix \ref{sec:response} are 
\begin{equation}
    \mathcal{S}^{A,E}_{n}(f) = |W(f,f^{*})|^{2}\left[(4+2\cos(f/f^{*}))\mathcal{S}^{LISA}_{opt}+8(1+\cos(f/f^{*}))+\cos^{2}(f/f^{*})\mathcal{S}^{LISA}_{acc}(f)\right].
\end{equation}
Combining the $A$ and $E$ modes, we reduce the noise power by a factor $\sqrt{2}$ to obtain
\citep{LISAxcosmo}
\begin{equation}
\mathcal{S}^{LISA}_{h}= \left[\left(\frac{\mathcal{R}_{A}}{\mathcal{S}^{A}_{n}}\right)^{2}+\left(\frac{\mathcal{R}_{E}}{\mathcal{S}^{E}_{n}}\right)^{2} \right]^{-1/2}.
\end{equation}
For BBO \citep{BBO_paper} we use
\begin{align}
    \mathcal{S}^{BBO}_{acc}(f) &= 2.3\times 10^{-52}(\SI{1}{\hertz}/f)^{4}\,\SI{}{\per\hertz}, \\
    \mathcal{S}^{BBO}_{opt} &= 8\times 10^{-50}\,\SI{}{\per\hertz},
\end{align}
and the noise model for one of the two identical triangular interferometers proposed in \citep{Smith_and_Caldwell_2016} 
\begin{equation}
    S^{BBO}_{n} = \frac{5}{2}\left[\mathcal{S}^{BBO}_{opt}(f)+2\mathcal{S}^{BBO}_{acc}(f)(1+\cos^{2}(f/f^{*}))\right].
\end{equation}
For DECIGO we use the noise model \citep{DECIGO_paper}:
\begin{equation}
    \mathcal{S}_{n}^{DECIGO}=\mathcal{S}^{DECIGO}_{shot}(f)+\mathcal{S}^{DECIGO}_{rad}(f) + \mathcal{S}^{DECIGO}_{acc}(f),
\end{equation}
with shot noise, radiation pressure noise and acceleration noise given by
\begin{align}
    \mathcal{S}^{DECIGO}_{shot}(f) &= \frac{\hbar  \pi\lambda}{P_{eff}}\left(\frac{1}{4fL}\right)^{2}\left[1+\left(\frac{f}{f^{*}}\right)^{2}  \right],\\
    \mathcal{S}^{DECIGO}_{rad}(f) &= \frac{\hbar P}{ \pi\lambda}\left(\frac{16F}{ML}\right)^{2}\left(\frac{1}{2\pi f}\right)^{4}\left[1+\left(\frac{f}{f^{*}}\right)^{2}  \right]^{-1},\\
    \mathcal{S}^{DECIGO}_{acc}(f) &= \frac{\hbar P}{ \pi\lambda}\left(\frac{16F}{3ML}\right)^{2}\left(\frac{1}{2\pi f}\right)^{4},
\end{align}
where $P=\SI{10}{\watt}$ is the laser output power, $\lambda=\SI{532}{\nano\metre}$ is the laser wavelenght, $M=\SI{100}{\kilo\gram}$ is the mirror mass, $R=\SI{0.5}{\meter}$ is the mirror radius, $F=10.18$ is the FP cavity finesse and $P_{eff}=\SI{6.68}{\watt}$ is the effective laser output power.

For DO we use the noise curves shown in Ref.~\citep{DO_paper} and kindly provided by Christopher Berry.
Also for $\mu$Ares we use the noise curves kindly provided by Alberto Sesana, as shown in Ref.~\citep{Ares_paper}.
For AEDGE we use the strain sensitivity curve shown in Ref.~\citep{AEDGE_paper} and kindly provided by the AEDGE collaboration. For ET we use the strain sensitivity curve available from Ref.~\citep{ET_paper} (see also website in footnote \ref{footnote2}).

\bibliographystyle{JHEP}
\bibliography{main}

\end{document}